\documentclass[11pt]{article}
\usepackage{jheppub}

\usepackage{epsfig}
\usepackage{amssymb}
\usepackage{amsmath}
\usepackage{multirow}
\usepackage{adjustbox}
\usepackage{tabularx}
\usepackage{longtable}
\usepackage{arydshln}

\usepackage{graphicx, subfigure, array, placeins, float}
\usepackage{epstopdf}
\usepackage{extarrows}

\usepackage{dsfont,bbm}
\usepackage{youngtab}

\newcommand{\evPr}{\mathcal{E}}  
\newcommand{\phPr}{\mathcal{P}}  
\newcommand{\gePr}{\mathcal{K}}  
\newcommand{\evfive}{\mathcal{U}} 
\newcommand{\kineF}{\mathfrak{F}}
\newcommand{\einv}{\omega}
\newcommand{\imI}{\mathbbm{i}}

\makeatletter \@addtoreset{equation}{section} \makeatother

\newcommand{\be}{\begin{equation}}
\newcommand{\ee}{\end{equation}}
\newcommand{\bea}{\begin{eqnarray}}
\newcommand{\eea}{\end{eqnarray}}

\allowdisplaybreaks[4]
\graphicspath{{./figs/}}

\title{Gluonic evanescent operators: classification and one-loop renormalization}

\author[a]{Qingjun Jin,}
\emailAdd{qjin@gscaep.ac.cn}
\author[b]{Ke Ren,}
\emailAdd{renke@itp.ac.cn}
\author[b,c,d]{and Gang Yang}
\emailAdd{yangg@itp.ac.cn}
\author[b]{Rui Yu,}
\emailAdd{yurui@itp.ac.cn}
\affiliation[a]{Graduate School of China Academy of Engineering Physics, No.~10 Xibeiwang East Road, Haidian District, Beijing, 100193, China}
\affiliation[b]{CAS Key Laboratory of Theoretical Physics, Institute of Theoretical Physics, \\Chinese Academy of Sciences, Beijing 100190, China}
\affiliation[c]{School of Fundamental Physics and Mathematical Sciences, Hangzhou Institute for Advanced Study, UCAS, Hangzhou 310024, China}
\affiliation[d]{International Centre for Theoretical Physics Asia-Pacific, Beijing/Hangzhou, China}

\abstract{
Evanescent operators are a special class of operators that vanish classically in four-dimensional spacetime,
while in general dimensions they are non-zero and are expected to have non-trivial physical effects at the quantum loop level in dimensional regularization.
In this paper we initiate the study of evanescent operators in pure Yang-Mills theory.
We develop a systematic method for classifying and constructing the $d$-dimensional Lorentz invariant evanescent operators,
which start to appear at mass dimension ten.
We also compute one-loop form factors for the dimension-ten operators
via the $d$-dimensional unitarity method  and obtain their one-loop anomalous dimensions.
These operators are necessary ingredients in the study of high dimensional operators in effective field theories involving a Yang-Mills sector.
}

\begin{document}

\maketitle

\setcounter{footnote}{0}

\renewcommand{\labelenumii}{\theenumii}
\renewcommand{\theenumii}{\theenumi.\alph{enumii}.}


\section{Introduction}

Gauge invariant local operators are important quantities in quantum field theories.
For example, composite local operators give the interaction vertices in effective Lagrangians and thus are central ingredients in the study of effective field theory (EFT), see \emph{e.g.}~\cite{Manohar:2018aog}.
In QCD, hadron states correspond to color-singlet operators composed of gluon and quark fields.
Similar gauge-invariant operators and their anomalous dimensions have been studied extensively in ${\cal N}=4$ SYM and played an important role in understanding the AdS/CFT correspondence and integrability \cite{Beisert:2010jr}.
When studying the renormalization of operators, the dimensional regularization scheme is often one of the most convenient choices to regularize the divergences \cite{tHooft:1972tcz}.
In this case, the spacetime dimension is generalized to $d=4-2\epsilon$ dimensions,
and a special subtlety appears that it is possible to construct a series of operators
that vanish in four dimensions but not in $d$ dimensions, which are known as evanescent operators.

The study of evanescent operators has been considered long time ago in the context of
four fermion interactions \cite{Buras:1989xd, Dugan:1990df, Herrlich:1994kh, Buras:1998raa}.
In general $d$-dimensional spacetime,
there are infinitely many operators involving anti-symmetric tensor of Dirac matrices $\gamma^\mu$ with ranks higher than 4:
\begin{equation}
\label{eq:4fermion}
{\cal O}_{\textrm{4-ferm}}^{(n)} = \bar\psi \gamma^{[\mu_1}\ldots\gamma^{\mu_n]} \psi \bar\psi \gamma_{[\mu_1}\ldots\gamma_{\mu_n]} \psi \,, \qquad n \geq 5\,.
\end{equation}
Such operators vanish in strictly four-dimensional spacetime but not in $d$ dimensions.
Similar evanescent operators also appear in the two-dimensional four-fermion models, see \emph{e.g.}~\cite{Bondi:1989nq,Vasiliev:1997sk, Gracey:2016mio}.
It was shown in those studies that these evanescent operators can not be ignored, as they affect the anomalous dimensions of physical operators.
In the scalar $\phi^4$ theory, the high-dimensional  evanescent operators were considered in \cite{Hogervorst:2015akt} and it was found that the evanescent operators can give rise to negative-norm states implying that the theory is not unitary in non-integer spacetime dimensions.
Counting scalar evanescent operators via Hilbert series was also considered in \cite{Cao:2021cdt}.

In this paper, we consider a new class of evanescent operators in the pure Yang-Mills theory,
which are composed of field strength $F_{\mu\nu}$ and covariant derivatives $D_\mu$.
A simple example of such operators can be given as
\begin{equation}
\label{eq:example-op}
\mathcal{O}_{\mathrm{e}} = \frac{1}{16}\delta^{\mu_1\mu_2\mu_3\mu_4\mu_5}_{\nu_1\,\nu_2 \, \nu_3 \, \nu_4 \,\nu_5}
{\rm tr}(D_{\nu_5}F_{\mu_1\mu_2} F_{\mu_3\mu_4} D_{\mu_5}F_{\nu_1\nu_2}
F_{\nu_3\nu_4})\,,
\end{equation}
where $\delta^{\mu_1..\mu_n}_{\nu_1...\nu_n}= {\rm det}(\delta^\mu_\nu)$ is the generalized Kronecker symbol
(see Section~\ref{sec:gluonicEvaOpe} for detail).
This operator is zero in four dimensions but has non-trivial matrix elements such as form factors in general $d$ dimensions. For example, its (color-ordered) minimal tree-level form factor can be given as
\begin{equation}
\label{eq:example-op2}
\mathcal{F}^{(0)}_{\mathcal{O}_{\mathrm{e} }}(1,2,3,4)
=2\delta^{e_1 e_2 p_1 p_2 p_3}_{e_3 e_4  p_3 p_4 p_1}
+2\delta^{e_1 e_4 p_1 p_4 p_2}_{e_2 e_3 p_2 p_3 p_4}
\,,
\end{equation}
which is a non-trivial function of Lorentz product of momenta and polarization vectors in $d$ dimensions.
The main goal of this paper is to study the classification of such operators and their one-loop renormalization.

Unlike the four fermion operators in \eqref{eq:4fermion}, due to the insertion of covariant derivatives and different ways of Lorentz contractions, the gluonic evanescent operators like \eqref{eq:example-op} exhibit richer structures. Moreover, at a given mass dimension,
the number of all the possible Lorentz contraction structures is finite, which means
that the gluonic evanescent operators are also finite, calling for a
systematic way to construct their independent basis.
To classify these operators, it will be convenient to apply the correspondence between local operators and form factors \cite{Zwiebel:2011bx, Wilhelm:2014qua}.
The main advantage is that form factors are on-shell matrix elements, thus the constraints from the equation of motion and Bianchi identities can be taken into account automatically, see \emph{e.g.}~\cite{Jin:2020pwh}.
Here, due to the special nature of evanescent operators, the usual spinor helicity formalism will be insufficient.
Instead, one needs to consider form factors consisting of $d$-dimensional Lorentz vectors (\emph{i.e.}~external momenta and polarization vectors)  such as in \eqref{eq:example-op2}.
Since the Yang-Mills operators contain non-trivial color factors, the form factor expressions provide also a useful framework to organize the color structures.
One can first classify function basis at the form factor level and then map back to basis operators.
We will apply a strategy to construct the basis evanescent operators along this line.

To study the quantum effect of evanescent operators, we perform one-loop
computation of their form factors.
The calculation is based on the unitarity method \cite{Bern:1994zx, Bern:1994cg} in $d$ dimensions.
Using the form factor results, we can study their renormalization and operator-mixing behaviors.
We provide explicit results of the one-loop renormalization matrices and the anomalous dimensions for the dimension-ten basis operators.
These one-loop results will be necessary ingredients for the two-loop renormalization of physical operators.

This paper is organized as follows.
In Section~\ref{sec:gluonicEvaOpe}, we first give the definition of evanescent operators and
then describe the systematic construction of the operator basis.
In Section~\ref{sec:oneloop} we first explain the  one-loop computation of full-color form factors using the unitarity method,
then we discuss the renormalization and obtain the anomalous dimensions of
the complete set of evanescent operators with dimension 10.
A summary and discussion are given in Section~\ref{sec:summary} followed by a series of appendices.
Several technique details in the operator construction are given in \ref{app:primi}-\ref{app:total-derivative}.
The basis of dimension-12 length-4 evanescent operators is given in Appendix \ref{app:eva-dim12}.
Efficient rules for calculating compact tree-level form factors are given in Appendix~\ref{app:tree-rule}.
The color-decomposition and one-loop infrared structure are discussed in  Appendix~\ref{app:colordecom}-\ref{app:IR}.
Finally, the full basis of dimension-10 physical operators as well as their one-loop renormalization are given in Appendix \ref{app:nevtoeva}.


\section{Gluonic evanescent operators}
\label{sec:gluonicEvaOpe}

In this section, we explain the classification and the basis construction for evanescent operators.
We first set up the conventions in Section~\ref{sec:setup} and then give the definition the evanescent operators in Section~\ref{sec:eva-define}.
Then we discuss the systematic construction of evanescent basis operators in Section \ref{sec:eva-cons}.
To be concrete and for simplicity,
our discussion will focus on the length-4 case, and the generalization to
high length cases will be given in Section~\ref{sec:len5}.
The full set of dim-10 evanescent operators including length-4 and length-5 ones
are summarized in Section~\ref{sec:dim10full}.

\subsection{Setup}
\label{sec:setup}

In this paper we consider the gauge invariant Lorentz scalar
operators in $\mathrm{SU}(N_c)$ pure Yang-Mills theory,
which are composed of field strength $F_{\mu\nu}$ and covariant derivatives $D_\mu$.
The field strength carries a color index as $F_{\mu\nu} = F^a_{\mu\nu} T^a$,
where $T^a$ are the generators of gauge group satisfying $[T^a, T^b] = \imI f^{abc} T^c$,
and the covariant derivative acts as
\begin{equation}
\label{eq:covDerivDef}
D_\mu \  \diamond = \partial_\mu \diamond + \imI g [A_\mu, \diamond \ ]\,,
\qquad [D_\mu, D_\nu] \ \diamond = -\imI g [F_{\mu\nu}, \diamond \ ] \,.
\end{equation}
A gauge invariant scalar operator can be given as
\begin{equation}
\label{eq:generalDF}
{\cal O}(x) \sim c(a_1, \ldots ,a_n) {\cal W}_{m_1}^{a_1} {\cal W}_{m_2}^{a_2} \ldots {\cal W}_{m_L}^{a_L} \,, \qquad
\textrm{with} \ \ {\cal W}_{m_i}^{a_i} = (D_{\mu_{i_1}}...D_{\mu_{i_{m_i}}}F_{\nu_i \rho_i})^{a_i} \,,
\end{equation}
where $c(a_1,...,a_n)$ is the color factor, and it can be written as products of
traces $\mathrm{Tr}(...T^{a_i}...T^{a_j}...)$.
Lorentz indices $\{\mu_i,\nu_i,\rho_i\}$ are contracted among different $\mathcal{W}_{m_i}^{a_i}$.%
\footnote{For simplicity, in this paper we will not distinguish upper and lower Lorentz indices.
For example, $\eta^{\mu\nu}$ and $\delta^{\mu}_\nu$ are regarded as equivalent. This will not cause any problem in flat spacetime.}
In this paper we focus on $d$-dim Lorentz covariant operators, so Lorentz indices in (\ref{eq:generalDF})
can only be contracted through metric.%
\footnote{ Since Levi-Civita tensor $\varepsilon_{\mu\nu\rho\sigma}$
 breaks $d$-dimensional covariance, we do not consider it in this paper and leave it in the future work.
 Such tensor corresponds to $P$-odd operators.}

For the convenience of classifying operators,
we define the \emph{length} of an operator $O$ as
the number of field strength $F$ (or equivalently, ${\cal W}$) in it.
For example, the length of the operator in (\ref{eq:generalDF}) is $L$.
We will classify operators according to a length hierarchy by setting
two operators to be equivalent if their difference can be written in terms of high length operators.
In other words,
if two operators of length $l$ differ by an
operator of higher length $L>l$:
\begin{align}
O_{l}-O'_{l}=O_{L>l}\,,
\end{align}
we say $O_l$ and $O'_l$ are equivalent at
the level of length $l$, and only one of them is kept in length-$l$ operator basis.
Following this length hierarchy, we first construct operator basis with lower length, then the higher length.
Note the commutator of two covariant derivatives produces a field strength $F$ as in \eqref{eq:covDerivDef},
therefore if we exchange the orders of two covariant derivatives $D$
lying in the same ${\cal W}$,
the newly obtained operator is equivalent to the original one up to a higher length operator:
\begin{equation}
\label{eq:orderDD}
\mathrm{tr}(D_{\mu_1} D_{\mu_2} ... D_{\mu_n} F_{\rho\sigma}...)
= \mathrm{tr}(D_{\mu_2} D_{\mu_1} ...  D_{\mu_n} F_{\rho\sigma}...) +  \textrm{(higher length operator)}\,.
\end{equation}

We introduce an important quantity, the
\emph{kinematic operator},  obtained by stripping off the color factor in the full operator,
which is a noncommutative product of $\mathcal{W}_i$.
We denote the kinematic operator of length-$L$ by
\begin{equation}
\lfloor \mathcal{W}_1, \mathcal{W}_2, \mathcal{W}_3, \ldots \mathcal{W}_L \rfloor \,.
\end{equation}
From a kinematic operator, one can create a full operator by dressing a trace color factor to it, denoted by ${\cal T}\circ\lfloor \ldots \rfloor$, for example,
\begin{equation}
\begin{aligned}
\label{eq:dress}
&\mathrm{tr}(T^{a_1}T^{a_3} T^{a_2} T^{a_4})\circ\lfloor \mathcal{W}_1,\mathcal{W}_2,\mathcal{W}_3,\mathcal{W}_4\rfloor
=\mathrm{tr}(\mathcal{W}_1\mathcal{W}_3\mathcal{W}_2\mathcal{W}_4),
 \\
&\mathrm{tr}(T^{a_1} T^{a_4})\mathrm(T^{a_2} T^{a_3})\circ\lfloor \mathcal{W}_1,\mathcal{W}_2,\mathcal{W}_3,\mathcal{W}_4\rfloor
=\mathrm{tr}(\mathcal{W}_1\mathcal{W}_4)\mathrm{tr}(\mathcal{W}_2\mathcal{W}_3)\,.
\end{aligned}
\end{equation}
We will often use the short notation $\mathrm{tr}(i j \cdots k) := \mathrm{tr}(T^{a_i}T^{a_j}\cdots T^{a_k})$ for the trace color factor.

For high dimensional composite operators, there usually exist
 a number of operators with the same canonical dimensions and quantum numbers.
Operators at a given dimension are in general not independent with each other,
for they can be related with each other through equations of motion (EoM) or Bianchi identities (BI):
\begin{align}
\label{eq:EoM}
\text{EoM}: \qquad  & D_\mu F^{\mu\nu} = 0 \,, \\
\text{BI}: \qquad  & D_\mu F_{\nu\rho}+D_\nu F_{\rho\mu}+D_\rho F_{\mu\nu} = 0 \,.
\end{align}
The redundancy caused by these operator relations makes the operator
counting    complicated.
To cure this problem, it is convenient to map operators to form factors, which are on-shell matrix elements.
For the kinematic operators, one can introduce an order-preserved mapping
between a kinematic operator and a polynomial composed of polarization vectors $\{e_i\}$ and
external momenta $\{p_i\}$:
\begin{equation}
\label{eq:OFmap-d-dim}
\lfloor \mathcal{W}_1, \mathcal{W}_2, \ldots,
\mathcal{W}_L
\rfloor\rightarrow
\kineF\big(
\lfloor \mathcal{W}_1, \mathcal{W}_2, \ldots,
\mathcal{W}_L
\rfloor
\big).
\end{equation}
The mapping images $\kineF\big(
\lfloor \mathcal{W}_1, \mathcal{W}_2, \ldots,
\mathcal{W}_L \rfloor\big)$ will be named as \emph{kinematic form factors}.

The rule of the map in general $d$-dimensional kinematics is
\begin{equation}
\label{eq:OFmap-d-dim2}
\kineF_{d\textrm{-dim}}(\mathcal{W}_i) :\quad
D^\mu \rightarrow \imI p_i^\mu,\quad
F^{\mu\nu}  \rightarrow
\imI (p_i^\mu e_i^\nu - p_i^\nu e_i^\mu )\,,
\end{equation}
and the kinematic form factor is given by the product $\prod_{i=1}^L \kineF_{d\textrm{-dim}}(\mathcal{W}_i)$.
It is easy to see the operator relations like EoM and BI automatically vanish
after the mapping (\ref{eq:OFmap-d-dim2}).
As a comparison, if we restrict to four-dimensional spacetime, it is usually convenient to decompose the field strength into self-dual and anti-self-dual components:
\begin{align}
\label{eq:Ftof}
F^{\mu\nu}\rightarrow F^{\alpha\dot{\alpha}\beta\dot{\beta}}
=\epsilon^{\alpha\beta}\bar{F}^{\dot{\alpha}\dot{\beta}}
+\epsilon^{\dot{\alpha}\dot{\beta}}F^{\alpha\beta}\,,
\end{align}
and in this case one can use the spinor helicity formalism to map $D$ and $F$
as \cite{Beisert:2010jq, Zwiebel:2011bx, Wilhelm:2014qua}:
\begin{equation}
\label{eq:OFmap-4-dim}
\kineF_{4\textrm{-dim}}(\mathcal{W}_i): \quad
D^{\dot{\alpha}\alpha}
\rightarrow \tilde{\lambda}_i^{\dot{\alpha}}\lambda_i^{\alpha}, \quad
F^{\alpha\beta} \rightarrow \lambda_i^\alpha\lambda_i^\beta \, \quad
\bar{F}^{\dot{\alpha}\dot{\beta}} \rightarrow
\tilde{\lambda}_i^{\dot{\alpha}}\tilde{\lambda}_i^{\dot{\beta}} \,.
\end{equation}
Similarly, EoM and BI (equivalent to Schouten identities) are also encoded in the map (\ref{eq:OFmap-4-dim}).
As a concrete example for these rules, for the kinematic operator is $\lfloor  F_{\mu\nu} , F^{\mu\nu} \rfloor$,
one can obtain its kinematic form factors in $D$-dim and 4-dim  respectively as
\begin{equation}
\begin{aligned}
&\kineF_{d\textrm{-dim}} = (p_{1\mu} e_{1\nu} - p_{1\nu} e_{1\mu}) (p_2^\mu e_2^\nu - p_2^\nu e_2^\mu) \,, \qquad
\kineF_{4\textrm{-dim}} =
\left
\{
\begin{array}{lcl}
 \langle 12\rangle^2\,,    &      & (-,-)  \\[0.3em]
\, {[12]} \,,    &      & (+,+)
\end{array}
\right.
\,,
\end{aligned}
\end{equation}
where $(\pm,\pm)$ label the helicities of external gluons in the four-dimension case.
To study the evanescent operators we will mostly use the $d$-dimensional rules.

The kinematic form factor is not yet a physical form factor for a full operator.
A physical form factor is defined as
a matrix element between an operator ${\cal O}(x)$ and $n$ on-shell states (see \emph{e.g.}~\cite{Yang:2019vag} for an introduction):
\begin{equation}
\mathbf{F}_{{\cal O},n} = \int d^4 x \, e^{-\imI q\cdot x} \langle 1, \ldots, n | {\cal O}(x) |0 \rangle \,.
\end{equation}
The simplest type of physical form factors are the so-called  \emph{minimal} tree-level form factors,
for which the number of external gluons is equal to the length of the operator.
One can establish a useful map between a length-$L$ operator and its tree-level minimal form factor as
\begin{equation}
\label{eq:1-to-1}
{\cal O}_L  \ \Leftrightarrow \ {\cal F}^{(0)}_{{\cal O}_L, L}(1,  \ldots, L) \,.
\end{equation}
To map a full operator to its minimal form factor, one first strips off the color factor
and reads the kinematic form factor from its kinematic operator,
and then takes cyclic symmetrization which is caused by trace factors.
As an example, consider the length-4 operator ${\cal O}_4 = {\rm tr}( {\cal W}_1 {\cal W}_2 {\cal W}_3 {\cal W}_4 )$.
One can first strip off the trace factor to obtain the kinematic operator $\lfloor {\cal W}_1, {\cal W}_2, {\cal W}_3, {\cal W}_4  \rfloor$,
then apply the above rules to obtain its kinematic form factor $\kineF(\lfloor {\cal W}_1, {\cal W}_2, {\cal W}_3, {\cal W}_4 \rfloor)$.
The color-ordered minimal form factor (with color factor ${\rm tr}(1234)$) is the cyclic symmetrization of kinematic form factor:
\begin{equation}
{\cal F}^{(0)}_{{\cal O}_4, 4}(1, 2,3,4) = \kineF(\lfloor {\cal W}_1, {\cal W}_2, {\cal W}_3, {\cal W}_4 \rfloor) +  \textrm{cyclic-perm.}(1,2,3,4) \,,
\end{equation}
and the full-color minimal form factor can be obtained as
\begin{equation}
\label{eq:fullcolorminiFF}
F^{(0)}_{{\cal O}_4, 4} = \sum_{{\sigma}\in S_3}
\mathrm{tr}\big(1{\sigma}(2)
{\sigma}(3){\sigma}(4)\big)
\mathcal{F}^{(0)}_{{\cal O}_4, 4} \big(
1,{{\sigma}(2)},{{\sigma}(3)},
{{\sigma}(4)}
\big) \,.
\end{equation}
We stress that for the gluonic operators considered in this paper,
the \emph{full-color} minimal form factor are invariant under full $S_L$ transformation.

Besides  the minimal form factor, there are higher point form factors
with the number $E$  of external gluons  larger than the length $L$ of the operator.
The next-to-minimal and next-next-to-minimal form factors correspond to
$E=L+1$ and $E=L+2$ respectively.

\subsection{Definition of evanescent operators}
\label{sec:eva-define}

Given the above preparation, we now introduce  evanescent operators.
An operator is called an \emph{evanescent} operator, if the tree-level matrix elements of this operator have non-trivial results
 in general $d$ dimensions but all vanish in four dimensions.
In terms of form factors, we can give a more practical definition:
for an evanescent operator ${\cal O}^{\rm e}_L$ of length-$L$, its tree-level form factors with arbitrary numbers of external on-shell states, are all zero in four dimensions, but it has a non-trivial minimal form factor in general $d$ dimensions, namely
\begin{equation}
\mathbf{F}^{(0)}_{{\cal O}^{\rm e}_L, n \geq L} \big|_{4\textrm{-dim}} = 0 \,, \qquad \mathbf{F}^{(0)}_{{\cal O}^{\rm e}_L, L} \big|_{d\textrm{-dim}} \neq 0 \,.
\end{equation}
Here we would like to emphasize that the vanishing of minimal form factors in four dimensions is not enough to fully characterize the property
of an evanescent operator, and its higher-point non-minimal form factors are also required to vanish in four dimensions.
If an operator is not an evanescent operator, \emph{i.e.}~its form factors
do not vanish in four dimensions, we call it a physical operator.

Let us review the example of evanescent operator mentioned in the introduction (\ref{eq:example-op}).
Using the map \eqref{eq:OFmap-d-dim}, one can obtain its color-ordered minimal form factor as \eqref{eq:example-op2},
which we reproduce here
\begin{align}
\mathcal{F}^{(0)}_{\mathcal{O}_{\mathrm{e} }}(1,2,3,4)
=2\delta^{e_1 e_2 p_1 p_2 p_3}_{e_3 e_4  p_3 p_4 p_1}
+2\delta^{e_1 e_4 p_1 p_4 p_2}_{e_2 e_3 p_2 p_3 p_4}
\,,
\end{align}
where the $\delta$ functions are Gram determinants defined as follows.
We define the generalized Kronecker symbol as
\begin{equation}
\delta^{\mu_1..\mu_n}_{\nu_1...\nu_n}= {\rm det}(\delta^\mu_\nu) =
\left|
\begin{matrix}
\delta^{\mu_1}_{\nu_1} & \ldots & \delta^{\mu_1}_{\nu_n} \\
\vdots &  & \vdots\\
\delta^{\mu_n}_{\nu_1} & \ldots & \delta^{\mu_n}_{\nu_n}
\end{matrix}
\right| .
\end{equation}
Given two lists of Lorentz vectors
$\{k_i\}$, $\{q_i\}$, $i=1,..,n$, the generalized $\delta$ function is defined as follows:
\begin{align}
\label{eq:delta}
\delta^{k_1,...,k_n}_{q_1,...,q_n}&=
\det (k_i\cdot q_j)\,.
\end{align}
It is easy to see that
\begin{enumerate}
\item
If there is a pair of $\{e_i,p_i\}$ contained
in $\{k_i\}$ or $\{q_i\}$,
(\ref{eq:delta}) is  invariant
under the gauge transformation
$e_i\rightarrow e_i + \alpha\ p_i$.
Thus \eqref{eq:example-op2} is manifestly gauge invariant.
\item
The rank of the matrix $k_i\cdot q_j$   is determined by the smaller one of the number $d$ and $n$.
If $d=4$, then (\ref{eq:delta}) vanishes for $n>4$.
Thus \eqref{eq:example-op2} is manifestly zero for $d=4$.
\end{enumerate}
Therefore, the minimal form factor \eqref{eq:example-op2}
is nonzero in general $d$ dimensions but vanish
in four dimensions.
Alternatively, one may also compute the minimal form factor using the four-dimensional rule \eqref{eq:OFmap-4-dim} and find it vanishing.
Furthermore, one can   show that the non-minimal form factors are also zero in four dimensions (which will be discussed later).
Thus we can conclude that $\mathcal{O}_{\rm e}$ is an evanescent
operator.

It is also instructive to express the gauge-invariant basis, such as used in  \cite{Boels:2018nrr}, in terms of generalized $\delta$ functions.
For the 4-gluon case, the gauge invariant basis can be chosen as products of
building blocks $A_{i;jk}$ and $B_{ij}$:\footnote{
The $B_{ij}$ in (\ref{eq:ADbasis}) is denoted by $D_{ij}$ in \cite{Boels:2018nrr}.
}
\begin{align}
\label{eq:ADbasis}
A_{i;jk}:=\delta^{e_i p_i}_{p_j p_k}\,,\quad
B_{ij}:= \frac{\delta^{e_i p_1 p_2 p_3 p_4}_{e_j p_1 p_2 p_3 p_4}}{
\mathrm{gr}_4}
\,,
\end{align}
where  $\mathrm{gr}_n:=\delta^{ p_1 p_2 ... p_n}_{ p_1 p_2 ...p_n}$ is a normalization factor.
Note that $B_{ij}$ involves a $\delta$ function with rank larger than 4 in the numerator and thus vanishes in four dimensions.
The gauge invariant basis for four-gluon form factors are in the form of $AAAA$, $AAB$ and $BB$, for example,
\begin{align}
\label{eq:ABbasis}
A_{1;23}A_{2;13}A_{3;12}A_{4;23},\quad
A_{1;23}A_{2;14}B_{34},\quad   B_{12}B_{34}\,.
\end{align}
The numbers of independent basis for the above three types are
16, 24, 3 relatively.

In the next section, we will explain how to construct the evanescent operators in a systematic way.
Before moving on, let us mention that
there are no evanescent operators in the length-two and length-three cases.
(The classification of length-2 and length-3 operators were considered in detail in \cite{Jin:2020pwh}.)
This is consistent with the fact that:
for 2-gluon or 3-gluon form factors,
the set $\{e_i\}$ and $\{p_i\}$ cannot form a generalized delta function as \eqref{eq:delta} with rank
larger than four.
Thus the shortest evanescent operators are of length four, and in the next subsection we consider
their basis construction.

\subsection{Construction of length-4 evanescent operators}

\label{sec:eva-cons}

In this section we explain the construction of evanescent basis operators.
As introduced in the previous subsection, one can first consider (color-stripped) kinematic operators, and then dress color factors to obtain full operators.
Following the same logic, we will first construct the basis of \emph{evanescent kinematic operators} in Section \ref{sec:step1}, and then we dress proper color factors to obtain the full basis of  evanescent operators in Section \ref{sec:step2}.

\subsubsection{Evanescent kinematic operators}

\label{sec:step1}

In this subsection, we first consider the kinematic operators $\lfloor \mathcal{W}_1, \mathcal{W}_2, \mathcal{W}_3, \mathcal{W}_4 \rfloor$ without color factors.
These operators are mapped to kinematic form factors
$\kineF\big(\lfloor \mathcal{W}_1, \mathcal{W}_2,  \mathcal{W}_4,  \mathcal{W}_4 \rfloor \big)$
according to \eqref{eq:OFmap-d-dim}.

To obtain the general high dimensional operator basis, one can introduce a finite set of  \emph{primitive kinematic operators}. Here the word ``primitive" is in the sense that higher dimensional operators can be constructed by inserting pairs of covariant derivatives $\{D^\mu, D_\mu\}$ into the primitive ones; for instance,
adding a $DD$ pair to the second and fourth sites of  $\lfloor \mathcal{W}_1, \mathcal{W}_2, \mathcal{W}_3, \mathcal{W}_4 \rfloor$ one has $\lfloor \mathcal{W}_1, D^\mu \mathcal{W}_2, \mathcal{W}_3, D_\mu\mathcal{W}_4 \rfloor$.

Denoting the set of primitive operators as $\{\gePr_i\}$,
the finiteness for the number of primitive kinematic operators are based on
the fact that any kinematic operator of dim $\Delta\geq12$ can always be
written as a sum like
\begin{align}
\label{eq:lessthan12}
\sum_{i}\sum_{\vec{n}_i}
c_{\vec{n}_i} \prod_{1\leq a<b\leq4} \mathbf{s}_{ab}^{n_{i,ab}}
\gePr_{i},\quad
\mathrm{dim}(\gePr_i)\leq12\,.
\end{align}
Here $\vec{n}_i =\{n_{i,12},n_{i,13},n_{i,23},n_{i,14},n_{i,24},n_{i,34}\}$
and $c_{\vec{n}_i}$ are rational numbers and
$ \mathbf{s}_{ab}^{n_{i,ab}}\gePr_i$ refers to inserting $n_{i,ab}$
pairs of  $\{D_\mu,D^\mu\}$ into the $j$-th and $k$-th sites
of the $i$-th kinematic operator $\gePr_i$. For example:\footnote{Changing the ordering of $\mathbf{s}_{12}^{n_{12}}...\mathbf{s}_{34}^{n_{34}}$
will not affect the minimal form factor of the operator, but in general
will affect the higher-point non-minimal form factors.
We will discuss this more in Appendix \ref{app:step3}.}
\begin{align}
\label{eq:insertSasDDpair}
\mathbf{s}_{13}\mathbf{s}_{23}\lfloor \mathcal{W}_1,\mathcal{W}_2,\mathcal{W}_3,\mathcal{W}_4\rfloor
=\lfloor D_{\nu}\mathcal{W}_1,D_{\mu}\mathcal{W}_2,D^{\nu\mu}\mathcal{W}_3,\mathcal{W}_4\rfloor\,.
\end{align}

To find out the minimal set of $\mathcal{K}_i$,
one can first enumerate all possible length-4 kinematic operators within dimension 12
and then find out the subset of kinematic operators that
are linearly independent, in the sense that
any linear combination of their kinematic form factors $\kineF(\gePr_i)$,
with possible polynomial $s_{ab}$ coefficients and total mass dimension no higher than 12, is nonzero.
These independent operators are the wanted primitive kinematic operators of length four,
and their number is 54.
More details about the proof of (\ref{eq:lessthan12}) is given in Appendix \ref{app:why12}.

The primitive kinematic operators $ \gePr_i $
can be organized to 27 evanescent ones
and 27 physical ones,
denoted by $\evPr_i$ and  $\phPr_i$ respectively.
In the limit of $d\rightarrow4$, $\kineF(\evPr_i) = 0$ and $\kineF(\phPr_i) \neq 0$.%
\footnote{One may   compare numbers of $\evPr_i$ and $\phPr_i$ with
the numbers of  gauge invariant basis mentioned in Section~\ref{sec:eva-define}.
The number of $\evPr_i$ is  equal to the number of $AAB$ and $BB$ basis,
while the number of $\phPr_i$  is not equal to that of $AAAA$ basis; this will be
discussed  further in Appendix \ref{app:indie}. }
Generic high-dimensional evanescent kinematic operators
can be expanded in terms of $\{\evPr_i\}$ as
\begin{align}
\label{eq:createK}
\sum_{i=1}^{27}\sum_{\vec{n}_i}
 c_{\vec{n}_{i}} \prod_{1\leq a<b\leq4} \mathbf{s}_{ab}^{ n_{i,ab}}
 \mathcal{E}_{i}\,.
\end{align}
We will denote the linear space of  dim-$\Delta$ evanescent basis kinematic operators
by $(\mathfrak{K}^e)_{\Delta}$.
Formula (\ref{eq:createK}) tells us $(\mathfrak{K}^e)_{\Delta}$
can be spanned by all the possible configurations of
$\mathbf{s}_{12}^{n_{i,12}}...\ \mathbf{s}_{34}^{n_{i,34}}\evPr_i$
that satisfy:
\begin{align}
\label{eq:dimeqn}
2(n_{i,12}+n_{i,13}+n_{i,14}+n_{i,23}+n_{i,24}+n_{i,34})=\Delta-\Delta_{\evPr_i}\,.
\end{align}
Here we point out that kinematic operators $\{\mathbf{s}_{12}^{n_{i,12}}...\ \mathbf{s}_{34}^{n_{i,34}}\evPr_i\}$
satisfying (\ref{eq:dimeqn}) are linearly independent with constant coefficients,
which is explained in Appendix \ref{app:indie}.
So they provide a basis of $(\mathfrak{K}^e)_{\Delta}$.

Below we discuss the 27 evanescent kinematic operators $ \evPr_i $ in more detail.
They can be constructed  by contracting generalized $\delta$ functions with tensor operators.
We classify them as the following four classes according to their dimensions and symmetry structures.


{\bf Class 1:  $\evPr_{\mathbf{1},i}$.}
The first class contains twelve linearly independent kinematic operators
$\evPr_{\mathbf{1},i}$
of mass dimension ten.
They can be chosen as
\begin{equation}
\label{eq:E1}
\evPr_{\mathbf{1},1} =\frac{1}{16}
\delta^{\mu_1\mu_2\mu_3\mu_4\rho}_{\nu_1\,\nu_2 \, \nu_3 \, \nu_4  \sigma}
\lfloor D_{\sigma}F_{\mu_1\mu_2}, F_{\mu_3\mu_4},
D_{\rho}F_{\nu_1\nu_2}, F_{\nu_3\nu_4} \rfloor \,,
\end{equation}
together with its $S_4$ permutations.
Here the permutation means changing the positions of $\mathcal{W}_i$,
\emph{i.e.}~
\begin{align}
\sigma \cdot \lfloor
\mathcal{W}_{1},\mathcal{W}_{2},
\mathcal{W}_{3},\mathcal{W}_{4}
\rfloor
=
\lfloor
\mathcal{W}_{\sigma(1)},\mathcal{W}_{\sigma(2)},
\mathcal{W}_{\sigma(3)},\mathcal{W}_{\sigma(4)}
\rfloor \,,\quad
\sigma\in S_4\,.
\end{align}
The full set of operators are explicitly given in (\ref{eq:KOdelta10}).
The kinematic form factor of $\evPr_{\mathbf{1},i}$ are also written
in terms of $\delta$ functions, \emph{e.g.}~
\begin{equation}
\label{eq:delta1}
\kineF(\evPr_{\mathbf{1},1})
=\delta^{ e_1e_2p_1p_2p_3}_{e_3e_4p_3p_4p_1}\,.
\end{equation}


{\bf Class 2:  $\evPr_{\mathbf{2},i}$.}
The second class contains two linearly independent kinematic operators of dimension 12:
\begin{align}
\label{eq:E2}
&\evPr_{\mathbf{2},1} =\frac{1}{16}
\delta^{\mu_1\mu_2\mu_3\mu_4\rho_1\rho_2}_{\nu_1\,\nu_2 \, \nu_3 \, \nu_4 \sigma_1\sigma_2}
\lfloor D_{\sigma_1}F_{\mu_1\mu_2}, D_{\sigma_2}F_{\mu_3\mu_4},
D_{\rho_1}F_{\nu_1\nu_2}, D_{\rho_2} F_{\nu_3\nu_4} \rfloor \,,
\nonumber\\
&\evPr_{\mathbf{2},2} =\mathcal{E}_{\mathbf{2},1}
\big|_{\lfloor\mathcal{W}_1,\mathcal{W}_2,\mathcal{W}_3,\mathcal{W}_4\rfloor
\rightarrow
\lfloor\mathcal{W}_1,\mathcal{W}_3,\mathcal{W}_2,\mathcal{W}_4\rfloor
 }\,.
\end{align}
Their kinematic form factors are
\begin{align}
\label{eq:delta2}
\kineF(\evPr_{\mathbf{2},1})=\delta^{e_1e_2p_1p_2p_3p_4}_{e_3e_4p_1p_2p_3p_4}\,,
\qquad
\kineF(\evPr_{\mathbf{2},2})=\delta^{e_1e_3p_1p_2p_3p_4}_{e_2e_4p_1p_2p_3p_4}\,.
\end{align}


{\bf Class 3:  $\evPr_{\mathbf{3},i}$.}
The third class contains twelve linearly independent kinematic operators
$\evPr_{\mathbf{3},i}$ of dimension 12.
They are given by
\begin{equation}
\label{eq:E3}
\evPr_{\mathbf{3},1} =\frac{1}{4}
\delta^{\mu_1\mu_2\mu_3\mu_4\mu_5}_{\nu_1\,\nu_2 \, \nu_3 \, \nu_4 \,\nu_5}
 \lfloor D_{\nu_3}F_{\mu_1\mu_2}, F_{\mu_5\rho},
D_{\mu_3}F_{\nu_1\nu_2}, D_{\mu_4\nu_4} F_{\nu_5\rho} \rfloor
  \,,
\end{equation}
together with its permutations, see (\ref{eq:KOsAAD}).
The kinematic form factor of $\evPr_{\mathbf{3},i}$ are also written
in terms of $\delta$ functions, \emph{e.g.}~
\begin{equation}
\label{eq:delta3}
\kineF(\evPr_{\mathbf{3},1})
=
\delta^{e_2 p_2}_{\mu\,\,\rho}\delta^{e_4 p_4}_{\nu\,\,\rho}
\delta^{e_1 p_1 p_3 p_4\mu}_{e_3 p_3 p_1 p_4\nu}
\,.
\end{equation}


{\bf Class 4:  $\evPr_{\mathbf{4}}$.}
The last class contains a single operator of dimension 12 which is invariant under $S_4$ permutations:
\begin{equation}
\label{eq:E4}
\evPr_{\mathbf{4}} =\frac{1}{4}
\delta^{\mu_1\mu_2\mu_3\mu_4\mu_5}_{\nu_1\,\nu_2 \, \nu_3 \, \nu_4 \,\nu_5}
\Big(
\lfloor D_{\nu_3}F_{\mu_1\mu_2}, D_{\mu_3}F_{\nu_1\nu_2},
D_{\mu_4}F_{\nu_4\rho}, D_{\mu_5}F_{\nu_5\rho} \rfloor
+(S_4\mbox{-permutations})
\Big) \,,
\end{equation}
whose kinematic form factor is
\begin{align}
\kineF(\evPr_{\mathbf{4}})=\delta^{e_3p_3}_{\mu\,\,\rho}
\delta^{e_4p_4}_{\nu\,\,\rho}
\delta^{e_1 p_1 p_2 p_3 p_4}_{e_2 p_2 p_1\mu\,\,\nu}
+(S_4\mbox{-permutations})\,.
\end{align}

In summary, evanescent primitive kinematic operators are given by
$\{\evPr_{\mathbf{1},i},\evPr_{\mathbf{2},i},\evPr_{\mathbf{3},i},
\evPr_{\mathbf{4}}\}$, and the total number is $12+2+12+1=27$.
In the following context,
if it is not necessary to give the concrete class or the explicit operator expression,
we will use $\evPr_i$ to represent these 27 elements for simplicity.
An important advantage of the above basis is that they manifest symmetry properties such that
primitive kinematic operators of each class  are closed under $S_4$ action.
Consequently, different ways of inserting  $DD$ pairs into
the primitive ones of each class also create a set of
kinematic operators $\{\mathbf{s}_{12}^{n_{12}}...\mathbf{s}_{34}^{n_{34}}\evPr_i\}$
that are closed under $S_4$ action.
It is worth mentioning that the choice of basis
is not unique, and we will discuss another choice of  $\evPr_i$ in Appendix \ref{app:basechoice2}
by including as many total derivative operators as possible.

The linear spaces  spanned by $\evPr_{\mathbf{1},i},
\evPr_{\mathbf{2},i},\evPr_{\mathbf{3},i},\evPr_{\mathbf{4}}$ are denoted
by $\mathfrak{K}_{1},\mathfrak{K}_2,\mathfrak{K}_3,\mathfrak{K}_4$ respectively.
All of them are closed representation spaces of $S_4$.
According to (\ref{eq:createK}), the linear space $(\mathfrak{K}^e)_{\Delta}$ of
dim-$\Delta$ evanescent kinematic operators can be
decomposed as
\begin{align}
\label{eq:union1}
(\mathfrak{K}^e)_{\Delta}=
\bigg[\mathfrak{M}(\frac{\Delta-10}{2};\mathbf{s}_{ij})\otimes \mathfrak{K}_{1}\bigg]
\oplus\bigg[\mathfrak{M}(\frac{\Delta-12}{2};\mathbf{s}_{ij})\otimes(\mathfrak{K}_{2}
\oplus\mathfrak{K}_{3}\oplus\mathfrak{K}_{4})\bigg],
\end{align}
where $\mathfrak{M}(N;\mathbf{s}_{ij})$ refers to the linear space spanned by
all the homogenous monomials $\mathbf{s}_{12}^{n_{12}}...\mathbf{s}_{34}^{n_{34}}$ with
total power $n_{12}+...+n_{34}=N$. They represent all the possible ways
to insert $N$ pairs of identical $D$s into the
primitive kinematic operators.
In Appendix \ref{app:indie} we show that different kinematic operators
$\mathbf{s}_{12}^{n_{12}}...\mathbf{s}_{34}^{n_{34}}\evPr_i$ with the same
mass dimension are independent with constant coefficients, so they can be
chosen as the  basis of  $(\mathfrak{K}^e)_{\Delta}$.
As mentioned before, they are closed under $S_4$ permutations.
The counting of basis operators for $(\mathfrak{K}^e)_{\Delta}$ with
$\Delta=10,...,24$ are given in Table \ref{tab:evakine}.


\begin{table}[!t]
\centering
\caption{\label{tab:evakine} Counting of the independent dim-$\Delta$
evanescent kinematic operators, which expand
linear space $(\mathfrak{K}^e)_{\Delta}$, $\Delta=10,...,24$.}
\vspace{0.4cm}
\begin{tabular}{|c|c|c|c|c|c|c|c|c|}
\hline
$\Delta$ & 10 & 12 & 14 & 16 & 18 & 20 & 22 & 24 \\
\hline
 $s^{\Delta-10}\otimes \mathfrak{K}_{1}$ & $12\times1$ & $12\times 6$ & $12\times 21$ & $12\times 56$ &
$12\times 126$ & $12\times 252$ & $12\times 462$ & $12\times 792$\\
\hline
 $s^{\Delta-12}\otimes\mathfrak{K}_{2}$ & 0 & $2\times 1$ & $2\times 6$ & $2\times 21$ & $2\times 56$ &
$2\times 126$ & $2\times 252$ & $2\times 462$  \\
\hline
$s^{\Delta-12}\otimes\mathfrak{K}_{3}$ & 0 & $12\times 1$ & $12\times 6$ & $12\times 21$ & $12\times 56$ &
$12\times 126$ & $12\times 252$ & $12\times 462$ \\
\hline
$s^{\Delta-12}\otimes\mathfrak{K}_{4}$ & 0 & $1\times 1$ & $1\times 6$ & $1\times 21$ & $1\times 56$ &
$1 \times 126$ & $1 \times 252$ & $1 \times 462$\\
\hline
$(\mathfrak{K}^e)_{\Delta}$ & 12 & 87 & 342 & 987 & 2352 & 4914 & 9324 & 16434\\
\hline
\end{tabular}

\end{table}

\subsubsection{Dressing color factors}

\label{sec:step2}

From  kinematic operators,
one can obtain a real gauge invariant operator by dressing a color factor according to (\ref{eq:dress}).
In this subsection, based on the evanescent kinematic operators obtained in the previous subsection, we will construct the basis of evanescent operators.

For the length-four operators, there are two types of color factors,
which are of single-trace and double-trace respectively:
\begin{equation}
\label{eq:def-length4cf}
\mathrm{tr}(T^{a_i}T^{a_j}T^{a_k}T^{a_l}) = : \mathrm{tr}(ijkl)  \,,
\qquad
\mathrm{tr}(T^{a_i}T^{a_j})\mathrm{tr}(T^{a_k}T^{a_l}) = : \mathrm{tr}(ij) \mathrm{tr}(kl)  \,.
\end{equation}
Correspondingly, one will obtain single-trace and double-trace operators.
The main problem here is how to obtain a set of independent gauge-invariant
operators by dressing color factors to the kinematic operators.

We first point out that the independence of the kinematic operators does not mean that their color-dressed operators are independent.
For example,
$\mathrm{tr}(1234)\circ \mathcal{E}_{\mathbf{1},1}$ and
$\mathrm{tr}(1243)\circ \mathcal{E}_{\mathbf{1},3}$ may look different but actually
they are the same operators.
This is related to the fact that, any $S_4$ permutation simultaneously acting on the color factor
and the kinematic operator does not change the operator, namely,
\begin{align}
\label{eq:fullopS4}
\mathcal{T}^a \circ \lfloor\mathcal{W}_1,\mathcal{W}_2,\mathcal{W}_3,\mathcal{W}_4\rfloor
=(\sigma\cdot \mathcal{T}^a)\circ \big(\sigma^{-1}\cdot \lfloor\mathcal{W}_{1},
\mathcal{W}_2,\mathcal{W}_3,\mathcal{W}_4\rfloor\big)\,,\quad
\sigma\in S_4\,,
\end{align}
where $\mathcal{T}^a$ represents a color factor in \eqref{eq:def-length4cf},
and for the above example, one has
\begin{equation}
\mathrm{tr}(1243)\circ \mathcal{E}_{\mathbf{1},3}=
(\sigma\cdot\mathrm{tr}(1234))\circ
(\sigma^{-1}\circ\mathcal{E}_{\mathbf{1},1}) \,, \qquad \sigma=(3\leftrightarrow 4) \,.
\end{equation}
Thus, one needs to avoid the over-counting and pick out the independent operators.
As mentioned in Section \ref{sec:step1},  the basis set of kinematic operator
space $(\mathfrak{K}^e)_{\Delta}$ can be chosen as
$\{\mathbf{s}_{12}^{n_{12}}...\mathbf{s}_{34}^{n_{34}}\evPr_i\}$
which is closed under $S_4$ permutations.
So the color dressed operator set
$\{\mathcal{T}^a\circ\mathbf{s}_{12}^{n_{12}}...\mathbf{s}_{34}^{n_{34}}\evPr_i\}$
is also  closed under $S_4$ action defined by (\ref{eq:fullopS4}).
Therefore we can identify different elements as the same operator
if they produce the same orbit under $S_4$ action, and we can keep only
one of them as the independent operator.


An alternative more systematic method to construct independent basis operators
is inspired by one-to-one correspondence between operators
and minimal form factors.
The problem of finding linearly independent operators is transformed to a problem of obtaining linearly independent full-color form factors.
It is convenient to consider form factors since the full-color minimal form factor of a length-4 operator has the $S_4$ symmetry by definition, as mentioned in \eqref{eq:fullcolorminiFF}.
We can use the representation analysis of $S_4$ to find  independent full-color form factors.

The space spanned by color factors $\{\mathcal{T}^{a}\}$ is denoted by $\mathbf{Span}\{\mathcal{T}^a\}$
and the space spanned by kinematic form factors $\{\kineF
\big(\lfloor\mathcal{W}_{1},\mathcal{W}_{2},
\mathcal{W}_{3},\mathcal{W}_{4}\rfloor\big)\}$
is denoted by $\mathbf{Span}\{\kineF\big(
\lfloor\mathcal{W}_1,\mathcal{W}_2,
\mathcal{W}_3,\mathcal{W}_4\rfloor\big)\} $.
Color factors and kinematic form factors
can be formally multiplied together to form a tensor product space:%
\footnote{
Here we demand $\mathcal{T}^a$ and $\kineF(\lfloor\mathcal{W}_1,\mathcal{W}_2,
\mathcal{W}_3,\mathcal{W}_4\rfloor)$
transform individually under $S_4$ action, so their formal product
 satisfies the transformation law of a tensor product.}
\begin{align}
\label{eq:cktensor}
\mathbf{Span}\{\mathcal{T}^a\kineF\big(
\lfloor\mathcal{W}_1,\mathcal{W}_2,
\mathcal{W}_3,\mathcal{W}_4\rfloor
\big)\}
=\mathbf{Span}\{\mathcal{T}^a\}
\otimes
\mathbf{Span}\{\kineF\big(
\lfloor\mathcal{W}_1,\mathcal{W}_2,
\mathcal{W}_3,\mathcal{W}_4\rfloor
\big)\}\,.
\end{align}

Since full-color form factors are invariant under $S_4$ permutation,
they must belong to the \emph{trivial representation} in the space of (\ref{eq:cktensor}).
This means, linearly independent form factors
correspond to different  trivial representations
of $\mathbf{Span}\{\mathcal{T}^a\kineF\big(
\lfloor\mathcal{W}_1,\mathcal{W}_2,
\mathcal{W}_3,\mathcal{W}_4\rfloor\big)\}$.
Having this picture in mind, one can now apply some techniques of group theory, which we now explain.


First, one can expand the color and kinematic form factor spaces into irreducible representations with the
representation decomposition as follows:
\begin{align}
\label{eq:kine-decom0}
\mathbf{Span}\{\mathcal{T}^a\}\sim   \oplus_i x_i R_i
\,,\qquad
\mathbf{Span}\{\kineF\big(
\lfloor\mathcal{W}_1,\mathcal{W}_2,
\mathcal{W}_3,\mathcal{W}_4\rfloor
\big)\} \sim
\oplus_j y_j R_j\,.
\end{align}
Here $R_i$ refer to irreducible inequivalent $S_4$ representations,
which can be represented by Young diagrams as
\begin{align}
\label{eq:young}
\Yvcentermath1
R_{[4]}= {\tiny\yng(4) } \,,\quad
R_{[3,1]}= {\tiny\yng(3,1) } \,,\quad
R_{[2,2]}= {\tiny\yng(2,2) } \,,\quad
R_{[2,1,1]}= {\tiny\yng(2,1,1) } \,,\quad
R_{[1,1,1,1]}= {\tiny\yng(1,1,1,1) }~.
\end{align}
Integers $x_i$ and $y_i$ refer to how many $R_i$ appear in
$\mathbf{Span}\{\mathcal{T}^a\}$ and $\mathbf{Span}\{\kineF\big(
\lfloor\mathcal{W}_1,\mathcal{W}_2,
\mathcal{W}_3,\mathcal{W}_4\rfloor \big)\}$.

The representation decomposition of the tensor product (\ref{eq:cktensor}) is
\begin{align}
\label{eq:tensor}
\mathbf{Span}\{\mathcal{T}^a\, \kineF\big(
\lfloor\mathcal{W}_1,\mathcal{W}_2,
\mathcal{W}_3,\mathcal{W}_4\rfloor\big)\}
&\sim   \oplus_k (\sum_{i,j} C^k_{ij} x_i y_j) R_k\,,
\end{align}
where coefficients $C^k_{ij}$ are known from representation theory \cite{fulton1997young,fulton2013representation}. Especially
for trivial representation  $R_{[4]}$ that we are interested,
$C^{[4]}_{ij}=\delta_{ij}$,
so the coefficient of $R_{[4]}$ in (\ref{eq:tensor}) is
\begin{align}
\label{eq:counting}
\sum_{i,j} C^{[4]}_{ij} x_i y_j
=\sum_i x_i y_i \,,
\end{align}
which counts the dimension of
trivial representation subspace and thus gives
the number of independent form factors, or equivalent, independent operators.

In Table \ref{tab:count}, we summarize the number of basis evanescent operators based on the above method, where
we have used the values of $x_i$ and $y_i$ given in (\ref{eq:cdecom}) (and Table \ref{tab:kineCG}) in Appendix~\ref{app:repofS4}.
We have also introduced $C$-even and $C$-odd spaces for the single-trace color factors, which have $+1$ or $-1$ sign change under reflection of the trace:
\begin{equation}
\label{eq:len4Cparity}
C\textrm{-even} \ (\mathcal{T}_{\mathrm{s}+}): \quad {\rm tr}(ijkl)+{\rm tr}(lkji) \,, \qquad\quad
C\textrm{-odd} \ (\mathcal{T}_{\mathrm{s}-}): \quad {\rm tr}(ijkl)-{\rm tr}(lkji) \,.
\end{equation}
The corresponding operators are also called $C$-even and $C$-odd single-trace operators.
They do not mix with each other under renormalization.

\begin{table}[!t]
\centering
\caption{\label{tab:count}
The number of independent length-4 minimally-evanescent operators
with dimension $10, \ldots ,24$. }
\vspace{0.4cm}
\begin{tabular}{|c|c|c|c|c|c|c|c|c|}
\hline
operator dimension & 10 & 12 & 14 & 16 & 18 & 20 & 22 & 24
\\
\hline
single trace $C$-odd & 1 & 9 & 38 & 114 & 278 & 589 & 1128 & 2001
\\
\hline
single trace $C$-even & 3 & 16 & 54 & 145 & 330 & 671 & 1248 & 2171
\\
\hline
double trace & 3 & 16 & 54 & 145 & 330 & 671 & 1248 & 2171
\\
\hline
\end{tabular}
\end{table}

To write down explicitly the basis operators, we note that
the basis of irreducible subspaces of a tensor product
representation $U\otimes V$ can be written as
products of the basis of irreducible subspaces of $U$ and $V$.
One can thus first consider the representation decomposition of
$\mathbf{Span}\{\mathcal{T}^a\}$ and
$\mathbf{Span}\{\kineF\big(
\lfloor\mathcal{W}_1,\mathcal{W}_2,
\mathcal{W}_3,\mathcal{W}_4\rfloor\big)\}$ separately,
and then write down the basis of the trivial representation
subspace of the tensor product, using
the method in the representation theory \cite{fulton1997young,fulton2013representation}.

For example, the space of $C$-even single-trace factors
$\mathbf{Span}\{\mathcal{T}_{\mathrm{s}+}^a\}$ is three dimensional and
has representation decomposition
\begin{align}
&\Yvcentermath1
\mathbf{Span}\{\mathcal{T}_{\mathrm{s}+}^a\}
\sim {\tiny\yng(4)\oplus\yng(2,2) }\,.
\end{align}
The basis belonging to the $R_{[r]}$-type sub-representation of
$\mathbf{Span}\{\mathcal{T}_{\mathrm{s}+}^a\}$  are denoted by $\mathcal{T}^{[r]}_{\mathrm{s}+}$.
The space $(\mathfrak{K}^e)_{10}$ spanned by the dim-10 evanescent kinematic operators
is 12 dimensional and has representation decomposition
\begin{align}
&\Yvcentermath1
(\mathfrak{K}^e)_{10}
\sim {\tiny\yng(4)\oplus\yng(3,1)
\oplus 2\ \yng(2,2) \oplus\yng(2,1,1)
\oplus\yng(1,1,1,1) }\,.
\end{align}
The basis belonging to the $R_{[r]}$-type sub-representation of $(\mathfrak{K}^e)_{10}$
are denoted by $\mathcal{E}^{[r]}_{\mathbf{1}}$.
Since the $S_4$ actions over $\lfloor
\mathcal{W}_{1},\mathcal{W}_{2},
\mathcal{W}_{3},\mathcal{W}_{4}
\rfloor$ and
$\kineF\big(
\lfloor
\mathcal{W}_1,\mathcal{W}_2,\mathcal{W}_3,
\mathcal{W}_4
\rfloor \big)$
are related as
\begin{align}
\kineF\big(
\lfloor
\mathcal{W}_{\sigma(1)},\mathcal{W}_{\sigma(2)},
\mathcal{W}_{\sigma(3)},\mathcal{W}_{\sigma(4)}
\rfloor \big)=
\sigma^{-1}\cdot \kineF\big(
\lfloor
\mathcal{W}_1,\mathcal{W}_2,\mathcal{W}_3,
\mathcal{W}_4
\rfloor \big)\,,
\end{align}
the kinematic form factors $\kineF(\mathcal{E}^{[r]}_{\mathbf{1}})$ also
belong to the $R_{[r]}$-type sub-representation of the kinematic
form factor space denoted by $\mathfrak{F}[(\mathfrak{K}^e)_{10}]$.

The tensor product of two irreducible subspaces
of the same type $R_{[r]}$ contains an element belonging to the trivial representation,
and the expression of this element is
\begin{equation}
\label{eq:tensorelem}
\mathcal{T}_{\mathrm{s}+}^{[r]}
\cdot M\cdot
\kineF(\evPr_{\mathbf{1}}^{[r]})\,.
\end{equation}
where  $\mathcal{T}_{\mathrm{s}+}^{[r]}$, $\kineF(\evPr_{\mathbf{1}}^{[r]})$
are given in (\ref{eq:strbasis}), (\ref{eq:K10basis}),
and matrix  $M$ is given in (\ref{eq:fuseM}).
They can be calculated according to the representation theory \cite{fulton1997young,fulton2013representation}.
Take  $R_{[4]}$ as an example where
$\mathcal{T}_{\mathrm{s}+}^{[4]}$ and $\kineF(\evPr_{\mathbf{1}}^{[4]})$
are of dimension one, (\ref{eq:tensorelem}) becomes
\begin{equation}
\label{eq:egOs1}
\mathcal{T}_{\mathrm{s}+}^{[4]}
\kineF(\evPr_{\mathbf{1}}^{[4]})\,.
\end{equation}

Plugging in
the concrete expressions of $\mathcal{T}_{\mathrm{s}+}^{[4]}$
and $\kineF(\evPr_{\mathbf{1}}^{[4]})$ which are
\begin{align}
&\mathcal{T}_{s+}^{[4]}  =
\frac{1}{6} \text{tr}(1234)+\frac{1}{6} \text{tr}(1243)+\frac{1}{6} \text{tr}(1324)
+\frac{1}{6} \text{tr}(1342)+\frac{1}{6} \text{tr}(1423)+\frac{1}{6} \text{tr}(1432)\,,
\nonumber\\
&\kineF(\evPr_{\mathbf{1}}^{[4]})=
\frac{1}{12}\sum_{i=1}^{12} \kineF(\evPr_{\mathbf{1},i})\,,
\nonumber
\end{align}
and only keeping the terms proportional to $\mathrm{tr}(1234)$,\footnote
{This is enough, since the coefficients of other $\sigma\cdot\mathrm{tr}(1234)$
give the same operator. Denote $\tau=\sigma^{-1}$:
\begin{align}
&\sigma\cdot \mathrm{tr}(1234) \ \sigma\cdot \kineF
\big(\lfloor\mathcal{W}_1,\mathcal{W}_2,\mathcal{W}_3,\mathcal{W}_4\rfloor\big)
= \mathrm{tr}(\sigma(1)\sigma(2)\sigma(3)\sigma(4))  \kineF
\big(\lfloor\mathcal{W}_{\tau(1)},\mathcal{W}_{\tau(2)},
\mathcal{W}_{\tau(3)},\mathcal{W}_{\tau(4)}\rfloor\big)
\nonumber\\
&\rightarrow
\mathrm{tr}(\sigma(1)\sigma(2)\sigma(3)\sigma(4))  \circ
 \lfloor\mathcal{W}_{\tau(1)},\mathcal{W}_{\tau(2)},
\mathcal{W}_{\tau(3)},\mathcal{W}_{\tau(4)}\rfloor
=\mathrm{tr}(\mathcal{W}_1\mathcal{W}_2\mathcal{W}_3\mathcal{W}_4).
\end{align}}
we obtain a sum
\begin{align}
\frac{1}{72} \mathrm{tr}(1234)\sum_{i=1}^{12}
\kineF(\evPr_{\mathbf{1},i})\,.
\end{align}
This can be understood as the form factor of  the following single-trace operator
\begin{align}
\label{eq:opeTKexample}
&\frac{1}{72} \sum_{i=1}^{12}
\mathrm{tr}(1234)\circ \evPr_{\mathbf{1},i}
=
\frac{1}{18} \mathrm{tr}(1234)\circ(
\frac{1}{2}\evPr_{\mathbf{1},1}+\frac{1}{2}\evPr_{\mathbf{1},4}
+ \evPr_{\mathbf{1},2}+ \evPr_{\mathbf{1},6}
)\,.
\end{align}
This is one of the three basis operators
of the $C$-even single-trace dim-10 evanescent sector.
In a similar way we can list all the basis operators
of single-trace $C$-even, single-trace $C$-odd
and double-trace sectors. Their expressions are summarized in
 (\ref{eq:str-dim10}) and (\ref{eq:dtr-dim10}).


A useful comment is that
numbers of $C$-even single trace operators and
double trace operators are always equal, as shown in
Table \ref{tab:count}.
This originates from the fact that in the case of length-4,
the $C$-even single-trace color factors and
the double-trace color factors form equivalent
$S_4$ representations and they have the same invariant subgroup.
Consider a pair of length-4 operators
\begin{align}
\label{eq:str-dtr}
\mathcal{O}_s&=
\mathrm{tr}(\mathcal{W}_{ 1} \mathcal{W}_{ 2}
\mathcal{W}_{ 3} \mathcal{W}_{ 4})
+
\mathrm{tr}(\mathcal{W}_{ 4} \mathcal{W}_{  3}
\mathcal{W}_{ 2} \mathcal{W}_{ 1})
\,,
\quad
\mathcal{O}_d =\frac{1}{2}
\mathrm{tr}(\mathcal{W}_{ 1} \mathcal{W}_{ 3})
\mathrm{tr}(\mathcal{W}_{ 2} \mathcal{W}_{ 4})\,.
\end{align}
The color-ordered tree form factor of $\mathcal{O}_s$ with color factor $\mathrm{tr}(1234)$
and the color-ordered tree form factor of $\mathcal{O}_d$  with color factor
$\mathrm{tr}(13)\mathrm{tr}(24)$ are equal:
\begin{align}
\mathcal{F}^{(0)}_{\mathcal{O}_s}(p_1,p_2,p_3,p_4)
=\mathcal{F}^{(0)}_{\mathcal{O}_d}(p_1,p_3\,|\, p_2,p_4).
\end{align}
In this way we can establish a one-to-one correspondence between
$C$-even single-trace operators and double-trace operators.
We demand the $i$-th double-trace operator is always
related to the $i$-th single-trace $C$-even operator through
(\ref{eq:str-dtr}), so once the bases of single-trace operators
are fixed, the bases of double-trace ones are also fixed.

In the above discussion, we consider the gauge group with general $N_c$, and the six single-trace
and three double-trace color factors can be taken as linearly independent.
When $N_c$ takes a special value such as
$N_c=2$, there exist extra linear relations among them,
which means basis operators reduce to a smaller set.
This fact provides a consistency check
for loop calculation, see Appendix \ref{app:Nequalto2}
for details.

\subsubsection*{Remark on non-minimal form factor}

As shown above, all the evanescent  operators are in the form of
generalized $\delta$ functions of rank higher than 4 contracting with   tensor operators, \emph{i.e.}
\begin{equation}
\label{eq:contraction1}
\mathcal{O}_e=\delta^{\mu_1\cdots\mu_n}_{\nu_1\cdots\nu_n}
\mathcal{T}_{\mu_1\cdots\mu_n}^{\nu_1\cdots\nu_n}\,,\qquad n\ge 5\,,
\end{equation}
where $\mathcal{T}_{\mu_1\cdots\mu_n}^{\nu_1\cdots\nu_n}$ stands for a full tensor operator composed of
$D_\mu$'s and $F_{\mu\nu}$'s.
The form factor of $\mathcal{O}_e$ for arbitrary $k$ also has the similar form:
\begin{equation}
\label{eq:contraction2}
\mathcal{F}_{k;\mathcal{O}_e}=\delta^{\mu_1\cdots\mu_n}_{\nu_1\cdots\nu_n}
(\mathcal{F}_{k;\mathcal{T}})_{\mu_1\cdots\mu_n}^{\nu_1\cdots\nu_n}\ \,.
\end{equation}
The   $\delta$ tensor guarantees that
$\mathcal{F}_{k;\mathcal{O}_e}$ vanishes in 4 dimensions, namely, $\mathcal{O}_e$ is an exactly evanescent operator.

Besides, the tree form factors of $\mathcal{O}_e$ do not explicitly depend on spacetime dimension $d$.
The reason is as following. Given a rank-$n$ $\delta$ function, $d$ can only come out from the  contraction
\begin{equation}
\label{eq:contraction2}
\delta^{\mu_1\cdots\mu_n}_{\nu_1\cdots\nu_n}\delta^{\nu_n}_{\mu_n}
=(d-n+1)\delta^{\mu_1\cdots\mu_{n-1}}_{\nu_1\cdots\nu_{n-1}}\ .
\end{equation}
While  $\mathcal{T}^{\nu_1\cdots\nu_n}_{\mu_1\cdots\mu_n}$ is composed of $D_\mu$ and $F_{\mu\nu}$,
there is no room for a $\delta^{\mu}_{\nu}$ in $\mathcal{F}_{k;\mathcal{T}}$  according to the Feynman rules.

One should be careful there exist operators whose minimal form factors vanish in four dimensions
but higher-point form factors do not.
In that case one can change them to exactly evanescent operators by
adding proper higher length operators, and the detailed discussion is given in Appendix \ref{app:step3}.

\subsection{Higher-length evanescent operators}

\label{sec:len5}

Our method can be generalized to construct higher-length operators.
At length-5, the Gram determinants of higher ranks such as $\delta^{p_1...p_5}_{p_1...p_5}$
appear, which increase the number of possible building blocks of
evanescent kinematic operators.
The analysis of color factor dressing also becomes more involving and one needs to consider the representations of $S_5$.
Alternatively, at a fixed canonical dimension that is not too high,
one can always find out the basis evanescent operators
by enumerating all the possible
configurations of site operators ${\cal W}_i$ and picking out the
linearly independent ones that vanish in four dimensions.%
\footnote{Enumeration in a brute-force way is also
applicable to construct length-4 evanescent basis operators, so it
is an alternative strategy apart from the method introduced in previous section.}
Below we briefly consider the length-5 operators.

For the case of dim-10, evanescent operators can only be of length-4 or length-5.
The length-4 ones have been constructed in previous subsections.
For the length-5 operators, we can first enumerate the kinematic operators.
There are six linearly independent
length-5 evanescent kinematic operators $\evfive_{10;i}$.
The first one is
\begin{equation}
\label{eq:example-op3}
\evfive_{10;1} =\frac{1}{8}
\delta^{\mu_1\mu_2\mu_3\mu_4\mu_5}_{\nu_1\,\nu_2 \, \nu_3 \, \nu_4 \,\nu_5}
\lfloor F_{\mu_1\mu_2}, F_{\nu_1\nu_2}, F_{\mu_3\mu_4}, F_{\nu_3\nu_4},
F_{\mu_5\nu_5} \rfloor \,.
\end{equation}
Other five ones are obtained via permutations of last three ${\cal W}_i$'s of $\evfive_{10;1}$:
\begin{equation}
\label{eq:len5kine}
\evfive_{10;1} |_{
\lfloor {\cal W}_i, {\cal W}_j, {\cal W}_k, {\cal W}_l, {\cal W}_m\rfloor
\rightarrow
\lfloor {\cal W}_i, {\cal W}_j, \sigma({\cal W}_k, {\cal W}_l, {\cal W}_m)\rfloor
} \,, \quad \sigma \in S_3 \,.
\end{equation}
See details in Appendix \ref{app:eva-kine5}.

Having the kinematic operator basis  $\evfive_{10;i}$, one can
dress color factors to obtain the full operators.
The color factors of length-5 operators are also classified into
$C$-even ones and $C$-odd ones.
Similar to (\ref{eq:len4Cparity}), they are defined as%
\footnote{The definition of $C$-even and $C$-odd are based on
the $C$-parity of Yang-Mills fields, where ``$C$'' stands for charge conjugation,
see \emph{e.g.}~\cite{Bardeen:1969md}. }
\begin{align}
C\textrm{-even} \ (\mathcal{T}_{\mathrm{s}+}): \  {\rm tr}(ijklm)-{\rm tr}(mlkji) \,, \qquad\qquad
 &C\textrm{-odd} \ (\mathcal{T}_{\mathrm{s}-}) : \  {\rm tr}(ijklm)+{\rm tr}(mlkji) \,,
 \nonumber\\
C\textrm{-even} \ (\mathcal{T}_{\mathrm{d}+}): \  \mathrm{tr}(ij){\rm tr}(klm)
-\mathrm{tr}(ij){\rm tr}(mlk ) \,, \quad
 & C\textrm{-odd} \ (\mathcal{T}_{\mathrm{d}-}) : \  \mathrm{tr}(ij){\rm tr}( klm)
+\mathrm{tr}(ij){\rm tr}(mlk )\,.
\end{align}

For dim-10 case, all the single-trace length-5 operators are $C$-even, for example:
\begin{align}
{\rm tr}(12345)\circ \evfive_{10;1}
=\frac{1}{2}\big( {\rm tr}(12345)-{\rm tr}(54321)\big)\circ \evfive_{10;1}\,.
\end{align}
It turns out there are two linearly independent single-trace operators and one double-trace operator.
They are the final evanescent length-5 operators at dimension 10 and their explicit expressions are given in
(\ref{eq:len5str}) and (\ref{eq:len5dtr}) in the next subsection.

\subsection{Complete set of dim-10 evanescent basis operators}
\label{sec:dim10full}

The complete set of dim-10 evanescent operators includes:
the four length-4 single-trace ones;
the three length-4 double-trace ones;
two length-5 single-trace ones;
one length-5 double-trace one.

Among four single-trace length-4 operators, three are  $C$-even
and one is $C$-odd. We label them as
$\tilde{\mathcal{O}}^e_{10;\mathrm{s}+;1}$, $\tilde{\mathcal{O}}^e_{10;\mathrm{s}+;2}$,
$\tilde{\mathcal{O}}^e_{10;\mathrm{s}+;3}$, $\tilde{\mathcal{O}}^e_{10;\mathrm{s}-;1}$.
Details of $\tilde{\mathcal{O}}^e_{10;\mathrm{s}+;1}$ has been given in \eqref{eq:egOs1}-\eqref{eq:opeTKexample}.
Similar construction applies to other operators.
Here we just list the explicit expressions of the operators:
\begin{align}
\label{eq:str-dim10}
\tilde{\mathcal{O}}^e_{10;\mathrm{s}+;1}
&=(-\imI g)^2 8\mathrm{tr}(1234)\circ(\frac{1}{2}\evPr_{\mathbf{1},1}+\frac{1}{2}\evPr_{\mathbf{1},4}
+\evPr_{\mathbf{1},2}+\evPr_{\mathbf{1},6})
 \,,
\nonumber\\
\tilde{\mathcal{O}}^e_{10;\mathrm{s}+;2}
&=(-\imI g)^2 8\mathrm{tr}(1234)\circ(\evPr_{\mathbf{1},2}-2\evPr_{\mathbf{1},6}
+\frac{1}{2}\evPr_{\mathbf{1},1}+\frac{1}{2}\evPr_{\mathbf{1},4})
 \,,
\nonumber\\
\tilde{\mathcal{O}}^e_{10;\mathrm{s}+;3}
&=(-\imI g)^2 8\mathrm{tr}(1234)\circ(\evPr_{\mathbf{1},6}-
\frac{1}{2}\evPr_{\mathbf{1},1}-\frac{1}{2}\evPr_{\mathbf{1},4})
 \,,
\nonumber\\
\tilde{\mathcal{O}}^e_{10;\mathrm{s}-;1}
&=(-\imI g)^2 4\mathrm{tr}(1234)\circ(\evPr_{\mathbf{1},1}-\evPr_{\mathbf{1},4})
\,.
\end{align}

The three double-trace length-4 operators are denoted by
$\tilde{\mathcal{O}}^e_{10;\mathrm{d}+;1}$, $\tilde{\mathcal{O}}^e_{10;\mathrm{d}+;2}$,
$\tilde{\mathcal{O}}^e_{10;\mathrm{d}+;3}$, which are explicitly given as
\begin{align}
\label{eq:dtr-dim10}
\tilde{\mathcal{O}}^e_{10;\mathrm{d}+;1}
&=(-\imI g)^2 4\mathrm{tr}(12)\mathrm{tr}(34)\circ(\evPr_{\mathbf{1},5}
+\evPr_{\mathbf{1},7}+\evPr_{\mathbf{1},1})
 \,,
\nonumber\\
\tilde{\mathcal{O}}^e_{10;\mathrm{d}+;2}
&=(-\imI g)^2 4\mathrm{tr}(12)\mathrm{tr}(34)\circ(\evPr_{\mathbf{1},7}
-2\evPr_{\mathbf{1},1}
+\evPr_{\mathbf{1},5})
 \,,
\nonumber\\
\tilde{\mathcal{O}}^e_{10;\mathrm{d}+;3}
&=(-\imI g)^2 4\mathrm{tr}(12)\mathrm{tr}(34)\circ(\evPr_{\mathbf{1},1}
-\evPr_{\mathbf{1},5})
 \,.
\end{align}

The two single-trace length-5 operators and one double-trace length-5 one
are all $C$-even, and they are labeled as $\Xi^e_{10;\mathrm{s}+;1}$,
$\Xi^e_{10;\mathrm{s}+;2}$, $\Xi^e_{10;\mathrm{d}+;1}$, explicitly defined as
\begin{align}
\label{eq:len5str}
\tilde{\Xi}^e_{10;\mathrm{s}+;1}&=(-\imI g)^3
\mathrm{tr}(12345)\circ
( \evfive_{10;1} +\evfive_{10;4})
\,,
\quad
\tilde{\Xi}^e_{10;\mathrm{s}+;2}=(-\imI g)^3
\mathrm{tr}(12345)\circ
(-2 \evfive_{10;1} + \evfive_{10;4})
\,,
\\
\label{eq:len5dtr}
\tilde{\Xi}^e_{10;\mathrm{d}+;1}&=(-\imI g)^3
\mathrm{tr}(12 )\mathrm{tr}( 345)\circ
\evfive_{10;1}\,,
\end{align}
where $\evfive_{10;i}$ are defined in \eqref{eq:len5kine}.

We point out that in our definition, the operators are multiplied by
a proper power of gauge coupling, as shown in (\ref{eq:str-dim10})-(\ref{eq:len5dtr}):
each length-$L$ operator carries a factor $(-\imI g)^{L-2}$.
Such a convention is in accordance with (\ref{eq:covDerivDef}) that
a length-$L$ operator containing $[D_i,D_j]$ can be rewritten as a length-$(L+1)$ operator with
an increasing $-ig$ factor. In such choice, the $n$-point form factors of all the operators are  of the same
order $\mathcal{O}(g^{n-2})$.
Besides, such an operator has canonical dimension $(n-2\epsilon)$ where $n$ is an integer,
so in an EFT its Wilson coefficient times a certain integer power of mass are dimensionless.
Unlike in a conformal field theory such as ${\cal N}=4$ SYM,  in QCD changing the definition of an operator by a factor of coupling will in general change its anomalous dimension due to the contribution of the beta function.

The final basis choice for dim-10 evanescent operators is a linear recombination of above operators:
\begin{align}
\label{eq:changebase}
\left(\begin{array}{c}
\mathcal{O}^e_{10;\mathrm{s}+;1}\\
\mathcal{O}^e_{10;\mathrm{s}+;2}\\
\mathcal{O}^e_{10;\mathrm{s}+;3}\\
\mathcal{O}^e_{10;\mathrm{s}-;1}\\
\mathcal{O}^e_{10;\mathrm{d}+;1}\\
\mathcal{O}^e_{10;\mathrm{d}+;2}\\
\mathcal{O}^e_{10;\mathrm{d}+;3}\\
\Xi^e_{10;\mathrm{s}+;1}\\
\Xi^e_{10;\mathrm{s}+;2}\\
\Xi^e_{10;\mathrm{d}+;1}\\
\end{array}
\right)=
\left(\begin{array}{cccc ccc|cc c}
1 & $\,$ 0 & $\,$ 0 & $\,$ 0 & $\,$ 0 & $\,$ 0 & $\,$ 0 $\,$ &
$\,\,$ 4 & $\,\,$ 0 & $\,\,$  0 \\
0 & $\,$ 1 & $\,$ 0 & $\,$ 0 & $\,$ 0 & $\,$ 0 & $\,$ 0 $\,$ &
$\,\,$ 0 & $\,\,$ 4 & $\,\,$  0 \\
0 & $\,$ 0 & $\,$ 1 & $\,$ 0 & $\,$ 0 & $\,$ 0 & $\,$ 0 $\,$ &
$\,\,$ 0 & $\,\,$ 0 & $\,\,$  0\\
0 & $\,$ 0 & $\,$ 0 & $\,$ 1 & $\,$ 0 & $\,$ 0 & $\,$ 0 $\,$ &
$\,\,$  0 & $\,\,$  0 & $\,\,$  0\\
0 & $\,$ 0 & $\,$ 0 & $\,$ 0 & $\,$ 1 & $\,$ 0 & $\,$ 0 $\,$ &
$\,\,$ 0 & $\,\,$  0 & $\,\,$  4\\
0 & $\,$ 0 & $\,$ 0 & $\,$ 0 & $\,$ 0 & $\,$ 1 & $\,$ 0 $\,$ &
$\,\,$ 0 & $\,\,$  0 & $\,\,$  4\\
0 & $\,$ 0 & $\,$ 0 & $\,$ 0 & $\,$ 0 & $\,$ 0 & $\,$ 1 $\,$ &
$\,\,$ 0 & $\,\,$  0 & $\,\,$  0\\
\hline\rule{0pt}{0.9\normalbaselineskip}
0 & $\,$ 0 & $\,$ 0 & $\,$ 0 & $\,$ 0 & $\,$ 0 & $\,$ 0 $\,$ &
$\,\,$1 & $\,\,$  0 & $\,\,$  0\\
0 & $\,$ 0 & $\,$ 0 & $\,$ 0 & $\,$ 0 & $\,$ 0 & $\,$ 0 $\,$ &
$\,\,$ 0 & $\,\,$  1 & $\,\,$  0\\
0 & $\,$ 0 & $\,$ 0 & $\,$ 0 & $\,$ 0 & $\,$ 0 & $\,$ 0 $\,$ &
$\,\,$ 0 & $\,\,$  0 & $\,\,$  1\\
\end{array}
\right)
\left(\begin{array}{c}
\tilde{\mathcal{O}}^e_{10;\mathrm{s}+;1}\\
\tilde{\mathcal{O}}^e_{10;\mathrm{s}+;2}\\
\tilde{\mathcal{O}}^e_{10;\mathrm{s}+;3}\\
\tilde{\mathcal{O}}^e_{10;\mathrm{s}-;1}\\
\tilde{\mathcal{O}}^e_{10;\mathrm{d}+;1}\\
\tilde{\mathcal{O}}^e_{10;\mathrm{d}+;2}\\
\tilde{\mathcal{O}}^e_{10;\mathrm{d}+;3}\\
\tilde{\Xi}^e_{10;\mathrm{s}+;1}\\
\tilde{\Xi}^e_{10;\mathrm{s}+;2}\\
\tilde{\Xi}^e_{10;\mathrm{d}+;1}\\
\end{array}
\right)
\,.
\end{align}
This choice intends to include as many total derivative operators as basis operators.
The detailed discussion is given in Appendix \ref{app:basechoice1}.

As mentioned in Section \ref{sec:step1}, the primitive evanescent
kinematic operators are grouped into four classes and for  dimension 10
all the evanescent operators are constructed by the first class kinematic operators.
The other three classes begin to appear for operator dimension 12.
In Appendix \ref{app:eva-dim12}, we also summarize dim-12 length-4 basis evanescent operators,
which are arranged according to the order of  primitive classes.

\section{One-loop renormalization of evanescent operators}
\label{sec:oneloop}

In this section, we compute the one-loop form factors of evanescent operators and obtain the renormalization matrix.
As an outline, we first explain the one-loop form factor calculation through the unitarity method in Section~\ref{sec:calc-full}, then we consider the IR and UV divergences in Section \ref{sec:IRUV},
and finally we discuss the renormalization matrix and anomalous dimensions in Section \ref{sec:Zmatrix}.

\subsection{One-loop full-color form factor}

\label{sec:calc-full}

To be explicit, we will mostly focus on the dim-10 evanescent basis operators given in (\ref{eq:changebase})
as concrete examples in this subsection, while it is straightforward to generalize the discussion to higher-dimensional cases as well as for physical operators.
For these operators we need to calculate the following three types of form factors:
\begin{align}
\label{eq:tobecalc}
&\text{
one-loop\ 4-gluon\ form\ factors\ of\ length-4\ operators,\ denoted\ by\ }
\mathbf{F}^{(1)}_{4;\mathcal{O}_{4;s}},\ \mathbf{F}^{(1)}_{4;\mathcal{O}_{4;d}}\,,
\nonumber \\
&\text{
one-loop\ 5-gluon\ form\ factors\ of\ length-5\ operators,\ denoted\ by\ }
\mathbf{F}^{(1)}_{5;\mathcal{O}_{5;s}},\ \mathbf{F}^{(1)}_{5;\mathcal{O}_{5;d}}\,,
\\
&\text{
one-loop\ 5-gluon\ form\ factors\ of\ length-4\ operators,\ denoted\ by\ }
\mathbf{F}^{(1)}_{5;\mathcal{O}_{4;s}},\ \mathbf{F}^{(1)}_{5;\mathcal{O}_{4;d}}\,,
\nonumber
\end{align}
where the first two lines are minimal form factors and the last line are next-to-minimal form factors.
The subscript `$s$' or `$d$' stands for single- or double-trace operators.

\begin{figure}
  \centering
  \includegraphics[scale=0.6]{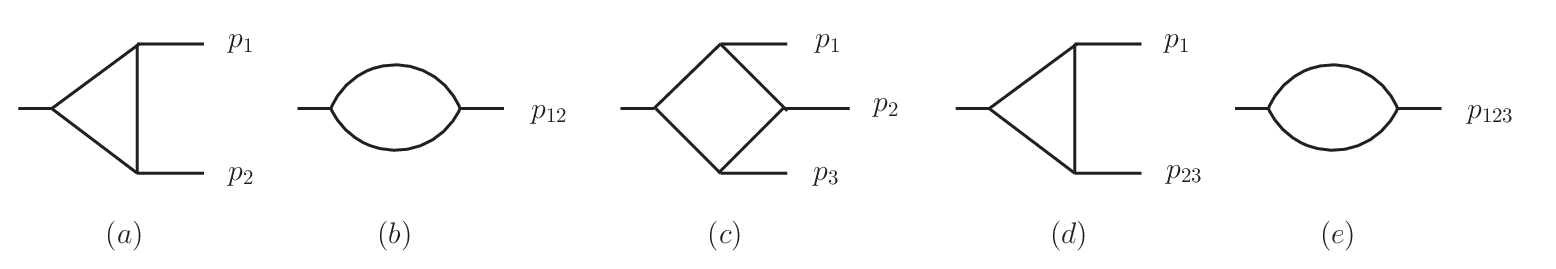}
  \caption{\label{fig:MIset} Up to particle permutations, the complete set of
  basis integrals contained in (\ref{eq:tobecalc}).  }
\end{figure}

These one-loop form factors can be expanded in a set of basis integrals $\{I_{i}\}$ like
\begin{align}
\label{eq:cI}
\mathcal{F}^{(1)}=\sum_i c_{i} I_{i}\,,
\end{align}
and we list the basis integrals in Figure \ref{fig:MIset}.
The truly physical information is contained in the coefficients $c_i$, and it is convenient to apply the unitarity method \cite{Bern:1994zx, Bern:1994cg} to compute them. 
The central idea of unitary method is that by putting internal propagators on-shell (\emph{i.e.}~by performing unitarity cuts), the loop form factors can be factorized as products of tree building blocks. In this way, one can use simpler tree-level form factors and amplitudes as input to reconstruct the loop form factor coefficients $c_i$.
The final form factor is guaranteed to be the correct physical result as long as it is consistent with all possible unitarity cuts.

\begin{figure}
  \centering
  \includegraphics[scale=0.6]{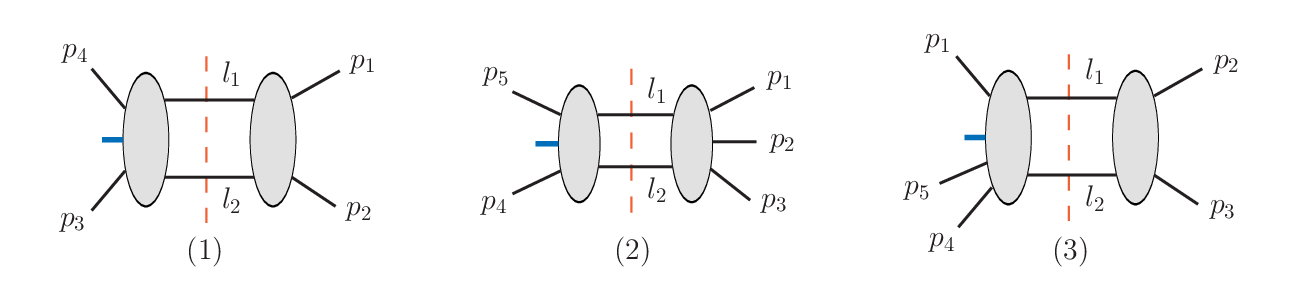}
  \caption{\label{fig:cut444554} Up to different choices of cut channels and
  color orders of tree blocks, there are three classes of cuts which can probe
  all the topologies of one-loop 4-gluon and 5-gluon
  form factors of length-4 and length-5 operators. }
\end{figure}

For the one-loop problem at hand, the complete set of cuts are shown in
Figure \ref{fig:cut444554}.
Concretely, the basis integrals for minimal form factors are (a) and (b),
which can be probed by cut (1) or (3) in Figure \ref{fig:cut444554}.
The integrals for next-to-minimal form factors are
(a)-(e), among which (a) and (b) are probed by cut (3) and
(e) is probed by cut (2).
Integrals (c) and (d) can be probed by both cut (2) and (3), and
their coefficients derived from these two different cuts
must be the same, which provides a consistency check of the computation.

Since we consider evanescent operators, it is crucial that we perform the computation in $d$ dimensions.
We use conventional dimensional regularization scheme which provides a Lorentz covariant representation for general $d$ dimensions.%
\footnote{
One may  consider the dimensional reduction scheme where the $4$-dimensional gauge fields may be decomposed in terms of $D$-dimensional ones plus the $\epsilon$-scalars \cite{Siegel:1979wq, Capper:1979ns, Jack:1993ws, Harlander:2006rj}.
One such example is the Konishi operator in ${\cal N}=4$ SYM, where to get the correct two-loop anomalous dimension, it is necessary to consider the scalars with $6+2\epsilon$ components \cite{Nandan:2014oga}.
One could also consider integer spacetime dimensions large than four,
for example, six-dimensional spinor helicity formalism has been used to compute form factors in pure YM theory in \cite{Huber:2019fea}, and operator renormalization has also been considered for gauge theories of six and eight dimensions \cite{Gracey:2015xmw, Gracey:2017nly}.
}
As mentioned before, in this paper we only consider operators with even $P$-parity,
while for $P$-odd operators such as the Weinberg-type operators \cite{Weinberg:1989dx}, the Levi-Civita tensor $\varepsilon_{\mu\nu\rho\sigma}$
enters  and breaks $d$-dimensional covariance, requiring other regularization schemes.%
\footnote{This is also related to the regularization of physical quantities involving $\gamma_5$ \cite{tHooft:1972tcz}.
For the 2-loop renormalization dealing with  evanescent operators and $\gamma_5$
see also \cite{Buras:1989xd,Schubert:1988ke}. }

We briefly outline the computational strategy as follows:
\begin{align}
\mathcal{F}^{(1)}\Big|_{\mbox{cut}} &
= \sum_{\rm helicity} \prod  (\textrm{tree blocks})  = \textrm{cut integrand}  \xlongrightarrow{\mbox{PV reduction}}
 \sum_{\mbox{cut\ permitted}}  c_i I_i \,. \nonumber
\end{align}
First, the input tree blocks in a cut channel can be obtained in terms of Lorentz products of momenta and polarization vectors via Feynman rules,
and thus they are valid in $d$ dimensions. We summarize efficient rules for computing tree-level form factors in Appendix~\ref{app:tree-rule}.
Next, one needs to perform the helicity sum for the cut legs, and we use the formula:
\begin{align}
\label{eq:polarsum}
\sum_{\mathrm{helicty}}
e^\mu(l_i)e^\nu(l_i)
=\eta^{\mu\nu}-\frac{q^\mu_i l_i^\nu+q_i^\nu l_i^\mu}{q_i\cdot l_i},
\quad i=1,2 \,,
\end{align}
where $q_i^\mu$ are arbitrary light-like reference momenta.
Furthermore,  for each cut integrand, we use the Passarino-Veltman reduction method
to perform integral reduction \cite{Passarino:1978jh}.
As long as the cut is allowed by the topology of  $I_{i}$,
its coefficients $c_{i}$ can be obtained from the given cut channel.
Running through the complete set of cuts can probe all  basis integrals,
and in this way, one gets all the $\{c_{i}\}$ and obtains the full
form factor results.

Since we would like to obtain the full-color one-loop form factors, there is some technical complication with respect to the color factors, which we explain in some details below.  The one-loop form factor can be decomposed into single- and double-trace color basis, similar to one-loop amplitudes \cite{Bern:1990ux}.
For example, the one-loop minimal form factor
of a length-4 single-trace operator has the following color decomposition:
\begin{equation}
\begin{aligned}
\label{eq:egF1ofstr4}
\mathbf{F}^{(1)}_{4;\mathcal{O}_s} &=
\sum_{\tau\in S_3}N_c\mathrm{tr}\big(1\tau(2)\tau(3)\tau(4)\big)\
 \mathcal{F}_{4;\mathcal{O}_s}^{(1),\mathrm{s}} \big(p_1,p_{\tau(2)},p_{\tau(3)},p_{\tau(4)}
 \big)
\\
&\quad +\sum_{\tilde{\tau}\in Z_3}\mathrm{tr}\big(1\tilde{\tau}(2)\big)
\mathrm{tr}\big(\tilde{\tau}(3)\tilde{\tau}(4)\big)
 \mathcal{F}_{4;\mathcal{O}_s}^{(1),\mathrm{d}} \big(p_1,p_{\tilde{\tau}(2)}\, |\,
 p_{\tilde{\tau}(3)},p_{\tilde{\tau}(4)} \big) \,,
\end{aligned}
\end{equation}
where $\mathcal{F}_{4;\mathcal{O}_s}^{(1),\mathrm{s}}  (p_1,p_{\tau(2)},p_{\tau(3)},p_{\tau(4)} )$
and $\mathcal{F}_{4;\mathcal{O}_s}^{(1),\mathrm{d}} (p_1,p_{\tilde{\tau}(2)}\, |\,
p_{\tilde{\tau}(3)},p_{\tilde{\tau}(4)} )$
are the leading single-trace and sub-leading double-trace color-ordered form factors respectively.
The double-trace form factors in the second line contain the mixing information with double-trace operators.

To apply unitarity method, let us consider the $s_{12}$-cut (cut (1) in Figure \ref{fig:cut444554})
for the form factor $\mathbf{F}^{(1)}_{4;\mathcal{O}_s}$ in (\ref{eq:egF1ofstr4}) as a concrete example.
The corresponding cut integrand is given by the product of a 4-gluon form factor
$\mathbf{F}^{(0)}_{4;\mathcal{O}_s}$ and a 4-gluon amplitude $\mathbf{A}^{(0)}_4$,
\begin{equation}
\label{eq:egF1ofstr4-cut}
\mathbf{F}^{(1)}_{4;\mathcal{O}_s} \big|_{s_{12}\textrm{-cut}} =
\int d\mathrm{PS}_{l_1,l_2} \sum_{\mathrm{helicty}}
\mathbf{F}^{(0)}_{4;\mathcal{O}_s} (-l_1,-l_2,p_3,p_4)
\times
\mathbf{A}^{(0)}_4(p_1,p_2, l_2,l_1) \,.
\end{equation}
These two full-color tree blocks can be expanded in color bases as \cite{DelDuca:1999rs}
\begin{equation}
\begin{aligned}
\label{eq:CDofF4str}
\mathbf{F}^{(0)}_{4;\mathcal{O}_s}&=
\sum_{\tilde{\sigma}\in S_3}
\mathrm{tr}\big(l_1\tilde{\sigma}(l_2)
\tilde{\sigma}(3)\tilde{\sigma}(4)\big)
\mathcal{F}^{(0)}_{4;\mathcal{O}_s} \big(
l_1,p_{\tilde{\sigma}(l_2)},p_{\tilde{\sigma}(3)},
p_{\tilde{\sigma}(4)}
\big)\,,
 \\
\mathbf{A}^{(0)}_4
& =\sum_{\sigma\in S_3}
 \mathrm{tr}\big(l_2 \sigma(l_1) \sigma(1) \sigma(2)\big)
\mathcal{A}_4^{(0)}\big(l_2, p_{\sigma(l_1)}, p_{\sigma(1)}, p_{\sigma(2)}\big) \\
& =f^{2 l_2 b} f^{l_1 1 b} \mathcal{A}_4^{(0)}\big(l_2, l_1, p_1, p_2\big) + f^{2 l_1 b} f^{l_2 1 b} \mathcal{A}_4^{(0)}\big(l_1, l_2, p_1, p_2\big) \,.
\end{aligned}
\end{equation}
By comparing \eqref{eq:egF1ofstr4} and \eqref{eq:egF1ofstr4-cut}
and extracting the terms with wanted color factors,
one can obtain the cut parts of one-loop color-ordered form factors in terms
of sums of products of tree-level color-ordered blocks.
For example, the color-ordered tree product
\begin{align}
\label{eq:sewing}
\int d\mathrm{PS}_{l_1,l_2} \sum_{\mathrm{helicty}}
\mathcal{F}^{(0)}_{4;\mathcal{O}_s} \big(
-l_1, -l_2, p_3, p_4
\big)
\mathcal{A}_4^{(0)}\big(l_2, l_1, p_1, p_2\big)
\end{align}
has the corresponding color factor product from \eqref{eq:CDofF4str}%
\footnote{We make use of the completeness relation of $\mathrm{SU}(N_c)$
Lie algebra
\begin{align}
\sum_a T^a_{ij} T^{a}_{kl}=\frac{1}{2}\delta_{il}\delta_{jk}
-\frac{1}{2N_c}\delta_{ij}\delta_{kl}\,.
\nonumber
\end{align}
}{}
\begin{align}
\mathrm{tr}\big(l_1 l_2 3 4\big)
f^{2 l_2 b} f^{l_1 1 b} =
\frac{1}{4}\Big( N_c \mathrm{tr}(1234)
+\mathrm{tr}(12)\mathrm{tr}(34)
\Big)\,.
\end{align}
Comparing with the one-loop color structure in \eqref{eq:egF1ofstr4},
one can see that (\ref{eq:sewing}) contributes to the $s_{12}$-cut
of both $\mathcal{F}^{(1),\mathrm{s}}_{4;\mathcal{O}_s}
(p_1,p_2,p_3,p_4)$ and $\mathcal{F}^{(1),\mathrm{d}}_{4;\mathcal{O}_s}
(p_1,p_2\,|\,p_3,p_4)$.
To be complete, these two color-stripped form factors also receive contributions from
other color orderings which have nonzero components of $\mathrm{tr}(1234)$ and $\mathrm{tr}(12)\mathrm{tr}(34)$.
Further details of color decomposition are provided in Appendix \ref{app:colordecom}.

In the remaining part of this subsection, we briefly discuss some features of the evanescent form factor results.
Consider the single-trace operator $\mathcal{O}^e_{10;\mathrm{s}+;1}$
defined in (\ref{eq:changebase}).
Its color-ordered 4-gluon form factors
(associated with $\mathrm{tr}(1234)$ and $\mathrm{tr}(12)\mathrm{tr}(34)$ respectively) are
\begin{align}
\label{eq:44strstr}
& \mathcal{F}^{(1),\mathrm{s}}_{4;\mathcal{O}^e_{10;\mathrm{s}+;1}}
 (p_1,p_2,p_3,p_4) =
N_c\Big[ c_{b,1}(s_{12}) I_{b}(s_{12})
+  c_{t,1}(s_{12}) I_{t}(s_{12})\Big]
+\mathrm{cyclic\  of\ }(1,2,3,4)\,,
\\
\label{eq:44strdtr}
&\mathcal{F}^{(1),\mathrm{d}}_{4;\mathcal{O}^e_{10;\mathrm{s}+;1}}
(p_1,p_2\, |\, p_3,p_4)
\nonumber\\
&=
\Big[(1+\mathrm{sgn}_{\mathcal{O}_s})
\Big(
c_{b,1}(s_{12}) I_{b}(s_{12})
+c_{t,1}(s_{12}) I_{t}(s_{12})
+(3\leftrightarrow 4)
\Big)+(1\leftrightarrow 4,2\leftrightarrow3)
\Big]
\nonumber\\
&+\Big[(1+\mathrm{sgn}_{\mathcal{O}_s})
\Big(
c_{b,2}(s_{23}) I_{b}(s_{23})
+c_{t,2}(s_{23}) I_{t}(s_{23})
+(1\leftrightarrow 3,2\leftrightarrow 4)
\Big)+(1\leftrightarrow 2 )
\Big]\,.
\end{align}
Here $I_{b}(s_{ij})$ and $I_{t}(s_{ij})$ refer to the
$s_{ij}$ bubble integral and the $s_{ij}$ one-mass triangle integral as in Figure~\ref{fig:MIset},
and $\mathrm{sgn}(\mathcal{O}_s)$ is the sign change of the operator
$\mathcal{O}_s$ under reflection.
Coefficients $c_{t,1}$ and $c_{b,1}$ can be computed using the unitarity cut (\ref{eq:sewing}),
which  contribute to both
$\mathcal{F}^{(1),\mathrm{s}}_{4;\mathcal{O}^e_{10;\mathrm{s}+;1}} (p_1,p_2,p_3,p_4)$
and $\mathcal{F}^{(1),\mathrm{d}}_{4;\mathcal{O}^e_{10;\mathrm{s}+;1}}
(p_1,p_2\, |\, p_3,p_4)$.
Coefficients $c_{t,2}$ and $c_{b,2}$ come from a product of
tree blocks with another choice of color ordering
in (\ref{eq:CDofF4str}), see details in (\ref{eq:color44-s}).

The coefficients $c_{i}$ are functions depending on the Lorentz product of polarization vector $\{e_j \}$
and external momenta $\{ p_j\}$, as well as the dimensional regularization parameter $\epsilon$.
The triangle integrals capture the universal IR divergences, and their coefficients are
proportional to tree-level evanescent form factors:%
\footnote{
The results of (\ref{eq:ct1}) and (\ref{eq:ct2})
only differ by an overall $s_{ij}$ factor, which is not a universal
property of general operators.
In this example, we choose the operator whose tree-level planar
amplitude is invariant under $S_4$ permutation, so
  $c_{t,1}(s_{12})$ and $c_{t,2}(s_{23})$ happen to be proportional.
}
\begin{align}
\label{eq:ct1}
c_{t,1}(s_{12})&=
- s_{12} \mathcal{F}^{(0)}_{4;\mathcal{O}^e_{10;\mathrm{s}+;1}}(1234)
=-8 s_{12} \sum_{i=1}^{12} \kineF(\evPr_{\mathbf{1},i})\,,
\\
\label{eq:ct2}
c_{t,2}(s_{23})&=
 \frac{s_{23}}{2} \Big(
\mathcal{F}^{(0)}_{4;\mathcal{O}^e_{10;\mathrm{s}+;1}}(1243)
+\mathcal{F}^{(0)}_{4;\mathcal{O}^e_{10;\mathrm{s}+;1}}(1342)
\Big)
=8 s_{23}\sum_{i=1}^{12} \kineF(\evPr_{\mathbf{1},i}) \,,
\end{align}
where $\kineF(\evPr_{\mathbf{1},i})$ are given in (\ref{eq:delta1}).
The leading $\mathcal{O}(\epsilon^0)$ order of the coefficients
$c_{b,1}$ and $c_{b,2}$ can also be written as linear combinations of
$\kineF(\evPr_{\mathbf{1},i})$ and therefore also vanish in four dimensions:
\begin{align}
c_{b,1}(s_{12})&=
8 \big(\sum_{i=1}^4 \kineF(\evPr_{\mathbf{1},i})
-3\sum_{j=5}^{12} \kineF(\evPr_{\mathbf{1},j})\big)
+\mathcal{O}(\epsilon)\,,
\\
c_{b,2}(s_{23})&=
8
\big(
-\sum_{i=9}^{12} \kineF(\evPr_{\mathbf{1},i})
+3\sum_{j=1}^{8} \kineF(\evPr_{\mathbf{1},j})
\big)
+\mathcal{O}(\epsilon)\,.
\end{align}
The $\mathcal{O}(\epsilon)$ order terms of  bubble coefficients which are not shown above
also have physical meaning, for they capture the \emph{finite} mixing from evanescent operators
to the physical ones and are expected to be important for two-loop calculation.
They give rise to a finite part of the one-loop form factors of the evanescent operators.
In the limit of dimension four, such finite contributions are equal to
linear combinations of the tree-level form factors of physical operators.


\subsection{IR subtraction and UV renormalization}

\label{sec:IRUV}

In this subsection we discuss the one-loop renormalization
of the gluonic operators, including both evanescent ones and physical ones.
The bare form factors contain both IR and UV divergences.
The renormalization $Z$-matrix can be obtained from the UV divergences of form factors.
It is convenient to obtain the UV divergence by subtracting the universal IR divergences
from the total divergence of a form factor.
Below we first give some details about the structure of the IR divergences.

The IR divergence of a one-loop form factor
can be given as a one-loop correction function acting on its tree-level form factor \cite{Catani:1998bh}:
\begin{equation}
\mathbf{F}^{(1)}_{\mathcal{O},\mathrm{IR}}  =
\mathbf{I}_{\rm IR}^{(1)} (\epsilon)
\mathbf{F}^{(0)}_{\mathcal{O}} , \qquad
\mathbf{I}_{\rm IR}^{(1)}(\epsilon)= \frac{e^{\gamma_E\epsilon}}{\Gamma(1-\epsilon)}
\big(\frac{1}{\epsilon^2}+\frac{\beta_0}{2 C_A\epsilon}\big)
\sum_{i<j}(-s_{ij})^{-\epsilon} \mathbf{T}_i\cdot \mathbf{T}_j\,,
\label{eq:catani}
\end{equation}
where $\beta_0=11C_A/3$ is the one-loop beta function, and $\mathbf{T}_i\cdot \mathbf{T}_j$  acts over the color factor
through taking Lie bracket of the adjoint vector $T^a$ of the $i$-th and $j$-th gluon, for instance
\begin{equation}
\begin{aligned}
\mathbf{T}_i\cdot \mathbf{T}_j\ \mathrm{tr}(X T^{a_i} Y T^{a_j} Z)
&=\sum_b\mathrm{tr}(X [T^b,T^{a_i}] Y [T^b,T^{a_j}] Z)\,.
\end{aligned}
\end{equation}
Here $X,\ Y,\ Z$ represent strings of adjoint vectors which do not
involve the $i$-th or $j$-th gluon.
Take the minimal form factor of a single-trace length-4 operator $\mathcal{O}_s$
as an example.
The color decomposition of its tree level form factors is given in
(\ref{eq:CDofF4str}), so together with (\ref{eq:catani}) one has
\begin{align}
\mathbf{I}_{\rm IR}^{(1)}(\epsilon)\mathbf{F}^{(0)}_{4;\mathcal{O}_s}
= \frac{e^{\gamma_E\epsilon}}{\Gamma(1-\epsilon)}
(\frac{1}{\epsilon^2}+\frac{\beta_0}{2C_A\epsilon})
\mathbf{C}(1,2,3,4)
\mathcal{F}^{(0)}_{4;\mathcal{O}_s}\big(p_1,p_{2},p_{3},
p_{4}\big)
+\text{Perm.}\{2,3,4\}\,.
\end{align}
Here $\mathbf{C}(1,2,3,4)$ is a sum of trace bases
with coefficients dependent on $s_{ij}$ and $\epsilon$:
\begin{align}
\mathbf{C}(1,2,3,4)&=-N_c \sum_{i=1}^4 (-s_{i,i+1})^{-\epsilon}
 \mathrm{tr} (1234 )
  \\
&+
\Big(-(-s_{12})^{-\epsilon}-(-s_{34})^{-\epsilon}
+(-s_{13})^{-\epsilon}+(-s_{24})^{-\epsilon}\Big)
\mathrm{tr}(12)\mathrm{tr}(34)
\nonumber\\
&+
\Big(-(-s_{14})^{-\epsilon}-(-s_{23})^{-\epsilon}
+(-s_{13})^{-\epsilon}+(-s_{24})^{-\epsilon}\Big)
\mathrm{tr}(14)\mathrm{tr}(23)\,.
\nonumber
\end{align}
Take series expansion of $\mathbf{C}(1,2,3,4)$ in $\epsilon$.
The leading order of the coefficient in the first line is of
$\mathcal{O}(\epsilon^0)$ and the leading orders of the coefficients in
the second and third line are of $\mathcal{O}(\epsilon)$.
This means that the IR divergence of $N_c$ leading form factors like
$\mathcal{F}^{(1),\mathrm{s}}_{4;\mathcal{O}_s}(p_1,p_2,p_3,p_4)$ are of
$\mathcal{O}(\epsilon^{-2})$ while the IR divergences of
$N_c$ sub-leading form factors like
$\mathcal{F}^{(1),\mathrm{d}}_{4;\mathcal{O}_s}
(p_1,p_2\,|\,p_3,p_4)$
are of $\mathcal{O}(\epsilon^{-1})$.
Similar analysis for double-trace operators and next-to-minimal form factors
are given in Appendix \ref{app:IR}, and the structure about the
$\epsilon$-expasion of the IR divergences of
$N_c$ leading and sub-leading form factors are the same.
We have checked that our results are consistent with these properties.


After subtracting the IR divergences,
the remaining UV divergences require renormalization of the operator and the coupling constant. The renormalized form factor
(with $n$ external gluons and for a length-$L_i$ operator multiplied with the bare $g$ to the power $m_i$)
can be given in the following form as
(see \emph{e.g.}~\cite{Jin:2020pwh} for detailed discussions)
\begin{align}
\label{eq:1loopRG}
\mathbf{F}^{(1)}_{\mathcal{O}_i,R}=\mathbf{F}^{(1)}_{\mathcal{O}_i,B}
+\sum_j \big(Z^{(1)}\big)_i^{\ j}\mathbf{F}^{(0)}_{\mathcal{O}_j,B}
-\frac{n-L_i+m_i}{2}\frac{\beta_0}{\epsilon}
\mathbf{F}^{(0)}_{\mathcal{O}_i,B}\,.
\end{align}
Subtracting the IR divergence from the bare form factor, (\ref{eq:1loopRG}) reads
\begin{align}
\label{eq:1loopRG2}
0=
(\mathbf{F}^{(1)}_{\mathcal{O}_i,B}
-\mathbf{F}^{(1)}_{\mathcal{O}_i,\mathrm{IR}})\big|_{\mathrm{div}}
+\sum_j  \big(Z^{(1) }\big)_i^{\ j} \mathbf{F}^{(0)}_{\mathcal{O}_j,B}
-\frac{n-L_i+m_i}{2}\frac{\beta_0}{\epsilon}
\mathbf{F}^{(0)}_{\mathcal{O}_i,B}\,.
\end{align}
By expanding one-loop and tree-level form factors in trace color bases,
(\ref{eq:1loopRG}) can be decomposed to different color ordered components.

In the $\overline{\mathrm{MS}}$ scheme \cite{Bardeen:1978yd}, there is no mixing from evanescent operators
to the physical ones, and $Z^{(1)}$ has the general form
\begin{align}
\label{eq:UVZ1}
Z^{(1)}=
\left(\begin{array}{c|c}
Z^{(1)}_{\mathrm{p}\rightarrow \mathrm{e}} $\,$ &
$\,$ Z^{(1) }_{\mathrm{p}\rightarrow \mathrm{e}} \\[0.1em]
\hline\rule{0pt}{1\normalbaselineskip}
0 & $\,$ Z^{(1) }_{\mathrm{e}\rightarrow \mathrm{e}}
\end{array}
\right)\,,
\end{align}
where the sub-matrices are all of order $\mathcal{O}(\epsilon^{-1})$, and subscripts
$\mathrm{p}$ and $\mathrm{e}$ refer to physical and evanescent operators respectively.
Note that although at one-loop level evanescent operators do not mix to physical ones at the order of $\mathcal{O}(\epsilon^{-1})$,
physical operators in general do mix to evanescent ones.

To be concrete, consider the form factors $\mathbf{F}^{(1)}_{4;\mathcal{O}_{4;s;i}}$
of single-trace length-4 operators as (\ref{eq:egF1ofstr4}),
the renormalization formulae of
leading  and sub-leading  color-ordered components
can be given as
\begin{align}
\mathcal{F}^{(1),\mathrm{s}}_{4;\mathcal{O}_{4;s;i},R}(p_1,p_2,p_3,p_4)
&=\mathcal{F}^{(1),\mathrm{s}}_{4;\mathcal{O}_{4;s;i},B}(p_1,p_2,p_3,p_4)
+\sum_j \big(Z^{(1)}_{4,\mathrm{s}\rightarrow 4,\mathrm{s}}\big)_i^{\ j}
\mathcal{F}^{(0)}_{4;\mathcal{O}_{4;s;j};B}(p_1,p_2,p_3,p_4)\,,
\nonumber \\
\mathcal{F}^{(1),\mathrm{d}}_{4;\mathcal{O}_{4;s;i},R}(p_1,p_2\, |\, p_3,p_4)
&=\mathcal{F}^{(1), \mathrm{d}}_{4;\mathcal{O}_{4;s;i},B}(p_1,p_2\, |\, p_3,p_4)
+\sum_k \big(Z^{(1)}_{4,\mathrm{s}\rightarrow 4,\mathrm{d}}\big)_i^{\ k}
\mathcal{F}^{(0)}_{4;\mathcal{O}_{4;d;k};B}(p_1,p_2\, |\, p_3,p_4)\,,
\end{align}
where $Z^{(1)}_{4,s\rightarrow 4,s}$ and $Z^{(1)}_{4,s\rightarrow 4,d}$
represent the mixing from length-4 single-trace operators
to length-4 single-trace and double-trace operators respectively.
In Table \ref{tab:FtoZ} we show that the correspondence between renormalization matrix elements and form factors
(for example, for the dimension-10 evanescent operators).
In particular, matrix elements of $Z^{(1)}_{4,s\rightarrow 4,s}$,
$Z^{(1)}_{4,s\rightarrow 4,d}$, $Z^{(1)}_{4,d\rightarrow 4,s}$
and $Z^{(1)}_{4,d\rightarrow 4,d}$ can be obtained from two different
form factors, and this provides a consistency check of our calculation.

\begin{table}[!t]
\centering
\caption{\label{tab:FtoZ} Matrix elements of $Z^{(1)}$ read from
each type of form factors.}
\vspace{0.4cm}
\begin{tabular}{|c|c|c|c|c|c|c|}
\hline
form factor &
$\mathbf{F}^{(1)}_{4;\mathcal{O}_{4;s}}$ & $\mathbf{F}^{(1)}_{4;\mathcal{O}_{4;d}}$ & $\mathbf{F}^{(1)}_{5;\mathcal{O}_{4;s}}$ & $\mathbf{F}^{(1)}_{5;\mathcal{O}_{4;d}}$ &
$\mathbf{F}^{(1)}_{5;\mathcal{O}_{5;s}}$ & $\mathbf{F}^{(1)}_{5;\mathcal{O}_{5;d}}$
\\
\hline
$Z^{(1)}_{4,s\rightarrow 4,s}$, $Z^{(1)}_{4,s\rightarrow 4,d}$ &
$\surd$ & $\times$ & $\surd$ & $\times$ & $\times$ & $\times$
\\
\hline
$Z^{(1)}_{4,d\rightarrow 4,s}$, $Z^{(1)}_{4,d\rightarrow 4,d}$ &
$\times$ & $\surd$ & $\times $ & $\surd$ & $\times$ & $\times$
\\
\hline
$Z^{(1)}_{4,s\rightarrow 5,s}$, $Z^{(1)}_{4,s\rightarrow 5,d}$ &
$\times$ & $\times$ & $\surd$ & $\times$ & $\times$ & $\times$
\\
\hline
$Z^{(1)}_{4,d\rightarrow 5,s}$, $Z^{(1)}_{4,d\rightarrow 5,d}$ &
$\times$ & $\times$ &  $\times$ & $\surd$ & $\times$ & $\times$
\\
\hline
$Z^{(1)}_{5,s\rightarrow 5,s}$, $Z^{(1)}_{5,s\rightarrow 5,d}$ &
$\times$ & $\times$ & $\times$ &  $\times$ & $\surd$ & $\times$
\\
\hline
$Z^{(1)}_{5,d\rightarrow 5,s}$, $Z^{(1)}_{5,d\rightarrow 5,d}$ &
$\times$ & $\times$ & $\times$ & $\times$ & $\times$ & $\surd$
\\
\hline
\end{tabular}
\end{table}

Using the renormalization matrix, it is straightforward to obtain the dilation operator,
 defined as
\begin{align}
\mathbb{D}:=-\frac{d\log Z}{d\log \mu} =\sum_{l=1}^{\infty}
\Big( \frac{\alpha_s}{4\pi}
\Big)^l \mathbb{D}^{(l)}\,,
\end{align}
and the anomalous dimensions are given as eigenvalues of the dilatation operator.
The one-loop correction $\mathbb{D}^{(1)}$ of the set of dim-10 evanescent operators is
\begin{align}
\label{eq:Dila1}
\mathbb{D}^{(1)}=2\epsilon Z^{(1)}\,,
\end{align}
where $Z^{(1)}$ includes $Z^{(1)}_{4\rightarrow 4}$, $Z^{(1)}_{4\rightarrow 5}$,
$Z^{(1)}_{5\rightarrow 5}$.%
\footnote{
We remark that the $\alpha_s$ expansion of $\mathbb{D}$ depends on the coupling-power in the definition of operators.
In our definition,
we multiply $L-2$ powers of bare gauge coupling $g$ to each length-$L$ operator
as shown in \eqref{eq:str-dim10}-\eqref{eq:len5dtr}.
In this way, matrix elements
$Z^{(1)}_{4\rightarrow 4}$, $Z^{(1)}_{4\rightarrow 5}$ and
$Z^{(1)}_{5\rightarrow 5}$ contained in $Z^{(1)}$  are all of order $\mathcal{O}(\alpha_s)$.
}

The above renormalization matrix $Z^{(1)}$ is defined in  $\overline{\mathrm{MS}}$ scheme by considering only the UV divergence of the one-loop form factors.
Alternatively one can choose another scheme to absorb the mixing from
evanescent operators to the physical ones which takes place at the finite
order of one-loop form factors and arises from the contribution of $\mathcal{O}(\epsilon)$ terms in
bubble coefficients as discussed in previous subsection.
This corresponds to the finite renormalization scheme which was used for the renormalization of four-fermion evanescent
operators, see \emph{e.g.}~\cite{Bondi:1989nq, Buras:1989xd, Dugan:1990df, Herrlich:1994kh, Vasiliev:1996rd, DiPietro:2017vsp}.
Such a scheme choice will be useful in the discussion of two-loop calculation,
but it does not affect the one-loop order anomalous dimensions.
The scheme change is realized by appending to $Z^{(1)}$ (which is of order
$\mathcal{O}(\epsilon^{-1})$) a finite term:
\begin{align}
\label{eq:schemchange}
Z^{(1)}\rightarrow  Z^{(1)} +
\left(\begin{array}{c|c}
0 & $\,\,\,$ 0 $\,\,\,$ \\[0.1em]
\hline\rule{0pt}{1\normalbaselineskip}
 Z^{(1),\mathrm{fin}}_{\mathrm{e}\rightarrow \mathrm{p}} $\,$ & $\,\,\,$ 0 $\,\,\,$
\end{array}
\right)\,.
\end{align}
The sub-matrix $Z^{(1),\mathrm{fin}}_{\mathrm{e }\rightarrow \mathrm{p}} $ is of
order $\mathcal{O}(\epsilon^0)$ and extracted from
\begin{align}
\label{eq:1loopRG3}
0=(\mathbf{F}^{(1)}_{\mathcal{O}^e_{i},B}
-\mathbf{F}^{(1)}_{\mathcal{O}^e_{i},\mathrm{IR}} )\big|_{\mathrm{rational,4d}}
+\sum_j \big(Z^{(1),\mathrm{fin}}_{\mathrm{e}\rightarrow \mathrm{p}}\big)_i^{\ j}
\mathbf{F}^{(0)}_{\mathcal{O}^p_{j},B}\big|_{\mathrm{4d}}\,,
\end{align}
where $\mathcal{O}^e_{i}$ and $\mathcal{O}^p_{j}$   refer to
evanescent and physical operators.

\subsection{Renormalization matrices and anomalous dimensions}
\label{sec:Zmatrix}

In this subsection we discuss in detail the results of the one-loop renormalization matrix of dim-10 evanescent basis operators,
\emph{i.e.} the sub-matrix $Z^{(1)}_{\mathrm{e}\rightarrow\mathrm{e}}$ in (\ref{eq:UVZ1}).
The complete results including all dim-10 physical basis operators are given in Appendix~\ref{app:nevtoeva}.

The basis operators have been classified according to their $C$-even or $C$-odd properties, and
since $C$-even and $C$-odd sectors do not mix to each other, their Z-matrices can be written separately.
For $C$-even sector we arrange operators as
$\{ \mathcal{O}^e_{10;\mathrm{s}+;i} ,\mathcal{O}^e_{10;\mathrm{d}+;i},
\Xi^e_{10;\mathrm{s}+;i}, \Xi^e_{10;\mathrm{d}+;1}\}$.
For $C$-odd sector there is only one operator $ \mathcal{O}^e_{10;\mathrm{s}-;1}$.

The full $Z$-matrices of dim-10 evanescent operators are
\begin{align}
\label{eq:dim10evaZeven}
&Z^{(1)}_{10;\mathrm{e}+\rightarrow \mathrm{e}+}
=\frac{N_c}{\epsilon}\times
\left(
\begin{array}{ccc;{2pt/2pt}ccc|cc;{2pt/2pt}c}
$\,\,$ 3 & -\frac{8}{3} $\,$  & 0 $\,$ & $\,\,\,$ 0 & \frac{16}{N_c} & $\,$ 0 $\,$ &
$\,\,\,\,$  0 & $\,\,\,$ 0 & $\,\,$   0  \\[0.4em]
 -\frac{10}{3} & $\,\,$ \frac{14}{3} & 0 $\,$ & $\,$ -\frac{20}{N_c} & 0 & $\,$ 0 $\,$ &
$\,\,\,\,$ 0 & $\,\,\,$ 0 & $\,\,$  0  \\[0.4em]
$\,\,$ 2 & $\,\,$ 1 & \frac{19}{3} $\,$ & $\,\,\,\,$ \frac{12}{N_c} & 0 & $\,$ 0 $\,$ &
 $\,$ -\frac{112}{9} & - \frac{20}{9} & $\,$ -\frac{224}{3 N_c} \\[0.4em]
\hdashline[2pt/2pt] \rule{0pt}{0.9\normalbaselineskip}
$\,\,$ 0 & -\frac{8}{N_c} $\,$  & 0 $\,$ & $\,\,\,$ 3 & \frac{16}{3} & $\,$ 0 $\,$ &
 $\,\,\,\,$  0 & $\,\,\,$ 0  & $\,\,$  0 \\[0.4em]
$\,\,$ 0 & -\frac{3}{N_c} $\,$  & 0 $\,$ & $\,\,\,$ \frac{20}{3} &   \frac{5}{3}   & $\,$ 0 $\,$ &
 $\,\,\,\,$ 0 & $\,\,\,$ 0  & $\,\,$  0  \\[0.4em]
$\,\,$ 0 & $\,\,$ \frac{3}{N_c} & \frac{2}{N_c} $\,$  & -4 $\,$ & 4 & $\,$ \frac{25}{3} $\,$ &
 $\,\,\,\,$  0 & -\frac{56}{3 N_c} $\,$ & $\,\,$ 4 \\[0.4em]
 \hline \rule{0pt}{0.9\normalbaselineskip}
$\,\,$ 0 & 0 & 0  $\,$ & $\,\,\,$ 0 & 0 & $\,$ 0 $\,$ &
$\,\,\,\,$ \frac{32}{9} & - \frac{26}{9} & $\,\,\,$ \frac{52}{3 N_c} \\[0.4em]
$\,\,$ 0 & 0 & 0  $\,$ & $\,\,\,$ 0 & 0 & $\,$ 0 $\,$ &   -\frac{55}{9} & $\,\,\,$  \frac{52}{9}  & -\frac{110}{3 N_c} \\[0.4em]
\hdashline[2pt/2pt] \rule{0pt}{0.9\normalbaselineskip}
$\,\,$ 0 & 0 & 0  $\,$ & $\,\,\,$ 0 & 0 & $\,$ 0 $\,$ &
 $\,\,\,\,$ 0  & - \frac{26}{3 N_c} & $\,\,$ \frac{28}{3} \\[0.4em]
\end{array}
\right)
\,,
\\
\label{eq:dim10evaZodd}
&Z^{(1)}_{10;\mathrm{e}-\rightarrow \mathrm{e}-}
=\frac{5N_c}{ \epsilon}\,.
\end{align}
In the matrix \eqref{eq:dim10evaZeven}, the length-4 and length-5 operators are separated by solid lines,
and within each length the single-trace and double-trace operators
are separated by dashed lines. 
There is no mixing from $\mathcal{O}^e_{10;\mathbf{x}+;i}$ with $i<3$
to $\mathcal{O}^e_{10;\mathbf{x}+;3}$ ($\mathbf{x}=s,d$)
or length-5 operators. This is because that
$\mathcal{O}^e_{10;\mathbf{x}+;i}$ with $i<3$ are total derivatives of
lower dimensional tensor operators, see Appendix \ref{app:basechoice1} for further discussion on this point.

One-loop dilatation operator $\mathbb{D}^{(1)}$ for
$C$-even and $C$-odd sectors can be obtained by plugging
(\ref{eq:dim10evaZeven}) and (\ref{eq:dim10evaZodd}) into (\ref{eq:Dila1}).
We denote the eigenvalues, \emph{i.e.}~one-loop anomalous dimensions by $\hat{\gamma}^{(1)}_{+,e}$ for the C-even operators and $\hat{\gamma}^{(1)}_{-,e}$ for the C-odd operator.
For the single C-odd operator, the anomalous dimension is
\begin{equation}
\label{eq:einodd}
\hat{\gamma}^{(1)}_{-,e}=10 N_c\,.
\end{equation}
Below we focus on the more non-trivial C-even sector.
The 4 to 5 mixing matrix elements $Z^{(1)}_{4\rightarrow 5}$ does not effect the eigenvalues of length-4 and length-5
sectors because at 1-loop level $Z$ matrix is upper-triangular.

For the length-4 operators, the six eigenvalues are determined by the following equations
\begin{align}
0&=
\Big(\omega_1-\frac{50}{3}\Big) \Big(\omega_1-\frac{38}{3}\Big)  \\
& \quad \times \bigg[
\omega_1^4 -\frac{74 \omega_1^3}{3}
+\omega_1^2 \Big(\frac{124}{3}-\frac{640}{N_c^2}\Big)
+\omega _1 \Big(\frac{45448}{27}+\frac{6400}{3 N_c^2}\Big)
-\Big(\frac{202400}{81}+\frac{317440}{9 N_c^2}\Big)
\bigg]\,,
\nonumber
\end{align}
where $\einv_1:=\hat{\gamma}^{(1)}_{+,e}/N_c$.
Taking the large $N_c$ limit and expanding the anomalous dimensions up to
$\mathcal{O}(1/N_c)$, one has
\begin{align}
\label{eq:einevenL4}
\hat{\gamma}^{(1), L=4}_{+,e}&=N_c \bigg\{
-\frac{22}{3}-\frac{400}{21 N_c^2}, \ \frac{38}{3}, \, \frac{50}{3}, \ \frac{50}{3}+\frac{520}{3 N_c^2}, \
\frac{23\pm\sqrt{345}}{3} -\frac{4 (3105\pm211 \sqrt{345})}{161 N_c^2} \bigg\}.
\end{align}
We note that in the limit of large $N_c$ there is a double degeneracy of the eigenvalue
$50N_c/3$, which is broken by sub-leading $N_c$ correction.

Similarly, the three eigenvalues of length-5 operators are given as solutions of
\begin{align}
0&=
\omega_1^3-\frac{112 \omega_1^2}{3}
+\omega_1 \Big(360-\frac{11440}{9 N_c^2}\Big)
-\Big(\frac{5824}{27}-\frac{45760}{27 N_c^2}\Big)\,,
\end{align}
and the anomalous dimensions expanded up to
$\mathcal{O}(1/N_c)$ are
\begin{align}
\label{eq:einevenL5}
\hat{\gamma}^{(1), L=5}_{+,e}&= N_c \bigg\{
\frac{56}{3}+\frac{5720}{3N_c^2}, \ \
\frac{2(14\pm\sqrt{170})}{3} -\frac{286 (170\pm13 \sqrt{170})}{51 N_c^2}
\bigg\}\,.
\end{align}

We have also computed the one-loop form factors and performed the one-loop renormalization
for all the dim-10 physical basis operators in Appendix~\ref{app:nevtoeva}.
There are 20 (15) single-trace (double-trace)
physical operators with length four,
and 4 (3) single-trace (double-trace)
physical  operators with length five.
Their explicit definitions are given in \ref{app:dim10oper4d-len4} and \ref{app:dim10oper4d-len5}.
The operator mixing sub-matrices
$Z^{(1) }_{10;\mathrm{p} \rightarrow \mathrm{p}}$ and
$Z^{(1) }_{10;\mathrm{p}\rightarrow \mathrm{e}}$ in (\ref{eq:UVZ1}) and the one-loop anomalous dimensions of physical operators
are given in Appendix \ref{app:Znon-eva}.
As discussed in Section \ref{sec:IRUV}, one can choose a finite
renormalization scheme as (\ref{eq:schemchange})
to absorb the finite mixing from evanescent operators to the physical ones.
The sub-matrix $Z^{(1),{\rm fin}}_{\mathrm{e}\rightarrow \mathrm{p}}$ defined in (\ref{eq:schemchange})
is   given in Appendix \ref{app:finiteZ}.
These results will be needed for the two-loop renormalization.

\section{Summary and discussion}
\label{sec:summary}

In this paper we initiate the study of the evanescent operators in pure Yang-Mills theory.
Such operators vanish when the number of spacetime dimensions is four but have non-trivial results in general $d$ dimensions.
We provide a systematic construction of the gluonic evanescent operators based on the study of their form factors in general $d$ dimensions.
The gluonic evanescent operators start to appear at canonical dimension $\Delta_0 =10$, and they are expected to take an important part in the study of high dimensional operators of $\Delta_0 \geq10$ in any effective field theory that contains a Yang-Mills sector.
We also compute the one-loop form factors of gluonic evanescent operators via
unitarity method in $d$-dimensions.
The one-loop operator-mixing renormalization matrices are given explicitly for the  complete dimension-10 basis operators (including both evanescent and physical operators).

A concrete further study is to explore the physical effect of evanescent operators by studying the two-loop renormalization,
in which it is important to include the evanescent operators to obtain the correct two-loop physical anomalous dimensions.
Another interesting problem is to consider the gauge theory at the Wilson-Fisher fixed point \cite{Wilson:1971dc}, where the spacetime is in general non-integer dimensions and the evanescent operators are also physical operators; in such case it is interesting to see if the evanescent operators can render the gauge theory non-unitary, as observed for the scalar theory in \cite{Hogervorst:2015akt}.
We leave these studies to another work \cite{twolooptoappear}.
It would be also interesting to generalize the study of this work to evanescent operators in gravity theories;
some discussion on the evanescent effect in gravity has been considered in \cite{Bern:2015xsa, Bern:2017puu}.


\acknowledgments

It is a pleasure to thank J.P.~Ma for discussions.
We also thank Mikael Chala for correspondence.
This work is supported in part by the National Natural Science Foundation of China (Grants No.~12175291, 11935013, 11822508, 12047503),
and by the Key Research Program of the Chinese Academy of Sciences, Grant NO. XDPB15.
We also thank the support of the HPC Cluster of ITP-CAS.

\appendix

\section{Primitive kinematic operators}
\label{app:primi}

In this appendix we give some details on the
primitive kinematic operators defined in Section~\ref{sec:step1}.
Appendix \ref{app:why12} provides a proof of  (\ref{eq:lessthan12}), which guarantees
the completeness of primitive kinematic operators.
The expressions of the primitive evanescent kinematic operators
with length four are given in Appendix~\ref{app:eva-kine}.
Appendix~\ref{app:indie} explains why the kinematic operators
$\{ \mathbf{s}_{12}^{n_{i,12}}...\ \mathbf{s}_{34}^{n_{i,34}}\evPr_i \}$
with the same mass dimension are linearly independent,
which results in the counting  shown in Table~\ref{tab:evakine}.

\subsection{Proof of (\ref{eq:lessthan12})}

\label{app:why12}

It is stated in (\ref{eq:lessthan12}) that
a generic length-4 kinematic operator
$\lfloor \mathcal{W}_1, \mathcal{W}_2, \mathcal{W}_3,
\mathcal{W}_4
\rfloor$ of dim $\Delta\geq12$ can always be
written as a sum like
\begin{align}
\label{eq:primitiveExp}
\lfloor \mathcal{W}_1, \mathcal{W}_2, \mathcal{W}_3,
\mathcal{W}_4\rfloor=
\sum_{i}\sum_{\vec{n}_i}
c_{\vec{n}_i} \prod_{1\leq a<b\leq4} \mathbf{s}_{ab}^{n_{i,ab}}
\gePr_{i},\quad
\mathrm{dim}(\gePr_i)\leq12\,.
\end{align}
Here $c_{ \vec{n} }$ are rational numbers for each choice of
$\vec{n}_i=\{n_{i,12},n_{i,13},n_{i,23},n_{i,14},n_{i,24},n_{i,34}\}$,
and $\mathbf{s}_{ab}^{n_{i,ab}}\gePr_{i}$
refers to inserting $n_{i,ab}$ pairs of $\{D_\mu, D^\mu\}$
into the $a$-th and $b$-th sites of the $i$-th kinematic operator
$\mathcal{K}_i$, as in \eqref{eq:insertSasDDpair}.
This can be actually generalized to generic length $L$:
a length-$L$ kinematic operator $\lfloor \mathcal{W}_1, ...
\mathcal{W}_L\rfloor$ of dim $\Delta\geq 4L-4$ can always be
written as a sum like
\begin{align}
\lfloor \mathcal{W}_1, \mathcal{W}_2, \mathcal{W}_3,
\mathcal{W}_4\rfloor=
\sum_{i}\sum_{\vec{n}_i}
c_{\vec{n}_i} \prod_{1\leq a<b\leq L} \mathbf{s}_{ab}^{n_{i,ab}}
\gePr_{i},\quad
\mathrm{dim}(\gePr_i)\leq 4L-4\,,
\end{align}
where $\vec{n}=\{n_{i,12},n_{i,13},\cdots,n_{i,(L-1)L}\}$.
The statement can be proved as follows:
\begin{enumerate}
\item
It  suffices to only consider the kinematic operators free of $\{D_\mu, D^\mu\}$ pairs.
For the case of length-$L$, the maximal mass dimension of such operator is $4L$, where every Lorentz index in $F_{\mu\nu}$ is contracted with a $D_\rho$.

\item
Consider such a dim-$4L$ kinematic operator $\mathfrak{O}=\lfloor \mathcal{W}_1, ...
\mathcal{W}_L\rfloor$.
It contains $2L$ $D$ and all of them contract with $F$.
At least one of $L$ sites, say $\mathcal{W}_a$, contains $D$,
and therefore can be written as
$\mathcal{W}_a=(...)D_i F_{jk}$, where $(...)$ represents other $D$s or $1$ in front of $D_i$. Applying Bianchi identity, one has
\begin{align}
\label{eq:why12-1}
&\mathfrak{O}=\lfloor \mathcal{W}_1 ,...,\mathcal{W}_{a-1},(...)D_i F_{jk}
,\mathcal{W}_{a+1},...,\mathcal{W}_L\rfloor
\\
&=- \lfloor \mathcal{W}_1 ,..,\mathcal{W}_{a-1},(...)D_j F_{ki}
,\mathcal{W}_{a+1},..,\mathcal{W}_L\rfloor
- \lfloor \mathcal{W}_1 ,..,\mathcal{W}_{a-1},(...)D_k F_{ij}
,\mathcal{W}_{a+1},..,\mathcal{W}_L\rfloor\,.
\nonumber
\end{align}
Since the original $F_{jk}$ must contract with $D_j$, $D_k$ from other sites,
the first term on the r.h.s of (\ref{eq:why12-1}) contains
two  $D_j$ and  the second term contains two $D_k$.
After stripping of these identical $D$s, $\mathfrak{O}$ becomes a sum
of two kinematic operators with dimension $4L-2$.

\item
Consider a dim-$(4L-2)$ kinematic operator  $\mathfrak{O}'=\lfloor \mathcal{W}_1, ...
\mathcal{W}_L\rfloor$ which is free of $D_i$ pairs.
There exist two sites, say $\mathcal{W}_a$ and $\mathcal{W}_b$, whose $F$ share a pair of Lorentz indices,
and therefore can be written as $(...)F_{ij}$ and $(...)F_{jk}$.
They contract with $D_i$, $D_k$ from other sites.

\begin{enumerate}
\item
If one of the other $L-2$ sites, say $\mathcal{W}_c$, can be written as $(...)D_l F_{mn}$,
then following the same argument as \eqref{eq:why12-1},
the operator $\mathfrak{O}'$ can be reduced to dimension-$(4L-4)$ operators with a $DD$-pair insertion.

\item
If none of the other $L-2$ sites contains $D_\mu$, then all the $D$s
must be placed in  site $a$ or site $b$. Especially,
$D_i$ must be contained by $\mathcal{W}_b$ and
$D_k$ must be contained by $\mathcal{W}_a$.
Applying Bianchi identity, one has
\begin{align}
\label{eq:why12-2}
&\mathfrak{O}'=\lfloor \mathcal{W}_1 ,...,\mathcal{W}_{a-1},(...)D_k    F_{ij}
,\mathcal{W}_{a+1},...,\mathcal{W}_{b-1},(...)D_i F_{jk},\mathcal{W}_{b+1},...,\mathcal{W}_L\rfloor
\nonumber\\
&=-\lfloor \mathcal{W}_1 ,...,\mathcal{W}_{a-1},(...)D_i    F_{jk}
,\mathcal{W}_{a+1},...,\mathcal{W}_{b-1},(...)D_i F_{jk},\mathcal{W}_{b+1},...,\mathcal{W}_L\rfloor
\nonumber\\
&-\lfloor \mathcal{W}_1 ,...,\mathcal{W}_{a-1},(...)D_j    F_{ki}
,\mathcal{W}_{a+1},...,\mathcal{W}_{b-1},(...)D_i F_{jk},\mathcal{W}_{b+1},...,\mathcal{W}_L\rfloor
\nonumber\\
&=-\frac{1}{2}\lfloor \mathcal{W}_1 ,...,\mathcal{W}_{a-1},(...)D_i    F_{jk}
,\mathcal{W}_{a+1},...,\mathcal{W}_{b-1},(...)D_i F_{jk},\mathcal{W}_{b+1},...,\mathcal{W}_L\rfloor
\,.
\end{align}
The r.h.s of (\ref{eq:why12-2}) contains a pair of identical $D$s, so it can be also reduced to dimension $4L-4$.
\end{enumerate}
\end{enumerate}
Thus we prove that the primitive operators in \eqref{eq:primitiveExp} has
at most dimension 12.

To find the complete set of primitive operators $\gePr_{i}$,
one can first enumerate all possible length-4 kinematic operators within dimension 12
and then find out the subset which is linearly independent within dimension 12, \emph{i.e.}~
\begin{equation}
\label{eq:54kplow12}
\sum_{i} f_i(s_{ab})\kineF(\gePr_i)
\neq0,
\quad
\mbox{with\ }\dim(f_i\kineF(\gePr_i)\leq 12 \,,
\end{equation}
where $f_i(s_{ab})$ are polynomials of Mandelstam variables $s_{ab}$.

This independent subset is found in following way.
Start with dimension 8, the lowest operator dimension for length-4 operators,
and find the subset that is linearly independent with constant coefficients,
which contains six elements.
Then consider dimension 10, and find the subset that is
linearly independent of the dim-10 operators  generated by
inserting  $DD$ pairs into six dim-8 ones, which contains 42 elements.
Then consider dimension 12, and find  the subset that is linearly
independent of the dim-12 operators  generated by inserting  $DD$ pairs
into six dim-8 ones and 42 dim-10 ones, which contains six elements.
So finally one finds that there are in total 54 $\gePr_i$,
which can be grouped into 27 evanescent ones and 27  physical ones, denoted by $\evPr_i$ and  $\phPr_i$ respectively.

\subsection{Primitive evanescent length-4 kinematic operators}
\label{app:eva-kine}

Below we provide  explicit expressions of the evanescent primitive kinematic operators $\evPr_{1,i}$
and $\evPr_{3,i}$,
following the discussion in Section~\ref{sec:step1}.

The 12 kinematic operators in the first class are
\begin{align}
\label{eq:KOdelta10}
&\evPr_{\mathbf{1},1} =\frac{1}{16}
\delta^{\mu_1\mu_2\mu_3\mu_4\rho}_{\nu_1\,\nu_2 \, \nu_3 \, \nu_4 \sigma}
\lfloor D_{\sigma}F_{\mu_1\mu_2}, F_{\mu_3\mu_4},
D_{\rho}F_{\nu_1\nu_2}, F_{\nu_3\nu_4} \rfloor \,,
\nonumber\\
&\evPr_{\mathbf{1},2}=  \ \evPr_{\mathbf{1},1}\big|_{
\lfloor\mathcal{W}_1,\mathcal{W}_2,\mathcal{W}_3,\mathcal{W}_4\rfloor
\rightarrow \lfloor\mathcal{W}_2,\mathcal{W}_1,\mathcal{W}_3,\mathcal{W}_4\rfloor},\quad
\evPr_{\mathbf{1},3}= \evPr_{\mathbf{1},1} \big|_{
\lfloor\mathcal{W}_1,\mathcal{W}_2,\mathcal{W}_3,\mathcal{W}_4\rfloor
\rightarrow \lfloor\mathcal{W}_1,\mathcal{W}_2,\mathcal{W}_4,\mathcal{W}_3\rfloor},\quad
\nonumber\\
&\evPr_{\mathbf{1},4}= \evPr_{\mathbf{1},1} \big|_{
\lfloor\mathcal{W}_1,\mathcal{W}_2,\mathcal{W}_3,\mathcal{W}_4\rfloor
\rightarrow \lfloor\mathcal{W}_2,\mathcal{W}_1,\mathcal{W}_4,\mathcal{W}_3\rfloor},\quad
\evPr_{\mathbf{1},5}=  \evPr_{\mathbf{1},1} \big|_{
\lfloor\mathcal{W}_1,\mathcal{W}_2,\mathcal{W}_3,\mathcal{W}_4\rfloor
\rightarrow \lfloor\mathcal{W}_1,\mathcal{W}_3,\mathcal{W}_2,\mathcal{W}_4\rfloor},\quad
\nonumber\\
&\evPr_{\mathbf{1},6}= \evPr_{\mathbf{1},1}\big|_{
\lfloor\mathcal{W}_1,\mathcal{W}_2,\mathcal{W}_3,\mathcal{W}_4\rfloor
\rightarrow \lfloor\mathcal{W}_2,\mathcal{W}_3,\mathcal{W}_1,\mathcal{W}_4\rfloor},\quad
\evPr_{\mathbf{1},7}= \evPr_{\mathbf{1},1}\big|_{
\lfloor\mathcal{W}_1,\mathcal{W}_2,\mathcal{W}_3,\mathcal{W}_4\rfloor
\rightarrow \lfloor\mathcal{W}_1,\mathcal{W}_4,\mathcal{W}_2,\mathcal{W}_3\rfloor},\quad
\nonumber\\
&\evPr_{\mathbf{1},8}=  \evPr_{\mathbf{1},1}\big|_{
\lfloor\mathcal{W}_1,\mathcal{W}_2,\mathcal{W}_3,\mathcal{W}_4\rfloor
\rightarrow \lfloor\mathcal{W}_2,\mathcal{W}_4,\mathcal{W}_1,\mathcal{W}_3\rfloor},\quad
\evPr_{\mathbf{1},9}= \evPr_{\mathbf{1},1}\big|_{
\lfloor\mathcal{W}_1,\mathcal{W}_2,\mathcal{W}_3,\mathcal{W}_4\rfloor
\rightarrow \lfloor\mathcal{W}_1,\mathcal{W}_3,\mathcal{W}_4,\mathcal{W}_2\rfloor},\quad
\nonumber\\
&\evPr_{\mathbf{1},10}= \evPr_{\mathbf{1},1}\big|_{
\lfloor\mathcal{W}_1,\mathcal{W}_2,\mathcal{W}_3,\mathcal{W}_4\rfloor
\rightarrow \lfloor\mathcal{W}_2,\mathcal{W}_3,\mathcal{W}_4,\mathcal{W}_1\rfloor},\quad
\evPr_{\mathbf{1},11}=   \evPr_{\mathbf{1},1}\big|_{
\lfloor\mathcal{W}_1,\mathcal{W}_2,\mathcal{W}_3,\mathcal{W}_4\rfloor
\rightarrow \lfloor\mathcal{W}_1,\mathcal{W}_4,\mathcal{W}_3,\mathcal{W}_2\rfloor},\quad
\nonumber\\
&\evPr_{\mathbf{1},12}= \evPr_{\mathbf{1},1}\big|_{
\lfloor\mathcal{W}_1,\mathcal{W}_2,\mathcal{W}_3,\mathcal{W}_4\rfloor
\rightarrow \lfloor\mathcal{W}_2,\mathcal{W}_4,\mathcal{W}_3,\mathcal{W}_1\rfloor}\,.
\end{align}

The 12 kinematic operators in the third class are
\begin{align}
\label{eq:KOsAAD}
&\evPr_{\mathbf{3},1} =\frac{1}{4}
\delta^{\mu_1\mu_2\mu_3\mu_4\mu_5}_{\nu_1\,\nu_2 \, \nu_3 \, \nu_4 \,\nu_5}
 \lfloor D_{\nu_3}F_{\mu_1\mu_2}, F_{\mu_5\rho},
D_{\mu_3}F_{\nu_1\nu_2}, D_{\mu_4\nu_4} F_{\nu_5\rho} \rfloor
  \,,
\nonumber\\
&\evPr_{\mathbf{3},2}=  \evPr_{\mathbf{3},1}\big|_{
\lfloor\mathcal{W}_1,\mathcal{W}_2,\mathcal{W}_3,\mathcal{W}_4\rfloor
\rightarrow \lfloor\mathcal{W}_2,\mathcal{W}_1,\mathcal{W}_3,\mathcal{W}_4\rfloor },\quad
\evPr_{\mathbf{3},3}= \evPr_{\mathbf{3},1} \big|_{
\lfloor\mathcal{W}_1,\mathcal{W}_2,\mathcal{W}_3,\mathcal{W}_4\rfloor
\rightarrow \lfloor\mathcal{W}_1,\mathcal{W}_2,\mathcal{W}_4,\mathcal{W}_3\rfloor },\quad
\nonumber\\
&\evPr_{\mathbf{3},4}= \evPr_{\mathbf{3},1} \big|_{
\lfloor\mathcal{W}_1,\mathcal{W}_2,\mathcal{W}_3,\mathcal{W}_4\rfloor
\rightarrow \lfloor\mathcal{W}_2,\mathcal{W}_1,\mathcal{W}_4,\mathcal{W}_3\rfloor },\quad
\evPr_{\mathbf{3},5}= \evPr_{\mathbf{3},1}\big|_{
\lfloor\mathcal{W}_1,\mathcal{W}_2,\mathcal{W}_3,\mathcal{W}_4\rfloor
\rightarrow \lfloor\mathcal{W}_1,\mathcal{W}_3,\mathcal{W}_2,\mathcal{W}_4\rfloor },\quad
\nonumber\\
&\evPr_{\mathbf{3},6}= \evPr_{\mathbf{3},1} \big|_{
\lfloor\mathcal{W}_1,\mathcal{W}_2,\mathcal{W}_3,\mathcal{W}_4\rfloor
\rightarrow \lfloor\mathcal{W}_1,\mathcal{W}_3,\mathcal{W}_4,\mathcal{W}_2\rfloor },\quad
\evPr_{\mathbf{3},7}= \evPr_{\mathbf{3},1} \big|_{
\lfloor\mathcal{W}_1,\mathcal{W}_2,\mathcal{W}_3,\mathcal{W}_4\rfloor
\rightarrow  \lfloor\mathcal{W}_1,\mathcal{W}_4,\mathcal{W}_2,\mathcal{W}_3\rfloor},\quad
\nonumber\\
&\evPr_{\mathbf{3},8} = \evPr_{\mathbf{3},1}\big|_{
\lfloor\mathcal{W}_1,\mathcal{W}_2,\mathcal{W}_3,\mathcal{W}_4\rfloor
\rightarrow \lfloor\mathcal{W}_2,\mathcal{W}_4,\mathcal{W}_1,\mathcal{W}_3\rfloor },\quad
\evPr_{\mathbf{3},9}= \evPr_{\mathbf{3},1} \big|_{
\lfloor\mathcal{W}_1,\mathcal{W}_2,\mathcal{W}_3,\mathcal{W}_4\rfloor
\rightarrow \lfloor\mathcal{W}_1,\mathcal{W}_4,\mathcal{W}_3,\mathcal{W}_2\rfloor },\quad
\nonumber\\
&\evPr_{\mathbf{3},10}= \evPr_{\mathbf{3},1} \big|_{
\lfloor\mathcal{W}_1,\mathcal{W}_2,\mathcal{W}_3,\mathcal{W}_4\rfloor
\rightarrow \lfloor\mathcal{W}_4,\mathcal{W}_1,\mathcal{W}_2,\mathcal{W}_3\rfloor },\quad
\evPr_{\mathbf{3},11}= \evPr_{\mathbf{3},1}\big|_{
\lfloor\mathcal{W}_1,\mathcal{W}_2,\mathcal{W}_3,\mathcal{W}_4\rfloor
\rightarrow \lfloor\mathcal{W}_4,\mathcal{W}_2,\mathcal{W}_1,\mathcal{W}_3\rfloor },\quad
\nonumber\\
&\evPr_{\mathbf{3},12}= \evPr_{\mathbf{3},1} \big|_{
\lfloor\mathcal{W}_1,\mathcal{W}_2,\mathcal{W}_3,\mathcal{W}_4\rfloor
\rightarrow \lfloor\mathcal{W}_4,\mathcal{W}_1,\mathcal{W}_3,\mathcal{W}_2\rfloor }\,.
\end{align}

\subsection{Dimension 10 length-5 kinematic operators}

\label{app:eva-kine5}

In Section \ref{sec:len5} we show the six length-5 kinematic operators
of dimension 10, which are related with each other by $S_3$ permutations.
Here we clarify the permutation for each $\evfive_{10;i}$.
\begin{align}
\evfive_{10;2}&= \evfive_{10;1} |_{
\lfloor {\cal W}_i, {\cal W}_j, {\cal W}_k, {\cal W}_l, {\cal W}_m\rfloor
\rightarrow
\lfloor {\cal W}_i, {\cal W}_j,  {\cal W}_k, {\cal W}_m, {\cal W}_l \rfloor
} \,,
\nonumber\\
\evfive_{10;3}&= \evfive_{10;1} |_{
\lfloor {\cal W}_i, {\cal W}_j, {\cal W}_k, {\cal W}_l, {\cal W}_m\rfloor
\rightarrow
\lfloor {\cal W}_i, {\cal W}_j, {\cal W}_l, {\cal W}_k, {\cal W}_m \rfloor
} \,,
\nonumber\\
\evfive_{10;4}&= \evfive_{10;1} |_{
\lfloor {\cal W}_i, {\cal W}_j, {\cal W}_k, {\cal W}_l, {\cal W}_m\rfloor
\rightarrow
\lfloor {\cal W}_i, {\cal W}_j, {\cal W}_l, {\cal W}_m, {\cal W}_k\rfloor
} \,,
\nonumber\\
\evfive_{10;5}&= \evfive_{10;1} |_{
\lfloor {\cal W}_i, {\cal W}_j, {\cal W}_k, {\cal W}_l, {\cal W}_m\rfloor
\rightarrow
\lfloor {\cal W}_i, {\cal W}_j, {\cal W}_m, {\cal W}_k, {\cal W}_l \rfloor
} \,,
\nonumber\\
\evfive_{10;6}&= \evfive_{10;1} |_{
\lfloor {\cal W}_i, {\cal W}_j, {\cal W}_k, {\cal W}_l, {\cal W}_m\rfloor
\rightarrow
\lfloor {\cal W}_i, {\cal W}_j, {\cal W}_m, {\cal W}_l, {\cal W}_k\rfloor
} \,.
\end{align}

\subsection{Linearly independence of kinematic operators}

\label{app:indie}

We show in this subsection that kinematic operators
$\{ \mathbf{s}_{12}^{n_{i,12}}...\ \mathbf{s}_{34}^{n_{i,34}}\evPr_i \}$
with the same mass dimension are linearly independent.
To prove this, it is enough to show that the
the evanescent kinematic form factors $\kineF(\evPr_i)$ are linearly independent as %
\begin{align}
\label{eq:27indie}
\sum_{i=1}^{27} \mathcal{R}_i(s_{ab}) \kineF(\evPr_i) \neq0\,,
\end{align}
where $\mathcal{R}_i(s_{ab})$ are arbitrary nonzero rational functions of $s_{ab}$.
If \eqref{eq:27indie} is true, $\{\kineF(\evPr_i)\}$ are linearly independent with polynomial $s_{ab}$ coefficients
of arbitrarily high  mass dimensions,
and thus the kinematic operators
$\{\mathbf{s}_{12}^{n_{i,12}}...\ \mathbf{s}_{34}^{n_{i,34}}\evPr_i\}$
are linearly independent and provide a basis of $(\mathfrak{K}^e)_{\Delta}$.

The proof of \eqref{eq:27indie} proceeds as follows.
We consider $\sum_{i}\mathcal{R}_i \kineF(\evPr_i) = 0$ as an equation
of to-be-determined variables $\{\mathcal{R}_i\}$.
Kinematic form factors $\kineF(\evPr_i)$ are known functions of $s_{ab}$ together with
monomials of Lorentz products containing external polarization vectors
$e_1^{\mu},e_2^{\mu},e_3^{\mu},e_4^{\mu}$. 
There are 138 such monomials such as
\begin{align}
(e_1\cdot e_2)( e_3\cdot e_4),\
(e_1\cdot p_2) (e_2\cdot p_4) (e_3\cdot e_4),\
(e_1\cdot p_2) (e_2\cdot p_4)(e_3\cdot p_1) (e_4\cdot p_2),
\ \cdots .
\end{align}
Requiring the coefficients of all these monomials vanish,
one obtains 138 equations about 27 $\mathcal{R}_i$, with
polynomial coefficients of  $s_{ab}$:
\begin{align}
\sum_{i=1}^{27}  f_{n,i}(s_{ab}) \mathcal{R}_i=0,\quad
n=1,...,138,
\end{align}
where $f_{n,i}(s_{ab})$ are polynomials of $s_{ab}$ and determined
by the expressions of $\kineF(\evPr_i)$.
One finds that there is no nonzero solution of $\{\mathcal{R}_i\}$ satisfying
these 138 equations, which means
no rational functions $\{ \mathcal{R}_i\}$ would turn (\ref{eq:27indie}) to
an equality.

It is worth mentioning that linear independence with $s_{ab}$ rational function coefficients
as shown in (\ref{eq:27indie}) is also the requirement of gauge invariant basis given in (\ref{eq:ABbasis}).
Since $\kineF(\evPr_i)$ already satisfy such condition, they must be linearly equivalent to the
 gauge invariant basis composed of $AAB$ and $BB$, which explains why
their total number are both 27.

By construction, the 54 general kinematic form factors $\kineF(\gePr_i)$ (including both
physical and evanescent ones) only satisfy (\ref{eq:54kplow12}),
and there exist 11 linear relations among them in the form of
\begin{equation}
\label{eq:54kphigh12}
\sum_{i} f_i(s_{ab})\kineF(\gePr_i)
=0,
\quad
\mbox{with\ }\dim(f_i\kineF(\gePr_i)> 12 \,,
\end{equation}
that explains why the number of $\gePr_i$ is 11 larger than the number of general
gauge invariant basis.

\section{Comment on minimally evanescent operator}

\label{app:step3}

The vanishing of the minimal form factor of an operator does not mean that its higher-point form factors are also zero.
An example is the following operator
\begin{align}
\label{eq:notexact1}
&\Omega_1=
\frac{1}{2} \text{tr}( D_3 F_{12} F_{46} D_5 F_{12} D_{34} F_{56})
-\frac{1}{2} \text{tr}( D_3 F_{12} F_{56} D_4 F_{12} D_{34} F_{56})
-\text{tr}( D_2 F_{13} F_{46} D_5 F_{14} D_{23} F_{56})
\nonumber\\
&+\text{tr}( D_4 F_{13} F_{46} D_5 F_{12} D_{23} F_{56})
-\text{tr}( D_2 F_{14} F_{56} D_2 F_{13} D_{34} F_{56})
+\text{tr}( D_2 F_{15} F_{46} D_2 F_{13} D_{34} F_{56})
\nonumber\\
&+\text{tr}( D_2 F_{15} F_{46} D_3 F_{14} D_{23} F_{56})
-\text{tr}( D_4 F_{15} F_{46} D_3 F_{12} D_{23} F_{56})
+\text{tr}( D_1 F_{35} F_{46} D_1 F_{24} D_{23} F_{56}) \,.
\end{align}
For the convenience of notation, we use integer numbers to represent Lorentz indices in $\Omega_1$
and abbreviate   $D_i D_j ....$ as $D_{ij...}$,
The minimal form factor of $\Omega_1$ vanishes in four dimensions, but its next-to-minimal form factor does not,
so $\Omega_1$ is  minimally evanescent but not exactly evanescent.

A general minimally evanescent operator can always be modified to be an exactly evanescent one.
The main idea is that by adding higher-length operators properly, one can construct evanescent operators that all their higher-point form factors vanish in four dimensions.

\begin{figure}
\centering
\begin{equation}
\adjincludegraphics[valign=c,scale=0.63,trim={0.2cm 0.28cm  0.4cm 0.1cm},clip]{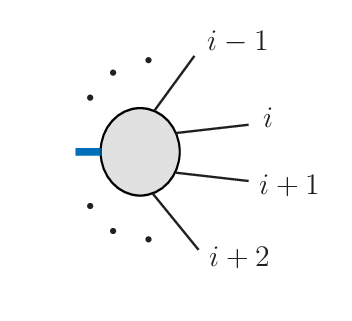}
\xlongrightarrow{s_{ii+1}\rightarrow 0}
\sum_{h_I}
\adjincludegraphics[valign=c,scale=0.63,trim={0cm 0.15cm  0.4cm 0.1cm},clip]{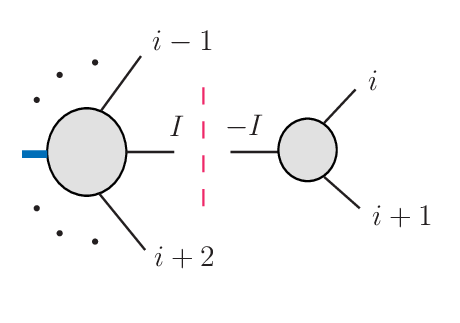}
\nonumber
\end{equation}
\vspace{-0.5cm}
\caption{\label{fig:factorize} In the limit of $s_{i(i+1)}\rightarrow 0$, the next-to-minimal
form factor factorizes into a minimal form factor times an 3-point amplitude, summed over
helicity $h_I$. }
\end{figure}

Let us start with a minimally-evanescent operator $\mathcal{O}$ of length-$L$ and of canonical dimension $\Delta>2L$.
We first show that its next-to-minimal form factor $\mathcal{F}_{\mathcal{O};L+1}$ in the four-dimensional limit
does not have any physical pole,
using proof by contradiction based on the unitarity-cut picture.
Let us assume that $\mathcal{F}_{\mathcal{O};L+1}$ has a physical pole in $s_{ij}$,
then in the limit of $s_{ij}\rightarrow 0$ the residue of the form factor should factorize as
a product of a minimal form factor and a 3-gluon amplitude,
as shown in Figure \ref{fig:factorize}. Since the minimal form factor vanishes in four dimensions, so does the residue.
Thus the 4-dim part of $\mathcal{F}_{\mathcal{O},L+1}$ is a gauge invariant quantity without any poles,
and therefore it can be taken as the minimal form factor of a length-$(L+1)$ operator, denoted by $\Xi_{L+1}$.
This means one should be able to construct a new operator $\mathcal{O}-\Xi_{L+1}$ whose
$L$- and $(L+1)$-gluon form factors  both vanish in four dimensions.

Similar analysis can be carried out iteratively to the higher point form factors
and after subtracting finitely many operators
one gets an operator $\tilde{\mathcal{O}}$  whose form factors
of $L,L+1,...,\Delta/2$ external gluons all vanish in four dimensions:
\begin{align}
\tilde{\mathcal{O}}=\mathcal{O}-\Xi_{n+1}-...-\Xi_{\Delta/2}
\quad \Longrightarrow \quad
\mathcal{F}_{\tilde{\mathcal{O}};n}\Big|_{4\text{-dim}}=0\,, \quad n=L,L+1,..,\Delta/2\,,
\end{align}
where $\Delta$ is the canonical dimension of the operator.
A nice and important point is that such an operator is already an exactly evanescent operator, and  there is no need to worry about the higher-point form factors.
Let us consider its $(\Delta/2+1)$-gluon form factor  $\mathcal{F}_{\tilde{\mathcal{O}};\Delta/2+1}$.
Using the similar factorization analysis one can show that it does not have poles in four dimensions,
so if it is nonzero, it should be equal to the minimal form factor of a
length-$(\Delta/2+1)$ operator.
However, there is no such operator because the highest length for mass dimension $\Delta$ is $\Delta/2$,
so we conclude that the form factor of $\tilde{\mathcal{O}}$
with external gluons more than $\Delta/2$ must vanish in four dimensions.

As an illustration, we can look into the former example (\ref{eq:notexact1}).
In four dimensions, the  next-to-minimal form factor of $\Omega_1$ can be
canceled by the minimal form factor of   length-5 operator
\begin{align}
\label{eq:notexact2}
\Omega_2=\frac{1}{2} \text{tr} (D_3 F_{12} F_{46} D_3 F_{12} [F_{54},F_{56}] )
-\frac{1}{2}\text{tr} (D_4 F_{12} F_{46} D_3 F_{12} [F_{53},F_{56}] )
-\text{tr} (D_2 F_{13} F_{46} D_2 F_{14} [F_{53},F_{56}] )\,.
\end{align}
One can check that $\Omega_1+\Omega_2$ is already an exactly evanescent operator, for
 there is no dim-10 operator with length higher than 5.


\section{Representation decomposition of color and kinematic factors}

\label{app:repofS4}

In this appendix we give the irreducible decomposition of the representations for
both the color factors ($\mathbf{Span}\{\mathcal{T}^a\}$) and
kinematic form factors ($\mathbf{Span}\{\kineF\lfloor\mathcal{W}_1,\mathcal{W}_2,
\mathcal{W}_3,\mathcal{W}_4\rfloor\}$) following the discussion
in Section~\ref{sec:step2}.

\subsection{Color factors}

We first organize basis color factors
so that all of them have sign change $+1$ or $-1$ under
reflection
\begin{align}
T^{\sigma(1)}T^{\sigma(2)}T^{\sigma(3)}T^{\sigma(4)}
\rightarrow
T^{\sigma(4)}T^{\sigma(3)}T^{\sigma(2)}T^{\sigma(1)}\,,
\end{align}
which we call $C$-even and $C$-odd sectors.
As defined in \eqref{eq:len4Cparity}, for the length-4 case, the basis color factors can be classified into
three types: single-trace $C$-even, single-trace $C$-odd,
double-trace $C$-even, which are denoted by
\begin{equation}
\begin{aligned}
&\{\mathcal{T}_{\mathrm{s}+}^a\}=\big\{
{\small\mathrm{tr}\big(1\sigma(2)\sigma(3)\sigma(4)\big)+
\mathrm{tr}\big(1\sigma(4)\sigma(3)\sigma(2)\big)}\big\}
\,,
\\
&\{\mathcal{T}_{\mathrm{s}-}^a\}=\big\{
{\small\mathrm{tr}\big(1\sigma(2)\sigma(3)\sigma(4)\big)-
\mathrm{tr}\big(1\sigma(4)\sigma(3)\sigma(2)\big)}\big\}
\,,
\\
&\{\mathcal{T}_{\mathrm{d}+}^a\}=\big\{
{\small \mathrm{tr}\big(1\sigma(2)\big)
\mathrm{tr}\big(\sigma(3)\sigma(4)\big)}\big\}\,,
\end{aligned}
\end{equation}
where $\sigma\in Z_3$.
Their representation decompositions are:
\begin{equation}
\begin{aligned}
\label{eq:cdecom}
&\Yvcentermath1
\mathbf{Span}\{\mathcal{T}_{\mathrm{s}+}^a\}
\sim {\tiny\yng(4)\oplus\yng(2,2) }\,,
\quad
\mathbf{Span}\{\mathcal{T}_{\mathrm{s}-}^a\}
\sim {\tiny \yng(2,1,1)}\,,
\quad
\mathbf{Span}\{\mathcal{T}_{\mathrm{d}+}^a\} \sim
{\tiny\yng(4)\oplus\yng(2,2)}\,.
\end{aligned}
\end{equation}

The bases which realize above irreducible representations are:
\begin{align}
\label{eq:strbasis}
\mathcal{T}_{s+}^{[4]} = & \
\Big\{ \frac{1}{6} \text{tr}(1234)+\frac{1}{6} \text{tr}(1243)+\frac{1}{6} \text{tr}(1324)
+\frac{1}{6} \text{tr}(1342)+\frac{1}{6} \text{tr}(1423)+\frac{1}{6} \text{tr}(1432) \Big\} \,,
\nonumber\\
\mathcal{T}_{s+}^{[2,2]} = & \
\Big\{ -\frac{1}{6} \text{tr}(1234)-\frac{1}{6} \text{tr}(1243)+\frac{1}{3} \text{tr}(1324)
-\frac{1}{6} \text{tr}(1342)+\frac{1}{3} \text{tr}(1423)-\frac{1}{6} \text{tr}(1432)\,,
\nonumber\\
& \quad \frac{1}{3} \text{tr}(1234)-\frac{1}{6} \text{tr}(1243)-\frac{1}{6} \text{tr}(1324)
-\frac{1}{6} \text{tr}(1342)-\frac{1}{6} \text{tr}(1423)+\frac{1}{3} \text{tr}(1432) \Big\} \,,
\nonumber \\
\mathcal{T}_{s-}^{[2,1,1]} = & \
\Big\{ \frac{1}{4} \text{tr}(1234)-\frac{1}{4} \text{tr}(1243)
+\frac{1}{4} \text{tr}(1342)-\frac{1}{4} \text{tr}(1432)\,,
\nonumber \\
& \quad \frac{1}{4} \text{tr}(1243)+\frac{1}{4} \text{tr}(1324)
-\frac{1}{4} \text{tr}(1342)-\frac{1}{4} \text{tr}(1423)\,,
\nonumber\\
& \quad \frac{1}{4} \text{tr}(1234)-\frac{1}{4} \text{tr}(1324)
+\frac{1}{4} \text{tr}(1423)-\frac{1}{4} \text{tr}(1432) \Big\}
\,, \nonumber\\
\mathcal{T}_d^{[4]} = & \
\Big\{ \frac{1}{3}\mathrm{tr}(12)\mathrm{tr}(34)
+\frac{1}{3}\mathrm{tr}(13)\mathrm{tr}(24)
+\frac{1}{3}\mathrm{tr}(14)\mathrm{tr}(23) \Big\}\,,
\nonumber\\
\mathcal{T}_d^{[2,2]} = & \
\Big\{\frac{2}{3}\mathrm{tr}(12)\mathrm{tr}(34)
-\frac{1}{3}\mathrm{tr}(13)\mathrm{tr}(24)
-\frac{1}{3}\mathrm{tr}(14)\mathrm{tr}(23)\,,
\nonumber\\
&\quad
-\frac{1}{3}\mathrm{tr}(12)\mathrm{tr}(34)
+\frac{2}{3}\mathrm{tr}(13)\mathrm{tr}(24)
-\frac{1}{3}\mathrm{tr}(14)\mathrm{tr}(23) \Big\}\,.
\end{align}

\subsection{Kinematic form factors}

Recall the space spanned by dim-$\Delta$ evanescent
kinematic operators is denoted by $(\mathfrak{K}^e)_\Delta$, with basis
$\{\mathbf{s}_{12}^{n_{12}}...\
\mathbf{s}_{34}^{n_{34}}\evPr_i\}$ and decomposition
(\ref{eq:union1}).
We denote the space spanned by kinematic form factors of these
kinematic operators  by $\kineF\big[(\mathfrak{K}^e)_\Delta\big]$,
with basis $\{\kineF\big(\mathbf{s}_{12}^{n_{12}}...\
\mathbf{s}_{34}^{n_{34}}\evPr_i\big)\}$ and decomposition
\begin{align}
\label{eq:union2}
\kineF\big[(\mathfrak{K}^e)_{\Delta}\big]=
\Big[\mathfrak{M}(\frac{\Delta-10}{2};s_{ij})
\otimes \kineF\big[\mathfrak{K}_{1}\big]\Big]
\oplus
\Big[\mathfrak{M}(\frac{\Delta-12}{2};s_{ij})
\otimes
\big(\kineF\big[\mathfrak{K}_{2}\big]
\oplus
\kineF\big[\mathfrak{K}_{3}\big]
\oplus
\kineF\big[\mathfrak{K}_{4}\big]\big)\Big]\,.
\end{align}
Therefore we can obtain the representation
decomposition of $\kineF\big[(\mathfrak{K}^e)_{\Delta}\big]$
through the representation decomposition of
$\mathfrak{M}(N;s_{ij})$ and $\kineF\big[\mathfrak{K}_{i}\big],\
i=1,2,3,4$.

The $S_4$ action over an kinematic operator $\lfloor\mathcal{W}_{1},\mathcal{W}_{2},
\mathcal{W}_{3},\mathcal{W}_{4}\rfloor$ and its kinematic form factor
$\kineF\big(\lfloor
\mathcal{W}_1,\mathcal{W}_2,\mathcal{W}_3,\mathcal{W}_4
\rfloor \big)$ are related as
\begin{align}
\kineF\big(
\lfloor
\mathcal{W}_{\sigma(1)},\mathcal{W}_{\sigma(2)},
\mathcal{W}_{\sigma(3)},\mathcal{W}_{\sigma(4)}
\rfloor \big)=
\sigma^{-1}\cdot \kineF\big(
\lfloor
\mathcal{W}_1,\mathcal{W}_2,\mathcal{W}_3,
\mathcal{W}_4
\rfloor \big)\,.
\end{align}
Therefore $\mathfrak{K}_i$ and $\mathfrak{F}[\mathfrak{K}_i]$
 have the same decompostion. The basis operators
that block-diagonalize $\mathfrak{K}_i$
also block-diagonalize $\mathfrak{F}[\mathfrak{K}_i]$
with their kinematic form factors. So in the following context we write
kinematic operators for simplicity and the results for
kinematic form factors are directly obtained by transforming $\mathcal{E}_{\mathbf{i},j}$
to $\mathfrak{F}(\mathcal{E}_{\mathbf{i},j})$.

The representation decomposition of  $\mathfrak{K}_{1}$,  $\mathfrak{K}_{2}$,
$\mathfrak{K}_{3}$, $\mathfrak{K}_{4}$:
\begin{align}
\Yvcentermath1
 \mathfrak{K}_{1} \sim {\tiny\yng(4)\oplus \yng(3,1)
\oplus 2\ \yng(2,2)\oplus \yng(2,1,1)\oplus \yng(1,1,1,1)}\,,
\quad
 \mathfrak{K}_{2} \sim  {\tiny \yng(2,2)} \,,
\nonumber\\
\Yvcentermath1
 \mathfrak{K}_{3} \sim {\tiny\yng(4)\oplus 2\ \yng(3,1)
\oplus \yng(2,2)\oplus \yng(2,1,1) }\,,
\quad
 \mathfrak{K}_{4} \sim {\tiny \yng(4) }\,.
\end{align}
The bases which realize above irreducible representations are summarized below.

For $ \mathfrak{K}_{1} $:
\begin{small}
\begin{align}
\label{eq:K10basis}
\evPr_{\mathbf{1}}^{[4]} = &\
\Big\{
\frac{1}{12}\left(\evPr_{\mathbf{1},1}+\evPr_{\mathbf{1},2}+\evPr_{\mathbf{1},3}
+\evPr_{\mathbf{1},4}+\evPr_{\mathbf{1},5}+\evPr_{\mathbf{1},6}+\evPr_{\mathbf{1},7}
+\evPr_{\mathbf{1},8}+\evPr_{\mathbf{1},9}+\evPr_{\mathbf{1},10}+\evPr_{\mathbf{1},11}
+\evPr_{\mathbf{1},12}\right)
\Big\}\,,
\nonumber\\
\evPr_{\mathbf{1}}^{[3,1]} =&\
\Big\{
\frac{1}{8}\left(\evPr_{\mathbf{1},1}+\evPr_{\mathbf{1},2}-\evPr_{\mathbf{1},3}-\evPr_{\mathbf{1},4}
+\evPr_{\mathbf{1},5}+\evPr_{\mathbf{1},6}-\evPr_{\mathbf{1},7}-\evPr_{\mathbf{1},8}
+\evPr_{\mathbf{1},9}-\evPr_{\mathbf{1},10}+\evPr_{\mathbf{1},11}-\evPr_{\mathbf{1},12}\right)\,,
\nonumber\\
& \quad \frac{1}{8}\left(-\evPr_{\mathbf{1},1}-\evPr_{\mathbf{1},2}+\evPr_{\mathbf{1},3}+\evPr_{\mathbf{1},4}
+\evPr_{\mathbf{1},5}-\evPr_{\mathbf{1},6}+\evPr_{\mathbf{1},7}-\evPr_{\mathbf{1},8}+\evPr_{\mathbf{1},9}
+\evPr_{\mathbf{1},10}-\evPr_{\mathbf{1},11}-\evPr_{\mathbf{1},12}\right)
\,,
\nonumber\\
& \quad  \frac{1}{8}\left(\evPr_{\mathbf{1},1}-\evPr_{\mathbf{1},2}+\evPr_{\mathbf{1},3}-\evPr_{\mathbf{1},4}
-\evPr_{\mathbf{1},5}-\evPr_{\mathbf{1},6}+\evPr_{\mathbf{1},7}+\evPr_{\mathbf{1},8}-\evPr_{\mathbf{1},9}
-\evPr_{\mathbf{1},10}+\evPr_{\mathbf{1},11}+\evPr_{\mathbf{1},12}\right)
\Big\}\,,
\nonumber\\
\evPr_{\mathbf{1}}^{[2,1,1]} =&\
\Big\{
\frac{1}{8}\left(\evPr_{\mathbf{1},1}+\evPr_{\mathbf{1},2}-\evPr_{\mathbf{1},3}-\evPr_{\mathbf{1},4}
-\evPr_{\mathbf{1},6}+\evPr_{\mathbf{1},7}+\evPr_{\mathbf{1},10}-\evPr_{\mathbf{1},11}\right)\,,
\nonumber\\
& \quad \frac{1}{8}\left(-\evPr_{\mathbf{1},2}+\evPr_{\mathbf{1},3}+\evPr_{\mathbf{1},5}+\evPr_{\mathbf{1},6}
-\evPr_{\mathbf{1},7}-\evPr_{\mathbf{1},8}-\evPr_{\mathbf{1},9}+\evPr_{\mathbf{1},12}\right)\,,
\nonumber\\
& \quad \frac{1}{8}\left(\evPr_{\mathbf{1},1}-\evPr_{\mathbf{1},4}-\evPr_{\mathbf{1},5}+\evPr_{\mathbf{1},8}
+\evPr_{\mathbf{1},9}+\evPr_{\mathbf{1},10}-\evPr_{\mathbf{1},11}-\evPr_{\mathbf{1},12}\right)
\Big\}\,,
\nonumber\\
\evPr_{\mathbf{1}}^{[1,1,1,1]} =&\
\Big\{
\frac{1}{12}\left(\evPr_{\mathbf{1},1}-\evPr_{\mathbf{1},2}-\evPr_{\mathbf{1},3}+\evPr_{\mathbf{1},4}
-\evPr_{\mathbf{1},5}+\evPr_{\mathbf{1},6}+\evPr_{\mathbf{1},7}-\evPr_{\mathbf{1},8}+\evPr_{\mathbf{1},9}
-\evPr_{\mathbf{1},10}-\evPr_{\mathbf{1},11}+\evPr_{\mathbf{1},12}\right)
\Big\}\,,
\nonumber\\
\evPr_{\mathbf{1}}^{[2,2],a} =&\
\Big\{
x_{11}\ \evPr_{\mathbf{1}}^{[2,2],1} +x_{12}\ \evPr_{\mathbf{1}}^{[2,2],3} \,,
\quad
x_{11}\ \evPr_{\mathbf{1}}^{[2,2],2} +x_{12}\ \evPr_{\mathbf{1}}^{[2,2],4}
\Big\}\,,
\nonumber\\
\evPr_{\mathbf{1}}^{[2,2],b} =&\
\Big\{
x_{21}\ \evPr_{\mathbf{1}}^{[2,2],1} +x_{22}\ \evPr_{\mathbf{1}}^{[2,2],3}\,,
\quad
x_{21}\ \evPr_{\mathbf{1}}^{[2,2],2} +x_{22}\ \evPr_{\mathbf{1}}^{[2,2],4}
\Big\}\,,
\end{align}
\end{small}
where
\begin{align}
\evPr_{\mathbf{1}}^{[2,2],1} &=
\frac{1}{6}\left(-\evPr_{\mathbf{1},5}+\evPr_{\mathbf{1},6}+\evPr_{\mathbf{1},7}-\evPr_{\mathbf{1},8}
-\evPr_{\mathbf{1},9}+\evPr_{\mathbf{1},10}+\evPr_{\mathbf{1},11}-\evPr_{\mathbf{1},12}\right)\,,
\nonumber\\
\evPr_{\mathbf{1}}^{[2,2],2} &=
\frac{1}{6}\left(-\evPr_{\mathbf{1},1}+\evPr_{\mathbf{1},2}+\evPr_{\mathbf{1},3}
-\evPr_{\mathbf{1},4}+\evPr_{\mathbf{1},9}-\evPr_{\mathbf{1},10}-\evPr_{\mathbf{1},11}+\evPr_{\mathbf{1},12}\right)\,,
\nonumber\\
\evPr_{\mathbf{1}}^{[2,2],3} &=
\frac{1}{6}\left(\evPr_{\mathbf{1},1}+\evPr_{\mathbf{1},2}+\evPr_{\mathbf{1},3}+\evPr_{\mathbf{1},4}
-\evPr_{\mathbf{1},5}-\evPr_{\mathbf{1},8}-\evPr_{\mathbf{1},9}-\evPr_{\mathbf{1},12}\right)\,,
\nonumber\\
\evPr_{\mathbf{1}}^{[2,2],4} &=
\frac{1}{6}\left(-\evPr_{\mathbf{1},1}-\evPr_{\mathbf{1},4}+\evPr_{\mathbf{1},5}+\evPr_{\mathbf{1},6}
+\evPr_{\mathbf{1},7}+\evPr_{\mathbf{1},8}-\evPr_{\mathbf{1},10}-\evPr_{\mathbf{1},11}\right)\,,
\end{align}
and $x_{11},x_{12},x_{21},x_{22}$ can be valued in any rational numbers as long as
$x_{11} x_{22}-x_{12} x_{21}\neq 0$.
Other requirement not related to $S_4$ representation helps to fix $x_{11}=1,x_{12}=-2$,
which we will explain Appendix~\ref{app:basechoice1}.
There is no further constrain on $x_{21},x_{22}$, and
in this paper we choose $x_{21}=0,x_{22}=1$.

For $ \mathfrak{K}_{2} $:
\begin{align}
\label{eq:K12basis}
\evPr_{\mathbf{2}}^{[2,2]}
=
\Big\{
\frac{1}{3}\big(
2\evPr_{\mathbf{2},2}-\evPr_{\mathbf{2},1}
\big)\,,\
\frac{1}{3}\big(
2\evPr_{\mathbf{2},1}-\evPr_{\mathbf{2},2}
\big)
\Big\}\,.
\end{align}

For $ \mathfrak{K}_{3} $:
\begin{small}
\begin{align}
\label{eq:Kaabasis}
\evPr_{\mathbf{3}}^{[4]}= & \
\Big\{\frac{1}{12}
\big(
\evPr_{\mathbf{3},1}+\evPr_{\mathbf{3},2}+\evPr_{\mathbf{3},3}
+\evPr_{\mathbf{3},4}+\evPr_{\mathbf{3},5}+\evPr_{\mathbf{3},6}
+\evPr_{\mathbf{3},7}+\evPr_{\mathbf{3},8}+\evPr_{\mathbf{3},9}
+\evPr_{\mathbf{3},10}+\evPr_{\mathbf{3},11}+\evPr_{\mathbf{3},12}
\big)\,,
\nonumber\\
\evPr_{\mathbf{3}}^{[2,2]}=& \
\Big\{
\frac{1}{12}\big(
-\evPr_{\mathbf{3},1}-\evPr_{\mathbf{3},2}-\evPr_{\mathbf{3},3}
-\evPr_{\mathbf{3},4}+2\evPr_{\mathbf{3},5}+2\evPr_{\mathbf{3},6}
-\evPr_{\mathbf{3},7}+2\evPr_{\mathbf{3},8}-\evPr_{\mathbf{3},9}
-\evPr_{\mathbf{3},10}+2\evPr_{\mathbf{3},11}-\evPr_{\mathbf{3},12}
\big)\,,
\nonumber\\
&\quad \frac{1}{12}\big(
2\evPr_{\mathbf{3},1}-\evPr_{\mathbf{3},2}-\evPr_{\mathbf{3},3}
+2\evPr_{\mathbf{3},4}-\evPr_{\mathbf{3},5}-\evPr_{\mathbf{3},6}
-\evPr_{\mathbf{3},7}-\evPr_{\mathbf{3},8}+2\evPr_{\mathbf{3},9}
+2\evPr_{\mathbf{3},10}-\evPr_{\mathbf{3},11}-\evPr_{\mathbf{3},12}
\big)
\Big\}\,,
\nonumber\\
\evPr_{\mathbf{3}}^{[2,1,1]}=&\
\Big\{
\frac{1}{8}\big(
-\evPr_{\mathbf{3},1}-\evPr_{\mathbf{3},2}+\evPr_{\mathbf{3},3}
+\evPr_{\mathbf{3},4}+2\evPr_{\mathbf{3},5}-2\evPr_{\mathbf{3},6}
-\evPr_{\mathbf{3},7}+\evPr_{\mathbf{3},9}-\evPr_{\mathbf{3},10}
+\evPr_{\mathbf{3},12}
\big)\,,
\nonumber\\
&\quad \frac{1}{8}\big(
2\evPr_{\mathbf{3},1}-\evPr_{\mathbf{3},2}-\evPr_{\mathbf{3},3}
-\evPr_{\mathbf{3},5}+\evPr_{\mathbf{3},6}+\evPr_{\mathbf{3},7}
+\evPr_{\mathbf{3},8}-2\evPr_{\mathbf{3},9}-\evPr_{\mathbf{3},11}
+\evPr_{\mathbf{3},12}
\big)\,,
\nonumber\\
&\quad \frac{1}{8}\big(
-\evPr_{\mathbf{3},1}+2\evPr_{\mathbf{3},3}-\evPr_{\mathbf{3},4}
+\evPr_{\mathbf{3},5}-\evPr_{\mathbf{3},6}-2\evPr_{\mathbf{3},7}
+\evPr_{\mathbf{3},8}+\evPr_{\mathbf{3},9}+\evPr_{\mathbf{3},10}
-\evPr_{\mathbf{3},11}
\big)
\Big\}\,,
\nonumber\\
\evPr_{\mathbf{3}}^{[3,1],a}= &\
\Big\{
x_{11}\ \evPr_{\mathbf{3}}^{[3,1],1}+
x_{12}\ \evPr_{\mathbf{3}}^{[3,1],4}\,,
\quad
x_{11}\ \evPr_{\mathbf{3}}^{[3,1],2}+
x_{12}\ \evPr_{\mathbf{3}}^{[3,1],5}\,,
\quad
x_{11}\ \evPr_{\mathbf{3}}^{[3,1],3}+
x_{12}\ \evPr_{\mathbf{3}}^{[3,1],6}
\Big\}\,,
\nonumber\\
\evPr_{\mathbf{3}}^{[3,1],b}=& \
\Big\{x_{21}\ \evPr_{\mathbf{3}}^{[3,1],1}+
x_{22}\ \evPr_{\mathbf{3}}^{[3,1],4}\,,
\quad
x_{21}\ \evPr_{\mathbf{3}}^{[3,1],2}+
x_{22}\ \evPr_{\mathbf{3}}^{[3,1],5}\,,
\quad
x_{21}\ \evPr_{\mathbf{3}}^{[3,1],3}+
x_{22}\ \evPr_{\mathbf{3}}^{[3,1],6}
\Big\}\,,
\end{align}
\end{small}
where
\begin{align}
&\evPr_{\mathbf{3}}^{[3,1],1}=
\frac{1}{8}\big(
2\evPr_{\mathbf{3},1}+2\evPr_{\mathbf{3},2}-\evPr_{\mathbf{3},3}
-\evPr_{\mathbf{3},4}+2\evPr_{\mathbf{3},5}-\evPr_{\mathbf{3},7}
-\evPr_{\mathbf{3},8} -\evPr_{\mathbf{3},10}-\evPr_{\mathbf{3},11}
\big)\,,
\nonumber\\
&\evPr_{\mathbf{3}}^{[3,1],2}=
\frac{1}{8}\big(
-\evPr_{\mathbf{3},1}-\evPr_{\mathbf{3},2}+2\evPr_{\mathbf{3},3}
+2\evPr_{\mathbf{3},4}+2\evPr_{\mathbf{3},6}-\evPr_{\mathbf{3},8}
-\evPr_{\mathbf{3},9}-\evPr_{\mathbf{3},11}-\evPr_{\mathbf{3},12}
\big)\,,
\nonumber\\
&\evPr_{\mathbf{3}}^{[3,1],3}=
\frac{1}{8}\big(
-\evPr_{\mathbf{3},2}-\evPr_{\mathbf{3},4}-\evPr_{\mathbf{3},5}
-\evPr_{\mathbf{3},6}+2\evPr_{\mathbf{3},7}+2\evPr_{\mathbf{3},8}
+2\evPr_{\mathbf{3},9}-\evPr_{\mathbf{3},10}-\evPr_{\mathbf{3},12}
\big)\,,
\nonumber\\
&\evPr_{\mathbf{3}}^{[3,1],4}=
\frac{1}{8}\big(
-\evPr_{\mathbf{3},3}-\evPr_{\mathbf{3},4}+2\evPr_{\mathbf{3},6}
-\evPr_{\mathbf{3},7}-\evPr_{\mathbf{3},8}+2\evPr_{\mathbf{3},9}
-\evPr_{\mathbf{3},10} -\evPr_{\mathbf{3},11}+2\evPr_{\mathbf{3},12}
\big)\,,
\nonumber\\
&\evPr_{\mathbf{3}}^{[3,1],5}=
\frac{1}{8}\big(
-\evPr_{\mathbf{3},1}-\evPr_{\mathbf{3},2}+2\evPr_{\mathbf{3},5}
+2\evPr_{\mathbf{3},7}-\evPr_{\mathbf{3},8}-\evPr_{\mathbf{3},9}
+2\evPr_{\mathbf{3},10} -\evPr_{\mathbf{3},11}-\evPr_{\mathbf{3},12}
\big)\,,
\nonumber\\
&\evPr_{\mathbf{3}}^{[3,1],6}=
\frac{1}{8}\big(
2\evPr_{\mathbf{3},1}-\evPr_{\mathbf{3},2}+2\evPr_{\mathbf{3},3}
-\evPr_{\mathbf{3},4}-\evPr_{\mathbf{3},5}-\evPr_{\mathbf{3},6}
-\evPr_{\mathbf{3},10}+2\evPr_{\mathbf{3},11}-\evPr_{\mathbf{3},12}
\big)\,,
\end{align}
and $x_{11},x_{12},x_{21},x_{22}$ can be valued in any rational numbers as long as
$x_{11} x_{22}-x_{12} x_{21}\neq 0$.
In this paper we choose $x_{11}=1,x_{12}=0,x_{21}=0,x_{22}=1$.

For $ \mathfrak{K}_{4} $:
\begin{align}
\label{eq:Kbbbasis}
\evPr_{\mathbf{4}}^{[4]}
=\evPr_{\mathbf{4}}\,.
\end{align}

\begin{table}[!t]
\renewcommand{\arraystretch}{1.3}
\centering
\caption{\label{tab:sCG} Representation decomposition of the linear
space $\mathfrak{M}(N;s_{ij})$
spanned by homogenous monomials $s_{12}^{n_{12}}...s_{34}^{n_{34}}$ with
total power $N$, which contribute to decomposition (\ref{eq:union1}).}
\vspace{0.2cm}
\begin{tabular}{|c|c|c|c|c|c|c|c| }
\hline
 $N$ & 1 & 2 & 3 & 4 & 5 & 6  & 7  \\
\hline
$\Yvcentermath1
  {\tiny\yng(4) }$       & 1 & 3  & 6 & 11 & 18 & 32 & 48   \\
\hline
$\Yvcentermath1
  {\tiny \yng(3,1)} $    & 1 & 3 & 8 & 17 & 34 & 61 & 104 \\
\hline
$\Yvcentermath1
  { \tiny\yng(2,2)}$     & 1 & 3 & 6 & 14 & 26 & 45 & 76  \\
\hline
$\Yvcentermath1
  {\tiny \yng(2,1,1)}$   & 0 & 1  & 4 & 11 & 24 & 47 & 84 \\
\hline
$\Yvcentermath1
  {\tiny \yng(1,1,1,1)}$ & 0 & 0  & 2 & 3 & 8 & 16 & 28  \\
\hline
total & 6 & 21 & 56 & 126 & 252 & 462 & 792 \\
\hline
\end{tabular}
\end{table}

The linear space  $\mathfrak{M}(N;s_{ij})$
is spanned by all the homogenous monomials
$s_{12}^{n_{12}}... s_{34}^{n_{34}}$ with total
power $N$.
We list the representation decomposition of $\mathfrak{M}(N;\mathbf{s}_{ij})$
in Table~\ref{tab:sCG} for $N=1,...,7$.
 As shown in (\ref{eq:union1}),
the representation information of $(\mathfrak{K}^e)_\Delta $,
the space of kinematic operators of dimension $\Delta$, can be read from
the representation decomposition of
$\mathfrak{K}_1$, $\mathfrak{K}_2$, $\mathfrak{K}_3$, $\mathfrak{K}_4$ and
$\mathfrak{M}( \frac{\Delta-10}{2} ;\mathbf{s}_{ij})$,
$\mathfrak{M}( \frac{\Delta-12}{2} ;\mathbf{s}_{ij})$.

\begin{table}[!t]
\renewcommand{\arraystretch}{1.3}
\centering
\caption{\label{tab:kineCG} Representation decomposition of $(\mathfrak{K}^e)_{\Delta}$, the
spaces  of dim-$\Delta$ evanescent kinematic operators, $\Delta=10,...,24$.
The integers give the counting of each irreducible representation corresponding to $y_i$ in \eqref{eq:kine-decom0}.}
\vspace{0.4cm}
\begin{tabular}{|c|c|c|c|c|c|c|c|c|}
\hline
  & $(\mathfrak{K}^e)_{10}$ & $(\mathfrak{K}^e)_{12}$ & $(\mathfrak{K}^e)_{14}$
  & $(\mathfrak{K}^e)_{16}$ & $(\mathfrak{K}^e)_{18}$ & $(\mathfrak{K}^e)_{20}$
  & $(\mathfrak{K}^e)_{22}$ & $(\mathfrak{K}^e)_{24}$  \\
\hline
$\Yvcentermath1
  {\tiny\yng(4) }$       & 1 & 6  & 19 & 51 & 114 & 231 & 426 & 739 \\
\hline
$\Yvcentermath1
  {\tiny \yng(3,1)} $    & 1 & 10 & 41 & 121 & 290 & 609 & 1158 & 2045\\
\hline
$\Yvcentermath1
  { \tiny\yng(2,2)}$     & 2 & 10 & 35 & 94 & 216 & 440 & 822 & 1432\\
\hline
$\Yvcentermath1
  {\tiny \yng(2,1,1)}$   & 1 & 9  & 38 & 114 & 278 & 589 & 1128 & 2001\\
\hline
$\Yvcentermath1
  {\tiny \yng(1,1,1,1)}$ & 1 & 4  & 16 & 43 & 102 & 209 & 396 & 693\\
\hline
\end{tabular}
\end{table}

We summarize the representation decomposition of $(\mathfrak{K}^e)_\Delta $,
the space of dim-$\Delta$ evanescent kinematic operators,  with $\Delta$ valued from 10 to dim 24 in Table~\ref{tab:kineCG}.
This is also the representation decomposition of $\kineF\big[(\mathfrak{K}^e)_\Delta\big]$,
the space of dim-$\Delta$ kinematic form factors.
When constructing $(\mathfrak{K}^e)_\Delta $
by means of inserting $DD$ pairs into primitive kinematic operators,
we have counted the dimension of $(\mathfrak{K}^e)_\Delta $
in Table \ref{tab:evakine}.
Such dimension can also be read from Table \ref{tab:kineCG}.
For example, the dimension of $(\mathfrak{K}^e)_{10} $ is known
from the first column of Table \ref{tab:kineCG}:
\begin{align}
\Yvcentermath1
1\times \dim\big(\,{\tiny\yng(4)\,}\big)
+1\times \dim\big(\,{\tiny\yng(3,1)\,}\big)
+ 2\times\dim\big(\,{\tiny\yng(2,2)\,}\big)
+ 1\times\dim\big(\,{\tiny\yng(2,1,1)\,}\big)
+1\times \dim\big(\,{\tiny\yng(1,1,1,1)\,}\big)=12\,,
\nonumber
\end{align}
in accord with the number given in Table \ref{tab:evakine}.

As mentioned in Section \ref{sec:step2},
the tensor product of two irreducible subspaces
of the same type $R_{[r]}$ contains an element belonging to the trivial representation.
This trivial element is given by
\begin{equation}
\mathcal{T}^{[r]}
\cdot M_{[r]}\cdot
\kineF(\evPr_{\mathbf{a}}^{[r]})\,,
\end{equation}
where $\mathcal{T}^{[r]}$ and $\kineF(\evPr_{\mathbf{a}}^{[r]})$
refer to the basis of $R_{[r]}$-type sub-representations of color space and kinematic form factor space
respectively.
The matrix $ M_{[r]}$ are known from the knowledge of representation, and here we give the result:
\begin{align}
\label{eq:fuseM}
&M_{[4]}=M_{[1,1,1,1]}=1\,,\quad
M_{[3,1]}=\frac{1}{4}
\left(
\begin{array}{ccc}
 2 & $\,\,$ 1 $\,\,$ & 1  \\
 1 & $\,\,$ 2 $\,\,$ & 1\\
 1 & $\,\,$ 1 $\,\,$ & 2 \\
\end{array}
\right)
\,,\quad
\nonumber\\
&M_{[2,2]}=\frac{1}{3}
\left(
\begin{array}{cc}
 2 & $\,$ 1    \\
 1 & $\,$ 2 \\
\end{array}
\right)\,,
\quad
M_{[2,1,1]}=-\frac{1}{12}
\left(
\begin{array}{ccc}
$\,\,$ 3 & $\,$ 1 $\,$ & -1  \\
$\,\,$ 1 & $\,$ 3 $\,$ &$\,\,$ 1\\
 -1 $\,$ & $\,$ 1 $\,$ & $\,\,$ 3 \\
\end{array}
\right)
\,.
\end{align}

\section{Basis operators as total derivatives}
\label{app:total-derivative}

In this section we discuss on an issue on the choice of operator basis,
especially focusing on the inclusion of total derivative operators.
In Section \ref{app:basechoice1} we give the detail on how to
include total derivative operators into the basis and rewrite
the total derivative basis operators in explicit forms.
In Section \ref{app:basechoice2} we give another choice of
primitive evanescent kinematic operators different from
what is given in Section \ref{sec:step1}, which includes more
total derivatives.

\subsection{Dim-10 length-4 evanescent operators}
\label{app:basechoice1}

We choose the dim-10 length-4 basis evanescent operators according to three rules.
The first is that full-color minimal form factors
all belong to trivial subspaces of $T^{R_i}\otimes \mathfrak{K}^{R_i}$ for some $R_i$ defined in (\ref{eq:young}).
The second is that the single-trace $C$-even operators and double-trace operators
are in pairs, \emph{i.e.}~$\mathcal{O}^e_{10;\mathrm{s}+;i}$
and $\mathcal{O}^e_{10;\mathrm{d}+;i}$ are related by (\ref{eq:str-dtr}).
Following the steps introduced in Section \ref{sec:step1} and \ref{sec:step2},
one can  construct a set of independent evanescent operators satisfying
the above two conditions, and let us call them the old basis operators.

The third condition left unfolded is that
we choose as many basis operators as total derivatives of lower dimensional
rank-2 or rank-1 tensor operators.
Total derivative operators are preferred because they intrinsically appear in the lower mass dimension
and consequently do not mix to those intrinsically appearing in the concerned mass dimension (dimension 10).

To select the total derivative operators from old basis operators,
let us first look into their $n$-point form factors.
Denote the total momentum of the form factor as $q$.
We can replace one of external momenta, \emph{e.g.}~$p_n$, with $q-p_1-p_2...-p_{n-1}$
and expand the form factor in powers of $q$.
If the operator is the $r$-th order derivative of a rank-$r$ operator like
$D_{\mu_1}...D_{\mu_r}\mathcal{T}^{\mu_1...\mu_r}$, then the lowest degree of
$q$ in its $n$-point form factor is  $r$.

Based on this observation, one can apply following steps.
1. Pick out all the independent combinations of old basis whose minimal form factors
and next-to-minimal form factors (which are sufficient
for dimension 10 operators) as polynomials
of $q$ are of degree $r$,
where $r$ is the highest possible order of derivatives.
2. Continue this step for the rest old basis operators and the derivative order $r-1$, and so on.

An operator with dim-10 and length-4 is at most the second derivative of a rank-2 tensor
operator with dimension 8, so $r=2$.
For length-4 dim-10 evanescent operators,
we  find two second order derivatives and one first order
derivative in single-trace sector and two second order derivatives in
double-trace sector.

The length-4 basis operators given in (\ref{eq:changebase}) are chosen as follows:
\begin{enumerate}
\item
The minimal form factors of $\mathcal{O}^e_{10;\mathrm{s}+;1}$ and
$\mathcal{O}^e_{10;\mathrm{d}+;1}$ belong to
$\mathcal{T}_{\mathrm{s}+}^{[4]}\otimes \kineF(\evPr_{\mathbf{1}}^{[4]})$.
Besides, they can be written as second order total derivatives:
\begin{align}
\label{eq:egrank2tens1}
&\mathcal{O}^e_{10;\mathrm{s}+;1}=D^{\mu\nu}
\mathrm{tr}(1234)\circ(4 \mathcal{P}_{\mu\nu}^{e1}+2 \mathcal{P}_{\mu\nu}^{e2})\,,
\quad
\mathcal{O}^e_{10;\mathrm{d}+;1}=D^{\mu\nu}
\mathrm{tr}(12)\mathrm{tr}(34)\circ(  \mathcal{P}_{\mu\nu}^{e1}+2 \mathcal{P}_{\mu\nu}^{e2})\,,
\end{align}
where
\begin{align}
\label{eq:P01P02}
&\mathcal{P}_{\mu\nu}^{e1}=\frac{1}{16}
\delta^{\rho_1\rho_2\,\rho_3\,\rho_4 \mu}_{\sigma_1 \sigma_2 \sigma_3\,\sigma_4 \nu}
\lfloor  F_{\sigma_3\sigma_4},F_{\sigma_1\sigma_2},F_{\rho_1\rho_2},F_{\rho_3\rho_4} \rfloor
\,,
\nonumber\\
&\mathcal{P}_{\mu\nu}^{e2}=
\mathcal{P}_{\mu\nu}^{e1}\Big|_{\lfloor\mathcal{W}_1,\mathcal{W}_2,
\mathcal{W}_3,\mathcal{W}_4 \rfloor\rightarrow \lfloor\mathcal{W}_1,\mathcal{W}_3,
\mathcal{W}_2,\mathcal{W}_4 \rfloor}\,.
\end{align}

\item
The minimal form factors of
$\mathcal{O}^e_{10;\mathbf{x}+;2}$  and
$\mathcal{O}^e_{10;\mathbf{x}+;3}$ ($\mathrm{x}=\mathrm{s},\mathrm{d}$)
belong to $\mathcal{T}_{\mathrm{s}+}^{[2,2]}\otimes \kineF(\evPr_{\mathbf{1}}^{[2,2]})$.
Besides, $\mathcal{O}^e_{10;\mathbf{x}+;2}$ are second order total derivatives:
\begin{align}
\label{eq:egrank2tens3}
&
\mathcal{O}^e_{10;\mathrm{s}+;2}=
D^{\mu\nu}\mathrm{tr}(1234)\circ(4 \mathcal{P}_{\mu\nu}^{e1}
-4 \mathcal{P}_{\mu\nu}^{e2})\,,
\quad
\mathcal{O}^e_{10;\mathrm{d}+;2}=
D^{\mu\nu}\mathrm{tr}(12)\mathrm{tr}(34)\circ(-2 \mathcal{P}_{\mu\nu}^{e1}
+2 \mathcal{P}_{\mu\nu}^{e2})\,.
\end{align}
The corresponding choice of $x_{11}$ and $x_{12}$ introduced in (\ref{eq:K10basis})
is $x_{11}=1,x_{12}=-2$.

There is no remaining degree of freedom of total derivative operators,
and $\mathcal{O}^e_{10;\mathbf{x}+;3}$ have to be chosen artificially.
It is permitted to take a shift
$\mathcal{O}^e_{10;\mathbf{x}+;3}\rightarrow \mathcal{O}^e_{10;\mathbf{x}+;3}+
y\, \mathcal{O}^e_{10;\mathbf{x}+;2}$ for arbitrary rational $y$.

\item
$\mathcal{O}^e_{10;\mathrm{s}-;1}$ is the
only dim-10 $C$-odd operator. Its minimal form factor belongs to
$\mathcal{T}_{\mathrm{s}-}^{[2,1,1]}\otimes \kineF(\evPr_{\mathbf{1}}^{[2,1,1]})$.
Besides, it can be written as a first order total derivative:
\begin{align}
\label{eq:egrank2tens2}
\mathcal{O}^e_{10;\mathrm{s}-;1}=
D^{\mu}\mathrm{tr}(1234)\circ(8\mathcal{Q}_\mu^{e1}
-4D^\nu\mathcal{P}_{\mu\nu}^{e1})\,,
\end{align}
where
\begin{align}
\label{eq:Q01}
&\mathcal{Q}_\mu^{e1}=\frac{1}{16}
\delta^{\rho_1\rho_2\,\rho_3\,\rho_4 \mu}_{\sigma_1 \sigma_2 \sigma_3\,\sigma_4 \nu}
\lfloor D_\nu F_{\rho_1\rho_2},F_{\rho_3\rho_4},F_{\sigma_1\sigma_2},F_{\sigma_3\sigma_4} \rfloor
\,.
\end{align}

\end{enumerate}

Another comment is that the operators satisfying the first condition stated in
Section~\ref{sec:dim10full} automatically
encode the reflection symmetry.
This is because the non-vanishing irreducible sub-representations
appearing in the decomposition of color factors are either $C$-even or $C$-odd.
To be concrete, $\Yvcentermath1 R_{[4]}={\tiny \yng(4)}$
and $\Yvcentermath1 R_{[2,2]}={\tiny \yng(2,2)}$ only appear in
$C$-even color factors and $\Yvcentermath1 R_{[2,1,1]}={\tiny \yng(2,1,1)}$
only appear in $C$-odd color factors, as shown in (\ref{eq:cdecom}).

\subsection{Evanescent primitive kinematic operators}
\label{app:basechoice2}

For the choice of primitive kinematic operators $\{\evPr_i\}$ introduced in Section \ref{sec:step1},
one can also consider total derivative operators.
Another choice of primitive kinematic operators is given here, which also maintains $S_4$ permutation symmetry
and in the form of $\delta$ functions contracting tensor operators.
The difference between this new choice and the $\{\evPr_i\}$ given in \ref{sec:step1} is that,
it includes as many total derivative kinematic operators as basis.

{\bf Class 1.}  The same as $\evPr_{\mathbf{1},i}$ in (\ref{eq:E1}).


{\bf Class 2.} The same as $\evPr_{\mathbf{2},i}$ in (\ref{eq:E2}), but  we rewrite them in another form:
\begin{align}
&\evPr_{\mathbf{2},1} =\frac{1}{4} D_{\mu}D_{\nu}
\delta^{\rho_1\rho_2\rho_3\,\rho_4 \rho_5\mu}_{\sigma_1\sigma_2 \sigma_3 \sigma_4 \sigma_5\nu}
\lfloor D_{\rho_5}F_{\rho_1\sigma_1}, F_{\rho_3\rho_4},
D_{\sigma_5}F_{\rho_2\sigma_2},   F_{\sigma_3\sigma_4} \rfloor \,,
\nonumber\\
&\evPr_{\mathbf{2},2} =\mathcal{E}_{\mathbf{2},1}
\big|_{\lfloor\mathcal{W}_1,\mathcal{W}_2,\mathcal{W}_3,\mathcal{W}_4\rfloor
\rightarrow
\lfloor\mathcal{W}_1,\mathcal{W}_3,\mathcal{W}_2,\mathcal{W}_4\rfloor
 }\,.
\end{align}


{\bf Class 3:  $\tilde{\evPr}_{\mathbf{3},i}$.}
The new class 3 kinematic operators are given by
\begin{equation}
\tilde{\evPr}_{\mathbf{3},1} =-\frac{1}{4}D_\mu
\delta^{\rho_1\rho_2 \rho_3\rho_4\rho_5}_{\sigma_1 \sigma_2 \sigma_3 \sigma_4\mu}
 \lfloor D_{\sigma_4}F_{\rho_1\rho_2}, F_{\sigma_3\lambda},
D_{\rho_5}F_{\sigma_1\sigma_2}, D_{\rho_4 } F_{\rho_3\lambda} \rfloor
  \,,
\end{equation}
together with its permutations.
They differ from $\evPr_{\mathbf{3},i}$ in (\ref{eq:E3}) by sums of
$DD$ insertions of $\evPr_{\mathbf{1},i}$, and
the permutation relations between
$\tilde{\evPr}_{\mathbf{3},i}$ and $\tilde{\evPr}_{\mathbf{3},1}$ are
the same as (\ref{eq:KOsAAD}).


{\bf Class 4. }
The same as $\evPr_{\mathbf{4}}$ in (\ref{eq:E4}), but  we rewrite them in another form:
\begin{equation}
\evPr_{\mathbf{4}} =
\frac{1}{4}D_\mu
\delta^{\rho_1\rho_2 \rho_3\rho_4\rho_5}_{\sigma_1 \sigma_2 \sigma_3 \sigma_4\mu}
\Big(\lfloor D_{\sigma_4}F_{\rho_1\rho_2}, F_{\rho_3\lambda},
D_{\rho_5}F_{\sigma_1\sigma_2}, D_{\rho_4 } F_{\sigma_3\lambda} \rfloor
- (\rho_3\leftrightarrow \sigma_3)
\Big)
+(S_4\text{-permutations})\,.
\end{equation}

\section{Dim-12 length-4 evanescent operators}
\label{app:eva-dim12}

In this section we give the complete basis evanescent operators with
mass dimension 12 and length four.
As counted in Table \ref{tab:count}, there are 25
single-trace ones and 16  double-trace ones.
The 25 single-trace and 16 double-trace minimal  evanescent
operators can be classified into four classes
according to the primitive kinematic operators from which they are generated.

Here we clarify the notation of operators:
$\mathcal{O}^e_{12;\mathrm{s};i}$ stands for the $i$-th
dim-12 length-4 evanescent single-trace operator,
and $\mathcal{O}^e_{12;\mathrm{d};i}$ stands for the $i$-th
dim-12 length-4 evanescent double-trace operator.
Same as declared in Section \ref{sec:setup},
we abbreviate products of covariant derivatives like
$D_i D_j D_k ...$ to $D_{ijk...}$ for simplicity.
We also omit the $(-\imI g)^2$ in front of each length-4 operator for short.

\subsection{25 single-trace operators}

Twenty single-trace operators are generated by $ \evPr_{\mathbf{1},i}$, $i=1,...,12$.
As declared in (\ref{eq:lessthan12}), the symbol $\mathbf{s}_{ab}$
refers to inserting one pair of identical $D$s into the $a$-th and $b$-th sites.
The inserted $D$s must be located in front of the $D$s of the primitive operators,
otherwise the operators will become not exactly evanescent but only minimally evanescent.
\begin{align}
\label{eq:25dim12str-1}
\mathcal{O}^e_{12;\mathrm{s};1}&=
\mathrm{tr}(1234)\circ \mathbf{s}_{24}\evPr_{\mathbf{1},1}\,,
\quad
\mathcal{O}^e_{12;\mathrm{s};2}=
\mathrm{tr}(4321)\circ \mathbf{s}_{24}\evPr_{\mathbf{1},1}\,,
\nonumber\\
\mathcal{O}^e_{12;\mathrm{s};3}&=
\mathrm{tr}(1234)\circ \mathbf{s}_{24}\evPr_{\mathbf{1},4}\,,
\quad
\mathcal{O}^e_{12;\mathrm{s};4}=
\mathrm{tr}(4321)\circ \mathbf{s}_{24}\evPr_{\mathbf{1},4}\,,
\nonumber\\
\mathcal{O}^e_{12;\mathrm{s};5}&=
\mathrm{tr}(1234)\circ \mathbf{s}_{14}\evPr_{\mathbf{1},1}\,,
\quad
\mathcal{O}^e_{12;\mathrm{s};6}=
\mathrm{tr}(4321)\circ \mathbf{s}_{14}\evPr_{\mathbf{1},1}\,,
\nonumber\\
\mathcal{O}^e_{12;\mathrm{s};7}&=
\mathrm{tr}(1234)\circ \mathbf{s}_{23}\evPr_{\mathbf{1},5}\,,
\quad
\mathcal{O}^e_{12;\mathrm{s};8}=
\mathrm{tr}(4321)\circ \mathbf{s}_{23}\evPr_{\mathbf{1},5}\,,
\nonumber\\
\mathcal{O}^e_{12;\mathrm{s};9}&=
\mathrm{tr}(1234)\circ \mathbf{s}_{34}\evPr_{\mathbf{1},2}\,,
\quad
\mathcal{O}^e_{12;\mathrm{s};10}=
\mathrm{tr}(4321)\circ \mathbf{s}_{34}\evPr_{\mathbf{1},2}\,,
\nonumber\\
\mathcal{O}^e_{12;\mathrm{s};11}&=
\mathrm{tr}(1234)\circ \mathbf{s}_{12}\evPr_{\mathbf{1},1}\,,
\quad
\mathcal{O}^e_{12;\mathrm{s};12}=
\mathrm{tr}(4321)\circ \mathbf{s}_{12}\evPr_{\mathbf{1},1}\,,
\nonumber\\
\mathcal{O}^e_{12;\mathrm{s};13}&=
\mathrm{tr}(1234)\circ \mathbf{s}_{13}\evPr_{\mathbf{1},2}\,,
\quad
\mathcal{O}^e_{12;\mathrm{s};14}=
\mathrm{tr}(4321)\circ \mathbf{s}_{13}\evPr_{\mathbf{1},2}\,,
\nonumber\\
\mathcal{O}^e_{12;\mathrm{s};15}&=
\mathrm{tr}(1234)\circ \mathbf{s}_{13}\evPr_{\mathbf{1},5}\,,
\quad
\mathcal{O}^e_{12;\mathrm{s};16}=
\mathrm{tr}(4321)\circ \mathbf{s}_{13}\evPr_{\mathbf{1},5}\,,
\nonumber\\
\mathcal{O}^e_{12;\mathrm{s};17}&=
\mathrm{tr}(1234)\circ \mathbf{s}_{23}\evPr_{\mathbf{1},2}\,,
\nonumber\\
\mathcal{O}^e_{12;\mathrm{s};18}&=
\mathrm{tr}(1234)\circ \mathbf{s}_{12}\evPr_{\mathbf{1},5}\,,
\nonumber\\
\mathcal{O}^e_{12;\mathrm{s};19}&=
\mathrm{tr}(1234)\circ \mathbf{s}_{14}\evPr_{\mathbf{1},2}\,,
\nonumber\\
\mathcal{O}^e_{12;\mathrm{s};20}&=
\mathrm{tr}(1234)\circ \mathbf{s}_{34}\evPr_{\mathbf{1},5}\,.
\end{align}

One single-trace operator is generated by $\evPr_{\mathbf{2},i}$, $i=1,2$.
\begin{align}
\label{eq:25dim12str-2}
&\mathcal{O}^e_{12;\mathrm{s};21}=
 \mathrm{tr}(1234)\circ\evPr_{\mathbf{2},1}
\,.
\end{align}

One single-trace operator is generated by kinematic operator $\evPr_{\mathbf{4}}$.
\begin{align}
\label{eq:25dim12str-3}
&\mathcal{O}^e_{12;\mathrm{s};22}=
\mathrm{tr}(1234)\circ\evPr_{\mathbf{4}}
\,.
\end{align}

Three single-trace operators are generated by kinematic operators
$\evPr_{\mathbf{3},i}$, $i=1,...,12$.
\begin{align}
\label{eq:25dim12str-4}
&\mathcal{O}^e_{12;\mathrm{s};23}=
\mathrm{tr}(1234)\circ\evPr_{\mathbf{3},2}\,,
\nonumber\\
&\mathcal{O}^e_{12;\mathrm{s};24}=
\mathrm{tr}(1234)\circ\evPr_{\mathbf{3},3}
\,,
\nonumber\\
&\mathcal{O}^e_{12;\mathrm{s};25}=
 \mathrm{tr}(1234)\circ(\evPr_{\mathbf{3},1}+\evPr_{\mathbf{3},4})
\,.
\end{align}

The 25 dim-12 single-trace length-4 basis evanescent operators can be recombined
to 16 $C$-even ones and 9 $C$-odd ones:
\begin{align}
 \mathrm{even}:\quad
&\mathcal{O}^e_{12;\mathrm{s};1}+\mathcal{O}^e_{12;\mathrm{s};2},\
\mathcal{O}^e_{12;\mathrm{s};3}+\mathcal{O}^e_{12;\mathrm{s};4},\
\mathcal{O}^e_{12;\mathrm{s};5}+\mathcal{O}^e_{12;\mathrm{s};6},\
\mathcal{O}^e_{12;\mathrm{s};7}+\mathcal{O}^e_{12;\mathrm{s};8},\
\nonumber\\
&\mathcal{O}^e_{12;\mathrm{s};9}+\mathcal{O}^e_{12;\mathrm{s};10},\
\mathcal{O}^e_{12;\mathrm{s};11}+\mathcal{O}^e_{12;\mathrm{s};12},\
\mathcal{O}^e_{12;\mathrm{s};13}+\mathcal{O}^e_{12;\mathrm{s};14},\
\mathcal{O}^e_{12;\mathrm{s};15}+\mathcal{O}^e_{12;\mathrm{s};16},\
 \nonumber\\
&\mathcal{O}^e_{12;\mathrm{s};17},\ \mathcal{O}^e_{12;\mathrm{s};18},\
\mathcal{O}^e_{12;\mathrm{s};19},\ \mathcal{O}^e_{12;\mathrm{s};20},\
\mathcal{O}^e_{12;\mathrm{s};21},\ \mathcal{O}^e_{12;\mathrm{s};22},\
\mathcal{O}^e_{12;\mathrm{s};23}+\mathcal{O}^e_{12;\mathrm{s};24},\
\mathcal{O}^e_{12;\mathrm{s};25};
\\
 \mathrm{odd}:\quad
&\mathcal{O}^e_{12;\mathrm{s};1}-\mathcal{O}^e_{12;\mathrm{s};2},\
\mathcal{O}^e_{12;\mathrm{s};3}-\mathcal{O}^e_{12;\mathrm{s};4},\
\mathcal{O}^e_{12;\mathrm{s};5}-\mathcal{O}^e_{12;\mathrm{s};6},\
\mathcal{O}^e_{12;\mathrm{s};7}-\mathcal{O}^e_{12;\mathrm{s};8},\
\nonumber\\
&\mathcal{O}^e_{12;\mathrm{s};9}-\mathcal{O}^e_{12;\mathrm{s};10},\
\mathcal{O}^e_{12;\mathrm{s};11}-\mathcal{O}^e_{12;\mathrm{s};12},\
\mathcal{O}^e_{12;\mathrm{s};13}-\mathcal{O}^e_{12;\mathrm{s};14},\
\mathcal{O}^e_{12;\mathrm{s};15}-\mathcal{O}^e_{12;\mathrm{s};16},\
 \nonumber\\
 &\mathcal{O}^e_{12;\mathrm{s};23}-\mathcal{O}^e_{12;\mathrm{s};24} .
\end{align}

\subsection{16 double-trace operators}

Twelve double-trace operators are generated by
$ \evPr_{\mathbf{1},i}$, $i=1,...,12$.
Similar to the single-trace case,
the inserted $D$s must be located in front of the $D$s of the primitive operators,
otherwise the operators will become not exactly evanescent but only minimally evanescent.
\begin{align}
\label{eq:16dim12dtr-1}
&\mathcal{O}^e_{12;\mathrm{d};1}=
\mathrm{tr}(12)\mathrm{tr}(34)\circ \mathbf{s}_{34}\evPr_{\mathbf{1},5}
\,,
\quad
 \mathcal{O}^e_{12;\mathrm{d};2}=
\mathrm{tr}(12)\mathrm{tr}(34)\circ \mathbf{s}_{12}\evPr_{\mathbf{1},5}
\,,
\nonumber\\
&\mathcal{O}^e_{12;\mathrm{d};3}=
\mathrm{tr}(12)\mathrm{tr}(34)\circ \mathbf{s}_{14}\evPr_{\mathbf{1},5}
\,,
\quad
 \mathcal{O}^e_{12;\mathrm{d};4}=
\mathrm{tr}(12)\mathrm{tr}(34)\circ \mathbf{s}_{14}\evPr_{\mathbf{1},1}
\,,
\nonumber\\
&\mathcal{O}^e_{12;\mathrm{d};5}=
\mathrm{tr}(12)\mathrm{tr}(34)\circ \mathbf{s}_{13}\evPr_{\mathbf{1},6}
\,,
\quad
 \mathcal{O}^e_{12;\mathrm{d};6}=
\mathrm{tr}(12)\mathrm{tr}(34)\circ \mathbf{s}_{13}\evPr_{\mathbf{1},5}
\,,
\nonumber\\
&\mathcal{O}^e_{12;\mathrm{d};7}=
\mathrm{tr}(12)\mathrm{tr}(34)\circ \mathbf{s}_{12}\evPr_{\mathbf{1},6}
\,,
\quad
 \mathcal{O}^e_{12;\mathrm{d};8}=
\mathrm{tr}(12)\mathrm{tr}(34)\circ \mathbf{s}_{12}\evPr_{\mathbf{1},1}
\,,
\nonumber\\
&\mathcal{O}^e_{12;\mathrm{d};9}=
\frac{1}{2}\mathrm{tr}(12)\mathrm{tr}(34)\circ \mathbf{s}_{23}\evPr_{\mathbf{1},6}
\,,
\quad
\mathcal{O}^e_{12;\mathrm{d};10}=
\frac{1}{2}\mathrm{tr}(12)\mathrm{tr}(34)\circ \mathbf{s}_{13}\evPr_{\mathbf{1},1}
\,,
\nonumber\\
&\mathcal{O}^e_{12;\mathrm{d};11}=
\frac{1}{2}\mathrm{tr}(12)\mathrm{tr}(34)\circ \mathbf{s}_{14}\evPr_{\mathbf{1},6}
\,,
\quad
\mathcal{O}^e_{12;\mathrm{d};12}=
\frac{1}{2}\mathrm{tr}(12)\mathrm{tr}(34)\circ \mathbf{s}_{24}\evPr_{\mathbf{1},1}
\,.
\end{align}

One double-trace operator is generated by $\evPr_{\mathbf{2},i}$, $i=1,2$.
\begin{align}
\label{eq:16dim12dtr-2}
&\mathcal{O}^e_{12;\mathrm{d};13}=
 \frac{1}{2}\mathrm{tr}(12)\mathrm{tr}(34)\circ
\evPr_{\mathbf{2},2}
\,.
\end{align}

One double-trace operator is generated by $\evPr_{\mathbf{4}}$.
\begin{align}
\label{eq:16dim12dtr-3}
&\mathcal{O}^e_{12;\mathrm{d};14}=
\frac{1}{2}\mathrm{tr}(12)\mathrm{tr}(34)\circ
\evPr_{\mathbf{4}}
\,.
\end{align}

Two double-trace operators are generated by $\evPr_{\mathbf{3},i}$, $i=1,..,12$.

\begin{align}
\label{eq:16dim12dtr-4}
&\mathcal{O}^e_{12;\mathrm{d};15}
=\mathrm{tr}(12)\mathrm{tr}(34)\circ \evPr_{\mathbf{3},7}
\,,
\nonumber\\
&\mathcal{O}^e_{12;\mathrm{d};16}
= \mathrm{tr}(12)\mathrm{tr}(34)\circ \evPr_{\mathbf{3},5}
\,.
\end{align}

The double-trace operators have one-to-one correspondence to
$C$-even single-trace operators.
The single-double pair like (\ref{eq:str-dtr})  are
\begin{align}
&(\mathcal{O}^e_{12;\mathrm{s};1}+\mathcal{O}^e_{12;\mathrm{s};2},
\mathcal{O}^e_{12;\mathrm{d};1})\ ,
(\mathcal{O}^e_{12;\mathrm{s};3}+\mathcal{O}^e_{12;\mathrm{s};4},
\mathcal{O}^e_{12;\mathrm{d};2})\ ,
(\mathcal{O}^e_{12;\mathrm{s};5}+\mathcal{O}^e_{12;\mathrm{s};6},
\mathcal{O}^e_{12;\mathrm{d};3})\ ,
\nonumber\\
&(\mathcal{O}^e_{12;\mathrm{s};7}+\mathcal{O}^e_{12;\mathrm{s};8},
\mathcal{O}^e_{12;\mathrm{d};4})\ ,
(\mathcal{O}^e_{12;\mathrm{s};9}+\mathcal{O}^e_{12;\mathrm{s};10},
\mathcal{O}^e_{12;\mathrm{d};5})\ ,
(\mathcal{O}^e_{12;\mathrm{s};11}+\mathcal{O}^e_{12;\mathrm{s};12},
\mathcal{O}^e_{12;\mathrm{d};6})\ ,
\nonumber\\
&(\mathcal{O}^e_{12;\mathrm{s};13}+\mathcal{O}^e_{12;\mathrm{s};14},
\mathcal{O}^e_{12;\mathrm{d};7})\ ,
(\mathcal{O}^e_{12;\mathrm{s};15}+\mathcal{O}^e_{12;\mathrm{s};16},
\mathcal{O}^e_{12;\mathrm{d};8})\ ,
(\mathcal{O}^e_{12;\mathrm{s};17} ,
\mathcal{O}^e_{12;\mathrm{d};9})\ ,
\nonumber\\
&(\mathcal{O}^e_{12;\mathrm{s};18} ,
\mathcal{O}^e_{12;\mathrm{d};10})\ ,
(\mathcal{O}^e_{12;\mathrm{s};19} ,
\mathcal{O}^e_{12;\mathrm{d};11})\ ,
(\mathcal{O}^e_{12;\mathrm{s};20} ,
\mathcal{O}^e_{12;\mathrm{d};12})\ ,
(\mathcal{O}^e_{12;\mathrm{s};21} ,
\mathcal{O}^e_{12;\mathrm{d};13})\ ,
\nonumber\\
&(\mathcal{O}^e_{12;\mathrm{s};22} ,
\mathcal{O}^e_{12;\mathrm{d};14})\ ,
(\mathcal{O}^e_{12;\mathrm{s};23}+\mathcal{O}_{12;e;\mathrm{s};24} ,
\mathcal{O}^e_{12;\mathrm{d};15})\ ,
(\mathcal{O}^e_{12;\mathrm{s};25} ,
\mathcal{O}^e_{12;\mathrm{d};16})\
\end{align}

\section{Tree-level form factors}
\label{app:tree-rule}

In this appendix, we give the compact expressions of tree-level  minimal,
next-to-minimal, and next-to-next-to-minimal form factors which are valid in general $d$ dimensions.
The operators we consider are in the form
of (\ref{eq:generalDF}):
\begin{align}
c(a_1,...,a_n) (D_{\mu_{11}}...D_{\mu_{1m_1}}F_{\nu_1\rho_1})^{a_1}...
(D_{\mu_{n1}}...D_{\mu_{nm_n}}F_{\nu_n\rho_n})^{a_n}\,,
\nonumber
\end{align}
with all $\mu,\nu,\rho$ contracted.
In each site, there is a site operator like ${\cal W} \sim D...DF$.
We will develop some useful rules for these site operators, such that the tree-level form factors can be efficiently read from them.

In following context, we write the sum $p_i+p_j+...+p_k$ as $p_{ij...k}$  for short.
Our results are for color-stripped form factors, where
the color decomposition is based on the convention
$f^{abc}=-2\imI \mathrm{tr}([t^a,t^b]t^c)$, and the normalization of trace is
$\mathrm{tr}(t^a t^b)=\frac{1}{2}\delta^{ab}$.

\subsection*{Minimal form factors}

How to read the minimal form factor from an operator has been
illustrated in (\ref{eq:OFmap-d-dim2}).
For convenience, let us denote the polynomial read from the site operator $\mathcal{W}_j$ which emits
gluon $i$ by $\mathfrak{F}_i(\mathcal{W}_j)$.
\begin{align}
\label{eq:rule-mini}
\mathfrak{F}_i\big(
D^{\rho_1}D^{\rho_2}...D^{\rho_n}
F^{\mu\nu}\big)
=(\imI p_i^{\rho_1})(\imI p_i^{\rho_2})...(\imI p_i^{\rho_n})
\varphi^{\mu\nu}_i
\end{align}
where $\varphi^{\mu\nu}_i$ is defined as
\begin{align}
\varphi^{\mu\nu}_i = \mathfrak{F}_i\big(
F^{\mu\nu}\big) =\imI (p_{i}^\mu e_{i}^\nu -
p_{i}^\nu e_{i}^\mu)\,.
\end{align}
In such notation,  (\ref{eq:OFmap-d-dim2}) corresponds to the special case where $i=j$.
The minimal form factor is obtained by summing over all the site-gluon distributions
that are permitted by the color factor.
For example,  the color-ordered minimal form factors of
single-trace and double-trace length-4 operators are
\begin{align}
\mathcal{F}^{(0)}_{4;\mathcal{O}_s}(p_1,p_2,p_3,p_4)
=\sum_{\sigma\in Z_4} \prod_{i=1}^4 \mathfrak{F}_{\sigma(i)}(\mathcal{W}_i)\,,\quad
\mathcal{F}^{(0)}_{4;\mathcal{O}_d}(p_1,p_2\,|\,p_3,p_4)
=\sum_{\sigma\in H_1} \prod_{i=1}^4 \mathfrak{F}_{\sigma(i)}(\mathcal{W}_i)\,.
\end{align}
Here $H_1$ is the order-8 subgroup of $S_4$ generated by
$(1\leftrightarrow 2)$, $(3\leftrightarrow 4)$ and
$(1\leftrightarrow 3, 2\leftrightarrow 4)$.

\subsection*{Next-to-minimal form factors}

For the case of next-to-minimal form factor,
one site emits two gluons and each of the other $L-1$ sites
emits one gluon.
Denote the polynomial read from the site operator $\mathcal{W}_k$ which emits
two gluons $\{i,j\}$ by $\mathfrak{F}_{i,j}(\mathcal{W}_k)$.
A site operator can emit two gluons through two ways, so  $\mathfrak{F}_{i,j}(\mathcal{W}_k)$
can be decomposed into two parts:
\begin{align}
\mathfrak{F}_{i,j}(\mathcal{W}_k)=
\mathfrak{F}_{i,j}^1(\mathcal{W}_k)+\mathfrak{F}_{i,j}^2(\mathcal{W}_k)\,.
\end{align}
\begin{enumerate}
\item
Polynomial $\mathfrak{F}_{i,j}^1(\mathcal{W}_k)$ is read from the
contribution where two gluons $i,j$ are both emitted by the field strength $F$ in $\mathcal{W}_k$:
\begin{align}
\label{eq:rule-next-1}
\mathfrak{F}_{i,j}^1\big(
D^{\rho_1}D^{\rho_2}...D^{\rho_n}
F^{\mu\nu}\big)=
(\imI p_{ij}^{\rho_1})(\imI  p_{ij}^{\rho_2})...(\imI  p_{ij}^{\rho_n})
\ \varphi_{(ij)}^{\mu\nu}\,,
\end{align}
where $\varphi_{(ij)}^{\mu\nu}$ is defined as
\begin{equation}
\begin{aligned}
\varphi_{(ij)}^{\mu\nu}&:=
(-\imI g)\Big(
-2 \,p_{ij}^\mu e^\nu_{(ij)}
+2 \,p_{ij}^\nu e^\mu_{(ij)}+
(e^\mu_i e^\nu_j-e^\nu_ie^\mu_j)
\Big)\,,
 \\
e^\mu_{(ij)}&:=\frac{1}{p_{ij}^2}
\Big[
(e_i\cdot p_j)e^\mu_j-(e_j\cdot p_i)e^\mu_i
+\frac{1}{2}\,e_i\cdot e_j (p_i-p_j)^\mu
\Big]\,.
\end{aligned}
\end{equation}
So $\varphi_{(ij)}^{\mu\nu}$ is antisymmetric  under
$\mu\leftrightarrow\nu$ and $i\leftrightarrow j$.

\item
Polynomial $\mathfrak{F}_{i,j}^2(\mathcal{W}_k)$ is read from the
contribution where one of the two gluons $i,j$ is emitted by the $F$ and the other is emitted by one $D$
in $\mathcal{W}_k$:
\begin{align}
\label{eq:rule-next-2}
\mathfrak{F}_{i,j}^2\big(
D^{\rho_1}D^{\rho_2}...D^{\rho_n}
F^{\mu\nu}\big)=
(-\imI g)\sum\limits_{m=1}^n\Big(
(\imI p_{ij}^{\rho_1})...(\imI  p_{ij}^{\rho_{m-1}}) e_i^{\rho_m}
(\imI p_{j}^{\rho_{m+1}})...(\imI  p_{j}^{\rho_n})
\varphi_{j}^{\mu\nu}
-(i\leftrightarrow j)
\Big)\,.
\end{align}
\end{enumerate}
The next-to-minimal form factor is obtained by
distributing $L+1$ gluons to
$L$ sites, and summing over all the distributions
permitted by color factors.
For example, the color-ordered next-to-minimal form factors of
single-trace and double-trace length-4 operators are
\begin{align}
&\mathcal{F}^{(0)}_{5;\mathcal{O}_s}(p_1,p_2,p_3,p_4,p_5)
=\sum_{\sigma\in Z_5} \sum_{i=1}^4
\mathfrak{F}_{\sigma(i),\sigma(i+1)}(\mathcal{W}_i)
 \Big[\prod_{j=1}^{i-1}\mathfrak{F}_{\sigma(j)}(\mathcal{W}_j)\Big]
 \Big[\prod_{j=i+1}^{4}\mathfrak{F}_{\sigma(j+1)}(\mathcal{W}_j)\Big]\,,
\nonumber\\
&\mathcal{F}^{(0)}_{5;\mathcal{O}_d}(p_1,p_2\,|\,p_3,p_4,p_5)
\nonumber\\
&=\sum_{\tau\in Z_2}\sum_{\sigma\in Z_3}\Big(
\mathfrak{F}_{\tau(1)}(\mathcal{W}_3)
\mathfrak{F}_{\tau(2)}(\mathcal{W}_4)
\Big[
\mathfrak{F}_{\sigma(3),\sigma(4)}(\mathcal{W}_1)
\mathfrak{F}_{\sigma(5)}(\mathcal{W}_2)
+
\mathfrak{F}_{\sigma(3),\sigma(4)}(\mathcal{W}_2)
\mathfrak{F}_{\sigma(5)}(\mathcal{W}_1)
\Big]
\nonumber\\
&+\mathfrak{F}_{\tau(1)}(\mathcal{W}_1)
\mathfrak{F}_{\tau(2)}(\mathcal{W}_2)
\Big[
\mathfrak{F}_{\sigma(3),\sigma(4)}(\mathcal{W}_3)
\mathfrak{F}_{\sigma(5)}(\mathcal{W}_4)
+
\mathfrak{F}_{\sigma(3),\sigma(4)}(\mathcal{W}_4)
\mathfrak{F}_{\sigma(5)}(\mathcal{W}_3)
\Big]\Big)\,.
\end{align}

\subsection*{Next-to-next-to-minimal form factors}

For the case of next-to-next-to-minimal form factor,
the possible site-gluon distributions can be classified into two types:
\begin{enumerate}
\item
One site emits three gluons and each of the other $L-1$ sites
emits one gluon.

\item
Two sites emits two gluons and each of the other $L-2$ sites
emits one gluon.
\end{enumerate}

Denote the polynomial read from the site operator $\mathcal{W}_m$ which emits
three gluons $\{i,j,k\}$ by $\mathfrak{F}_{i,j,k}(\mathcal{W}_m)$.
A site operator can emit three gluons through four ways, so  $\mathfrak{F}_{i,j,k}(\mathcal{W}_n)$
can be decomposed into four parts:
\begin{align}
\mathfrak{F}_{i,j,k}(\mathcal{W}_m)
=\mathfrak{F}_{i,j,k}^1(\mathcal{W}_m)+
\mathfrak{F}_{i,j,k}^2(\mathcal{W}_m)+
\mathfrak{F}_{i,j,k}^3(\mathcal{W}_m)\,.
\end{align}

\begin{enumerate}
\item
Polynomial $\mathfrak{F}_{i,j,k}^1(\mathcal{W}_m)$ is read from the contribution where
all three gluon $i,j,k$ are emitted by the field strength $F$ in $\mathcal{W}_m$:
\begin{align}
\label{eq:rule-nn-1}
\mathfrak{F}_{i,j,k}^1\big(
D^{\rho_1}D^{\rho_2}...D^{\rho_n}F^{\mu\nu}\big)
=
(\imI p_{ijk}^{\rho_1})(\imI p_{ijk}^{\rho_2})...(\imI p_{ijk}^{\rho_n})
\ \varphi_{(ijk)}^{\mu\nu}\,,
\end{align}
where $\varphi_{(ijk)}^{\mu\nu }$ is defined as
\begin{align}
\varphi_{(ijk)}^{\mu\nu }&:=
-\imI g^2\Big[
p^\mu_{ijk} e^\nu_{(ijk)}-p^\nu_{ijk} e^\mu_{(ijk)}
+2\Big(
 e_i^\mu e^\nu_{(jk)}- e_i^\nu e^\mu_{(jk)}
-e_k^\mu e^\nu_{(ij)}+ e_k^\nu e^\mu_{(ij)}
\Big)\Big]\,,
\nonumber\\
e^\mu_{(ijk)}&:=
\frac{1}{p^2_{ijk}}\Big[
(e_i\cdot e_j )e_k^\mu+ (e_j\cdot e_k) e_i^\mu
-2 (e_i\cdot e_k) e_j^\mu
-\mathfrak{E}^\mu_{(ij;k)}+\mathfrak{E}^\mu_{(jk;i)}
\Big]\,,
\nonumber\\
\mathfrak{E}^\mu_{(ij;k)}&:=
\frac{1}{p^2_{ij}}\Big[
4 (e_j\cdot p_i\ e_k\cdot p_{ij}) e_i^\mu
-4 (e_i\cdot p_j\ e_k\cdot p_{ij}) e_j^\mu
-4 (e_i\cdot p_k\ e_j\cdot p_i
-e_i\cdot p_j\ e_j\cdot p_k) e_k^\mu
\nonumber\\
&+2 e_i\cdot e_j (p_i\cdot p_k-p_j\cdot p_k) e_k^\mu
+2 (e_i\cdot p_j\ e_j\cdot e_k-e_j\cdot p_i\ e_i\cdot e_k)
(p_i^\mu+p_j^\mu-p_k^\mu)
\nonumber\\
&-e_i\cdot e_j \Big(
e_k\cdot p_i (p_i^\mu-3 p_j^\mu+p_k^\mu)
-e_k\cdot p_j (p_j^\mu-3 p_i^\mu+p_k^\mu)
\Big)
\Big]\,.
\end{align}
So $\varphi^{\mu\nu}_{(ijk)}$ is antisymmetric under
$\mu\leftrightarrow\nu$ and $i\leftrightarrow k$.

\item
Polynomial $\mathfrak{F}_{i,j,k}^2(\mathcal{W}_m)$ is read from the contribution where
 two of gluon $i,j,k$ are emitted by the $F$ and the other
one is emitted by one $D$ in $\mathcal{W}_m$:
\begin{align}
\label{eq:rule-nn-2}
\mathfrak{F}_{i,j,k}^2\big(
D^{\rho_1}D^{\rho_2}...D^{\rho_n}
F^{\mu\nu}\big)&=
-\imI g\sum\limits_{m=1}^n
\Big( p_{ijk}^{\rho_1}...\ p_{ijk}^{\rho_{m-1}}
 e_i^{\rho_m}
p_{jk}^{\rho_{m+1}}...\ p_{jk}^{\rho_n}\
\varphi_{(jk)}^{\mu\nu}
-(k\rightarrow j \rightarrow i\rightarrow k)
\Big)\,.
\end{align}

\item
Polynomial $\mathfrak{F}_{i,j,k}^3(\mathcal{W}_m)$ is read from the contribution where
 one of gluon $i,j,k$ is emitted by the $F$ and the other
two are emitted by one $D$ in $\mathcal{W}_m$:
\begin{align}
\label{eq:rule-nn-3}
&\mathfrak{F}_{i,j,k}^3\big(
D^{\rho_1}D^{\rho_2}...D^{\rho_n}F^{\mu\nu}\big)
\\
&=-\imI g^2 \sum\limits_{m=1}^n
\Big( p_{ijk}^{\rho_1}...\ p_{ijk}^{\rho_{m-1}}
\ e_{(ij)}^{\rho_m}\
p_{k}^{\rho_{m+1}}...\ p_{k}^{\rho_n}
\varphi_{k}^{\mu\nu}
-(i\leftrightarrow j)
-(i\rightarrow j\rightarrow k\rightarrow i)
+(i\leftrightarrow k)
\Big)\,.\nonumber
\end{align}

\item
Polynomial $\mathfrak{F}_{i,j,k}^4(\mathcal{W}_m)$ is read from the contribution where
 one of gluon $i,j,k$ is emitted by the $F$ and the other
two are emitted by two $D$s in $\mathcal{W}_m$:
\begin{align}
\label{eq:rule-nn-4}
&\mathfrak{F}_{i,j,k}^4\big(
D^{\rho_1}D^{\rho_2}...D^{\rho_n}F^{\mu\nu}
\big)
\\
&=
-g^2 \sum\limits_{r<m}
\Big(p_{ijk}^{\rho_1}...\ p_{ijk}^{\rho_{r-1}} e_i^{\rho_r}
p_{jk}^{\rho_{r+1}}...\ p_{jk}^{\rho_{m-1}}
 e_j^{\rho_m} p_{k}^{\rho_{m+1}}...\ p_{k}^{\rho_n}\
\varphi_{k}^{\mu\nu}
-(j\leftrightarrow k)-(k\rightarrow j\rightarrow i\rightarrow k)
+(i\leftrightarrow k)
\Big)\,.\nonumber
\end{align}

\end{enumerate}

The next-next-to-minimal form factor is obtained by
distributing $L+2$ gluons to
$L$ sites, and summing over all the distributions
permitted by color factors.
For example, the color-ordered next-next-to-minimal form factor of
a single-trace  length-4 operator is
\begin{small}
\begin{align}
&\mathcal{F}^{(0)}_{6;\mathcal{O}_s}(p_1,p_2,p_3,p_4,p_5,p_6)
=\sum_{\sigma\in Z_6} \sum_{i=1}^4
\mathfrak{F}_{\sigma(i),\sigma(i+1),\sigma(i+2)}(\mathcal{W}_i)
 \Big[\prod_{j=1}^{i-1}
 \mathfrak{F}_{\sigma(j)}(\mathcal{W}_j)\Big]
 \Big[\prod_{j=i+1}^{4}
 \mathfrak{F}_{\sigma(j+2)}(\mathcal{W}_j)\Big]
\nonumber\\
&+\sum_{\sigma\in Z_6} \sum_{j=2}^4\sum_{i=1}^{j-1}
\mathfrak{F}_{\sigma(i),\sigma(i+1)}(\mathcal{W}_i)
\mathfrak{F}_{\sigma(j+1),\sigma(j+2)}(\mathcal{W}_j)
\Big[\prod_{k=1}^{i-1}
\mathfrak{F}_{\sigma(k)}(\mathcal{W}_k)\Big]
\Big[\prod_{m=i+1}^{j-1}
\mathfrak{F}_{\sigma(m+1)}(\mathcal{W}_m)\Big]
\Big[\prod_{n=j+1}^{4}
\mathfrak{F}_{\sigma(n+2)}(\mathcal{W}_n)
\Big]\,.
\end{align}
\end{small}

For a double-trace length-4 operator, there are two types
of color factors appearing in next-next-to-minimal form factors.
One is like $\mathrm{tr}(12)\mathrm{tr}(3456)$ and the other is
like $\mathrm{tr}(123)\mathrm{tr}(456)$.
The corresponding color ordered form factors are
\begin{small}
\begin{align}
&\mathcal{F}^{(0)}_{6;\mathcal{O}_d}(p_1,p_2\,|\,p_3,p_4,p_5,p_6)
\\
&=\sum_{\sigma\in Z_4}\sum_{\tau\in Z_2}\Big(
\Big[
\mathfrak{F}_{\sigma(3),\sigma(4),\sigma(5)}(\mathcal{W}_1)
\mathfrak{F}_{\sigma(6)}(\mathcal{W}_2)
+
\mathfrak{F}_{\sigma(3),\sigma(4),\sigma(5)}(\mathcal{W}_2)
\mathfrak{F}_{\sigma(6)}(\mathcal{W}_1)
\Big]
\mathfrak{F}_{\tau(1)}(\mathcal{W}_3)
\mathfrak{F}_{\tau(2)}(\mathcal{W}_4)
\nonumber\\
&+\Big[
\mathfrak{F}_{\sigma(3),\sigma(4),\sigma(5)}(\mathcal{W}_3)
\mathfrak{F}_{\sigma(6)}(\mathcal{W}_4)
+
\mathfrak{F}_{\sigma(3),\sigma(4),\sigma(5)}(\mathcal{W}_4)
\mathfrak{F}_{\sigma(6)}(\mathcal{W}_3)
\Big]
\mathfrak{F}_{\tau(1)}(\mathcal{W}_1)
\mathfrak{F}_{\tau(2)}(\mathcal{W}_2)
\Big)
\nonumber\\
&+\sum_{\sigma\in Z_4}\sum_{\tau\in Z_2}\Big(
\mathfrak{F}_{\sigma(3),\sigma(4)}(\mathcal{W}_1)
\mathfrak{F}_{\sigma(5),\sigma(6)}(\mathcal{W}_2)
\mathfrak{F}_{\tau(1)}(\mathcal{W}_3)
\mathfrak{F}_{\tau(2)}(\mathcal{W}_4)
+
\mathfrak{F}_{\sigma(3),\sigma(4)}(\mathcal{W}_3)
\mathfrak{F}_{\sigma(5),\sigma(6)}(\mathcal{W}_4)
\mathfrak{F}_{\tau(1)}(\mathcal{W}_1)
\mathfrak{F}_{\tau(2)}(\mathcal{W}_2)
\Big)
\,,\nonumber
\end{align}
\end{small}
and
\begin{small}
\begin{align}
&\mathcal{F}^{(0)}_{6;\mathcal{O}_d}(p_1,p_2,p_3\,|\,p_4,p_5,p_6)
\\
&=\sum_{\sigma\in Z_3}\sum_{\tau\in Z_3}
\Big(\Big[
\mathfrak{F}_{\sigma(1),\sigma(2)}(\mathcal{W}_1)
\mathfrak{F}_{\sigma(3)}(\mathcal{W}_2)
+\mathfrak{F}_{\sigma(1),\sigma(2)}(\mathcal{W}_2)
\mathfrak{F}_{\sigma(3)}(\mathcal{W}_1)
\Big]
\nonumber\\
&\times\Big[
\mathfrak{F}_{\sigma(4),\sigma(5)}(\mathcal{W}_3)
\mathfrak{F}_{\sigma(6)}(\mathcal{W}_4)
+\mathfrak{F}_{\sigma(4),\sigma(5)}(\mathcal{W}_4)
\mathfrak{F}_{\sigma(6)}(\mathcal{W}_3)
\Big]
+(1\leftrightarrow 4,2\leftrightarrow 5,3\leftrightarrow 6)
\Big)\,.\nonumber
\end{align}
\end{small}

\section{Color decomposition of one-loop form factors}

\label{app:colordecom}

In this appendix we provide the details of the color decomposition associated with the unitarity cuts.

The basic idea has been explained around \eqref{eq:egF1ofstr4}-\eqref{eq:egF1ofstr4-cut} in Section~\ref{sec:calc-full}.
Under each cut channel, we can analyze the color structures of tree products, which are obtained
by sewing up the color factors of tree blocks through completeness
relation of Lie algebra
\begin{align}
\sum_a T^a_{ij} T^{a}_{kl}=\frac{1}{2}\delta_{il}\delta_{jk}
-\frac{1}{2N_c}\delta_{ij}\delta_{kl}\,.
\nonumber
\end{align}
Such an explicit example is given in \eqref{eq:sewing}.
By comparing with the one-loop full result \eqref{eq:egF1ofstr4}, one can express the cut of the full-color form factor in terms of color-ordered tree blocks.

A word about the notation: in this section
we denote single-trace operators by $\mathcal{O}_{s}$ and
double-trace operators by $\mathcal{O}_{d}$.

One can further simplify the cut expressions by applying the relations among color-ordered tree blocks, such as Kleiss-Kuijf  relations \cite{Kleiss:1988ne}
and reflection symmetry, so that only a small set of independent color-ordered tree blocks are left in
the expressions of cut integrands.
To be concrete, the Kleiss-Kuijf  relations reduce the
independent color-ordered components of  full-color tree level
four gluon and five gluon amplitudes to two and six respectively:
\begin{align}
&\mathbf{A}_4^{(0)}= f^{12a}f^{34a}\mathcal{A}^{0}_4(p_1,p_2,p_3,p_4)
+f^{13a}f^{24a}\mathcal{A}^{0}_4(p_1,p_3,p_2,p_4)\,,
\nonumber\\
&\mathbf{A}_5^{(0)}= f^{51a}f^{23b}f^{4ab}\mathcal{A}^{0}_5(p_1,p_2,p_3,p_4,p_5) +(S_3\mbox{-perms.\ of\ }\{3,4,5\})\,.
\end{align}
Reflection symmetry identifies tree-level color ordered form factors in the following way:
\begin{align}
C\mbox{-even (odd)\ }\mathcal{O}_s:\quad &\mathcal{F}^{(0)}_{4;\mathcal{O}_s}(p_i,p_j,p_k,p_l)
=\pm\mathcal{F}^{(0)}_{4;\mathcal{O}}(p_l,p_k,p_j,p_i),\quad
\nonumber\\
&\mathcal{F}^{(0)}_{5;\mathcal{O}_s}(p_i,p_j,p_k,p_l,p_m)=\mp\mathcal{F}^{(0)}_{5;\mathcal{O}}(p_m,p_l,p_k,p_j,p_i)\,,
\nonumber\\
C\mbox{-even (odd)\ }\mathcal{O}_d:\quad &\mathcal{F}^{(0)}_{4;\mathcal{O}_d}(p_i,p_j\,|\,p_k,p_l)
=\pm\mathcal{F}^{(0)}_{4;\mathcal{O}}(p_j,p_i\,|\,p_l,p_k),
\quad
\nonumber\\
&\mathcal{F}^{(0)}_{5;\mathcal{O}_d}(p_i,p_j\,|\,p_k,p_l,p_m)
=\mp\mathcal{F}^{(0)}_{5;\mathcal{O}_d}(p_j,p_i\,|\,p_m,p_l,p_k)\,.
\end{align}
Moreover, if two cut integrands are
related by renaming loop momenta as $l_1\leftrightarrow l_2$, \emph{e.g.}~the products
$\mathcal{F}^{(0)}_{4;\mathcal{O}_s}(p_1,p_2,l_1,l_2)\mathcal{A}^{(0)}_4(p_3,p_4,l_2,l_1)$ and
$\mathcal{F}^{(0)}_{4;\mathcal{O}_s}(p_1,p_2,l_2,l_1)\mathcal{A}^{(0)}_4(p_3,p_4,l_1,l_2)$,
one only needs to compute one of the two.
After the above steps, one finds that there is only a small set of color-ordered cut
integrands which are needed to obtained the full-color form factors.
See also \cite{Lin:2020dyj} for similar applications.

In the following context,
we summarize the cuts of the full-color form factors in terms of these independent
color-ordered cut integrands.

We focus on the form factors listed in (\ref{eq:tobecalc}), and
the complete set of cuts are shown in Figure~\ref{fig:cut444554}.
For both $\mathbf{x}=s$ and $\mathbf{x}=d$,
$\mathbf{F}^{(1)}_{4;\mathcal{O}_{\mathbf{x}}}$  are fully probed by cut (1) in Figure~\ref{fig:cut444554},
$\mathbf{F}^{(1)}_{5;\mathcal{O}_{\mathbf{x}}}$  are fully probed by cut (2) and cut (3),
and $\mathbf{F}^{(1)}_{5;\Xi_{\mathbf{x}}}$ are  fully probed by cut (3).

Consider first the cut (1) in the $s_{12}$-channel. The minimal form factor of a single-trace length-4 operator satisfies the relation:
\begin{align}
\label{eq:color44-s}
\mathbf{F}^{(1)}_{4;\mathcal{O}_s}\Big|_{s_{12}}&=
\bigg(
C_{11}\times
\adjincludegraphics[valign=c,scale=0.63,trim={0.2cm 0.28cm  1.32cm 0.1cm},clip]{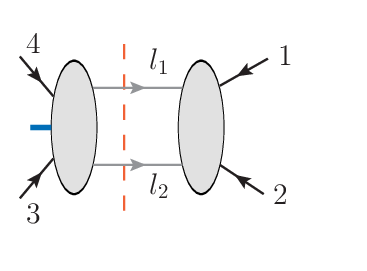}
+(3\leftrightarrow 4)
\bigg)
+ \bigg(
C_{12}\times
\adjincludegraphics[valign=c,scale=0.63,trim={0.18cm 0.15cm  0.9cm 0.1cm},clip]{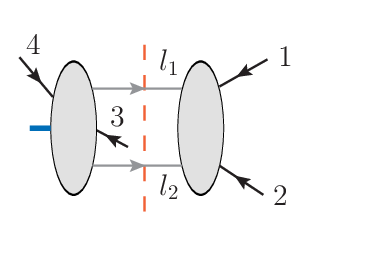}
\bigg) \,,
\end{align}
where
\begin{align}
\label{eq:cfofcuts12}
&C_{11}=(1+\mathrm{sgn}_{\mathcal{O}_s})\mathrm{tr}(12)\mathrm{tr}(34)
+N_c\big(
\mathrm{tr}(1234)+\mathrm{sgn}_{\mathcal{O}_s} \mathrm{tr}(1432)
\big)\,,
\nonumber\\
&C_{12}=-(1+\mathrm{sgn}_{\mathcal{O}_s})\big(
\mathrm{tr}(13)\mathrm{tr}(24)+
\mathrm{tr}(14)\mathrm{tr}(23)\big)\,,
\end{align}
and $\mathrm{sgn}_{\mathcal{O}_{\mathbf{x}}}$
is equal to $(-1)^{\mathrm{length}}$ times the sign change of the operator $\mathcal{O}_{\mathbf{x}}$ ($\mathbf{x}=s,d$)
under reflection of the trace, similar to (\ref{eq:44strdtr}).
Each grey blob in the figures in \eqref{eq:color44-s} represents a color-ordered tree form factor or amplitudes, for example, the last figure in \eqref{eq:color44-s} is
\begin{align}
\adjincludegraphics[valign=c,scale=0.53,trim={0.18cm 0.15cm  0.9cm 0.1cm},clip]{pure-cut44-b}
= \int d\mathrm{PS}_{l_1,l_2} \sum_{\mathrm{helicty}}
\mathcal{F}^{(0)}_{4;\mathcal{O}_s} \big(-l_1, p_3, -l_2, p_4\big)
\mathcal{A}_4^{(0)}\big(l_2, l_1, p_1, p_2\big)\,.
\end{align}
By comparing \eqref{eq:egF1ofstr4} and the permutation sum of \eqref{eq:color44-s}, one obtains
each color-ordered one-loop form factors.
As an example, we explain how the expression of $\mathcal{F}^{(1),\mathrm{d}}_{4;\mathcal{O}^e_{10;\mathrm{s}+;1}}
(p_1,p_2\, |\, p_3,p_4)$ given in (\ref{eq:44strdtr}) is obtained.
From (\ref{eq:cfofcuts12}) one finds that the terms contributing to
$\mathrm{tr}(12)\mathrm{tr}(34)$ are: the first graph in \eqref{eq:color44-s},
the first graph   after $(1\leftrightarrow 4,2\leftrightarrow 3)$,
the second graph  after $(2\leftrightarrow 3 )$ or $(1\leftrightarrow 4 )$,
the second graph   after $(1 \leftrightarrow 3 )$ or $(2\leftrightarrow 4 )$.
Summing above contributions together gives (\ref{eq:44strdtr}).

A similar relation for the minimal form factor of a double-trace length-4 operator is:
\begin{align}
\label{eq:color44-d}
\mathbf{F}^{(1)}_{4;\mathcal{O}_d}\Big|_{s_{12}}&=
\bigg(
C_{13}\times
\adjincludegraphics[valign=c,scale=0.63,trim={0.13cm 0.07cm  0.87cm 0},clip]{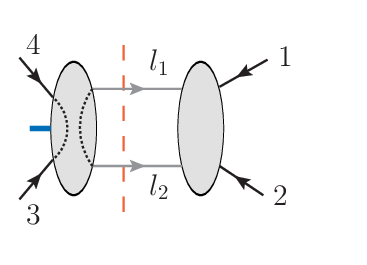}
\bigg)
+ \bigg(
C_{14}\times
\adjincludegraphics[valign=c,scale=0.63,trim={0.13cm 0  1.35cm 0.1cm},clip]{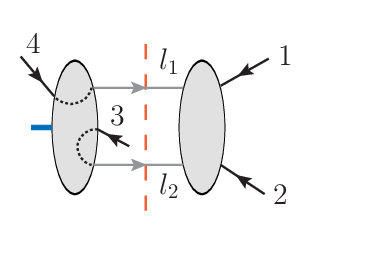}
+(3\leftrightarrow 4)
\bigg) \,,
\end{align}
where
\begin{align}
C_{13}=2N_c \mathrm{tr}(12)\mathrm{tr}(34)\,,\quad
C_{14}=\mathrm{tr}(1234)+\mathrm{tr}(1432)-\mathrm{tr}(1324)-\mathrm{tr}(1423)\,.
\end{align}
In the form factor blobs, the dotted lines indicate that the tree form factor is associated to double-trace color factors, such that: a dotted line connecting gluons $i$ and $j$ means that one of the double traces
is ranged from $i$ to $j$ clockwise. For example, the form factor tree blocks of
the first and the second diagrams in (\ref{eq:color44-d}) represent the color-stripped form factors
$\mathcal{F}_{4;\mathcal{O}_d}^{(0)}(l_1,l_2 | p_3,p_4)$ and $\mathcal{F}_{4;\mathcal{O}_d}^{(0)}(p_4,l_1 | p_3,l_2)$,
associated with $\mathrm{tr}(l_1 l_2)\mathrm{tr}(34)$ and $\mathrm{tr}(4 l_1)\mathrm{tr}(3l_2)$, respectively.

Next we consider for the cut (2) with channel $s_{123}$.
The next-to-minimal form factor of a single-trace length-4 operator satisfies the relation:
\begin{align}
\label{eq:color45-s}
&\mathbf{F}^{(1)}_{5;\mathcal{O}_s}\Big|_{s_{123}}=
\bigg(
C_{21}\times
\adjincludegraphics[valign=c,scale=0.63,trim={0.2cm 0.2cm  0.9cm 0},clip]{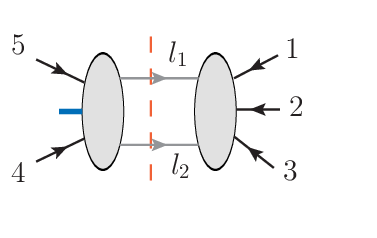}
+\ S_3\ \mathrm{perms\ of\ }\{1,2,3\}
\bigg)
\nonumber\\
&+\bigg(
C_{22}\times
\adjincludegraphics[valign=c,scale=0.63,trim={0.2cm 0.1cm  0.9cm 0},clip]{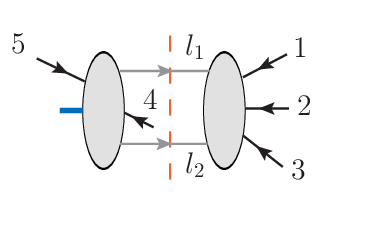}
+(2\leftrightarrow 3)+(1\leftrightarrow 2)
\bigg) \,,
\end{align}
where
\begin{align}
&C_{21}=N_c \mathrm{tr}(12345)
+\mathrm{tr}(12)\mathrm{tr}(345)+\mathrm{tr}(13)\mathrm{tr}(245)
+\mathrm{tr}(23)\mathrm{tr}(145)-\mathrm{tr}(45)\mathrm{tr}(132)
\nonumber \\
&-\mathrm{sgn}_{\mathcal{O}_s}\times \mathrm{reverse}\,,
\nonumber\\
&C_{22}=\mathrm{tr}(15)\mathrm{tr}(243)+\mathrm{tr}(25)\mathrm{tr}(143)
+\mathrm{tr}(35)\mathrm{tr}(142)-\mathrm{tr}(14)\mathrm{tr}(235)
-\mathrm{tr}(24)\mathrm{tr}(135)
\nonumber \\
&-\mathrm{tr}(34)\mathrm{tr}(125)
-\mathrm{sgn}_{\mathcal{O}_s}\times
\mathrm{reverse}\,.
\end{align}
For a double-trace length-4 operator, its next-to-minimal form factor satisfies the relation:
\begin{align}
\label{eq:color45-d}
&\mathbf{F}^{(1)}_{5;\mathcal{O}_d}\Big|_{s_{123}}=
\bigg(
C_{23}\times
\adjincludegraphics[valign=c,scale=0.63,trim={0.2cm 0.17cm  0.9cm 0},clip]{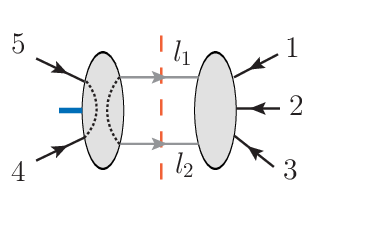}
+(2\leftrightarrow 3)+(1\leftrightarrow 2)
\bigg)
\nonumber\\
&+ \bigg(
C_{24}\times
\adjincludegraphics[valign=c,scale=0.63,trim={0.2cm 0.17cm  0.9cm 0},clip]{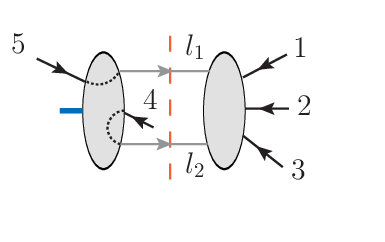}
+\ S_3\ \mathrm{perms\ of\ }\{1,2,3\}
\bigg) \,,
\end{align}
where
\begin{align}
&C_{23}=N_c\big(\mathrm{tr}(45)\mathrm{tr}(123)-
\mathrm{tr}(45)\mathrm{tr}(132)\big)\,,
\nonumber\\
&C_{24}=\mathrm{tr}(12345)-\mathrm{tr}(12435)-\mathrm{tr}(13425)
+\mathrm{tr}(14325)
-\mathrm{reverse}\,.
\end{align}

Finally, we consider the cut (3) with channel $s_{23}$.
The next-to-minimal (minimal) form factor of a single-trace length-4 (length-5) operator satisfies the relation:
\begin{align}
\label{eq:color54-s}
&\mathbf{F}^{(1)}_{5;\mathcal{O}_s}\Big|_{s_{23}}=
\bigg[\bigg(
C_{31}\times
\adjincludegraphics[valign=c,scale=0.63,trim={0.18cm 0.25cm  1.2cm 0},clip]{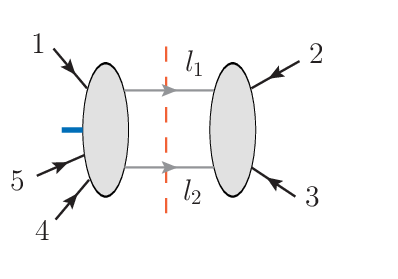}
+(1\leftrightarrow 5)+(4\leftrightarrow 5)
\bigg) +(2\leftrightarrow 3)\bigg]
\nonumber\\
&+\bigg[ \bigg(
C_{32}\times
\adjincludegraphics[valign=c,scale=0.63,trim={0.2cm 0.1cm  1.2cm 0},clip]{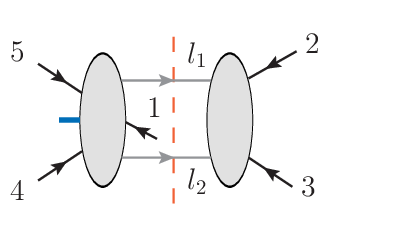}
+\ Z_3\mathrm{\ perms\ of\ }\{1,4,5\}
\bigg) +(2\leftrightarrow 3)\bigg]\,,
\end{align}
where
\begin{align}
&C_{31}=N_c\mathrm{tr}(12345)+\mathrm{tr}(23)\mathrm{tr}(451)
-\mathrm{sgn}_{\mathcal{O}_s}\times \mathrm{reverse}\,,
\nonumber\\
&C_{32}=\mathrm{tr}(45)\mathrm{tr}(132)-\mathrm{tr}(13)\mathrm{tr}(245)
-\mathrm{tr}(12)\mathrm{tr}(345)
-\mathrm{sgn}_{\mathcal{O}_s}\times \mathrm{reverse}\,.
\end{align}
For a double-trace length-4 (length-5) operator, the next-to-minimal (minimal) form factor satisfies the relation:
\begin{align}
\label{eq:color54-d}
&\mathbf{F}^{(1)}_{5;\mathcal{O}_d}\Big|_{s_{23}}=
\bigg(
C_{33}\times
\adjincludegraphics[valign=c,scale=0.63,trim={0.2cm 0.4cm  0.9cm 0},clip]{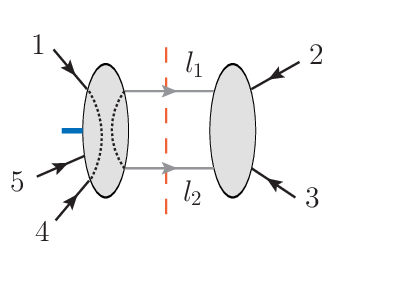}
\bigg)
+ \bigg(
C_{34}\times
\adjincludegraphics[valign=c,scale=0.63,trim={0.2cm 0.4cm  1.25cm 0},clip]{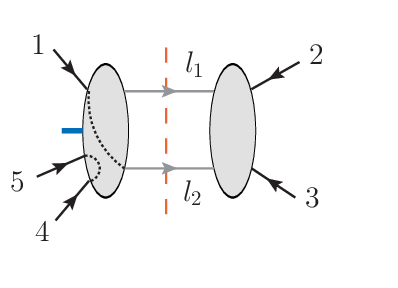}
+(1\leftrightarrow 4)+(1\leftrightarrow 5)
\bigg)
\nonumber\\
&+\bigg[ \bigg(
C_{35}\times
\adjincludegraphics[valign=c,scale=0.63,trim={0.2cm 0.17cm  1.25cm 0},clip]{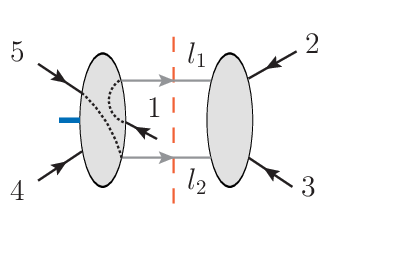}
+(1\leftrightarrow 5)+(1\rightarrow 4\rightarrow 5\rightarrow 1)
\bigg) +(2\leftrightarrow 3)\bigg]\,,
\end{align}
where
\begin{align}
&C_{33}= 2 N_c \big(
\mathrm{tr}(23)\mathrm{tr}(451)-\mathrm{sgn}_{\mathcal{O}_d}
\mathrm{tr}(23)\mathrm{tr}(541)
\big)\,,
\nonumber\\
&C_{34}=\mathrm{tr}(45)\mathrm{tr}(123)-\mathrm{sgn}_{\mathcal{O}_d}
\mathrm{tr}(45)\mathrm{tr}(132)\,,
\nonumber\\
&C_{35}=\mathrm{tr}(12345)-\mathrm{tr}(12453)-\mathrm{tr}(13452)
+\mathrm{tr}(14532)-\mathrm{sgn}_{\mathcal{O}_d}\times
\mathrm{reverse}\,.
\end{align}

\section{One-loop IR divergences of form factors}
\label{app:IR}

In this appendix, we provide some details of the infrared divergence used in Section \ref{sec:IRUV}.
In following context, the operator is considered to be either $C$-even or $C$-odd, \emph{i.e.}
it has sign change $\mathrm{sgn}_{\mathcal{O}}=(-1)^{\mathrm{length}}$ or $-(-1)^{\mathrm{length}}$ under reflection of the trace.

The IR formula of one-loop 4-gluon form factors is (see \emph{e.g.}~\cite{Catani:1998bh}):
\begin{align}
\mathbf{I}_{\rm IR}^{(1)} (\epsilon)\mathbf{F}^{(0)}_{4;\mathcal{O}}
&=\Big(\mathrm{tr}(1 2 3 4)
+\mathrm{sgn}_{\mathcal{O}}\mathrm{tr}(1 4 3 2)\Big)
\mathcal{F}^{(1),\mathrm{s}}_{4;\mathcal{O};\mathrm{IR}}(p_1,p_2,p_3,p_4)
+\text{cyclic\  perm.\ of \{2,3,4\}}
\nonumber\\
&+\mathrm{tr}(1 2)\mathrm{tr}(3 4)\,
\mathcal{F}^{(1),\mathrm{d}}_{4;\mathcal{O};\mathrm{IR}}(p_1,p_2\,|\,p_3,p_4)
+\text{cyclic\ perm.\ of \{2,3,4\}}\,.
\end{align}
If $\mathcal{O}$ is a single-trace operator:
\begin{align}
&\mathcal{F}^{(1),\mathrm{s}}_{4;\mathcal{O}_s;\mathrm{IR}} (p_1,p_2,p_3,p_4)
=-N_c\ \varphi(\epsilon)\
\sum_{i=1}^4(-s_{i  i+1})^{-\epsilon}
 \mathcal{F}^{(0)}_{4;\mathcal{O}_s}(p_1,p_2,p_3,p_4)\,,
\\
&\mathcal{F}^{(1),\mathrm{d}}_{4;\mathcal{O}_s;\mathrm{IR}} (p_1,p_2\,|\,p_3,p_4)
=\varphi(\epsilon) (1+\mathrm{sgn}_{\mathcal{O}_s})
\nonumber\\
&\qquad \times \Big[\big(
(-s_{13})^{-\epsilon}+(-s_{24})^{-\epsilon}
-(-s_{12})^{-\epsilon}-(-s_{34})^{-\epsilon}
\big)
\mathcal{F}^{(0)}_{4;\mathcal{O}_s}(p_1,p_2,p_3,p_4)
+ (3\leftrightarrow 4)
\Big]\,,
\nonumber
\end{align}
where
\begin{align}
\varphi(\epsilon)=\frac{e^{\gamma_E\epsilon}}{\Gamma(1-\epsilon)}
\big(\frac{1}{\epsilon^2}+\frac{\beta_0}{2 N_c\epsilon}\big).
\end{align}
If $\mathcal{O}$ is a double-trace operator, $\mathrm{sgn}_{\mathcal{O}}$ must be equal to 1, and then
\begin{align}
&\mathcal{F}^{(1),\mathrm{s}}_{4;\mathcal{O}_d;\mathrm{IR}}(p_1,p_2,p_3,p_4)
\nonumber\\
&=\varphi(\epsilon)\Big(
(-s_{13})^{-\epsilon}+(-s_{24})^{-\epsilon}
-(-s_{23})^{-\epsilon}-(-s_{14})^{-\epsilon}
\Big)\mathcal{F}^{(0)}_{4;\mathcal{O}_d}(p_1,p_2\,|\,p_3,p_4)
+(2\leftrightarrow 4)\,,
\nonumber\\
&\mathcal{F}^{(1),\mathrm{d}}_{4;\mathcal{O}_d;\mathrm{IR}}(p_1,p_2\,|\,p_3,p_4)
=-2N_c\ \varphi(\epsilon)\
\big(
(-s_{12})^{-\epsilon}+(-s_{34})^{-\epsilon}
\big)
\mathcal{F}^{(0)}_{4;\mathcal{O}_d }(p_1,p_2\,|\,p_3,p_4)\,.
\end{align}

The IR formula of one-loop 5-gluon form factors are given as:
\begin{align}
&\mathbf{I}_{\rm IR}^{(1)} (\epsilon)\mathbf{F}^{(0)}_{5;\mathcal{O}}=
\\
& \ \Big[\Big(
\big(\mathrm{tr}(12345)
-\mathrm{sgn}_{\mathcal{O}}\mathrm{tr}(15432)
\big)
\mathcal{F}^{(1),\mathrm{s}}_{5;\mathcal{O};\mathrm{IR}}(p_1,p_2,p_3,p_4,p_5)
+ (2\leftrightarrow 3)\Big)
+S_3\text{-perm.\ of \{3,4,5\}}\Big]
\nonumber\\
&+\Big[\Big(
\big(\mathrm{tr}(1 2)\mathrm{tr}(345)-\mathrm{sgn}_{\mathcal{O}}
\mathrm{tr}(1 2)\mathrm{tr}(354)\big)
\mathcal{F}^{(1),\mathrm{d}}_{5;\mathcal{O};\mathrm{IR}}(p_1,p_2\,|\,p_3,p_4,p_5)
\nonumber\\
&+(2\leftrightarrow 3)+(1\rightarrow 2\rightarrow 3\rightarrow 1)
+(1\leftrightarrow 3,2\leftrightarrow 4) \Big)+\mbox{cyclic\ perms.\ of\ }\{3,4,5\}
\Big]\,.
\nonumber
\end{align}
If $\mathcal{O}$ is a single-trace operator:
\begin{align}
&\mathcal{F}^{(1),\mathrm{s}}_{5;\mathcal{O}_s;\mathrm{IR}}(p_1,p_2,p_3,p_4,p_5)
=-N_c \varphi(\epsilon)
\sum_{i=1}^5(-s_{i  i+1})^{-\epsilon}
\mathcal{F}^{(0)}_{5;\mathcal{O}_s}(p_1,p_2,p_3,p_4,p_5)\,,
\nonumber\\
&\mathcal{F}^{(1),\mathrm{d}}_{5;\mathcal{O}_s;\mathrm{IR}}(p_1,p_2\,|\,p_3,p_4,p_5) =
\nonumber\\
& \varphi(\epsilon) \Big[
\big(
(-s_{13})^{-\epsilon}+(-s_{25})^{-\epsilon}
-(s_{12})^{-\epsilon}-(-s_{35})^{-\epsilon}
\big)\mathcal{F}^{(0)}_{5;\mathcal{O}_s}(p_1,p_2,p_3,p_4,p_5)
+\text{cyclic\ perm.\ of\ }\{3,4,5\}\Big]
\nonumber\\
&+(1\leftrightarrow 2)\,.
\end{align}
If $\mathcal{O}$ is a double-trace operator:
\begin{align}
&\mathcal{F}^{(1),\mathrm{s}}_{5;\mathcal{O}_d;\mathrm{IR}}(p_1,p_2,p_3,p_4,p_5) =
\nonumber\\
& \varphi(\epsilon) \big(
(-s_{13})^{-\epsilon}+(-s_{25})^{-\epsilon}
-(s_{15})^{-\epsilon}-(-s_{23})^{-\epsilon}
\big)\mathcal{F}^{(0)}_{5;\mathcal{O}_d}(p_1,p_2\,|\,p_3,p_4,p_5)
+\text{cyclic\ perm.\ of\ }\{1,2,3,4,5\}\,,
\nonumber\\
&\mathcal{F}^{(1),\mathrm{d}}_{5;\mathcal{O}_d;\mathrm{IR}}(p_1,p_2\,|\,p_3,p_4,p_5) =
\nonumber\\
& -  N_c\, \varphi(\epsilon)
\big(
2(-s_{12})^{-\epsilon}+
(-s_{34})^{-\epsilon}+(-s_{35})^{-\epsilon}
+(s_{45})^{-\epsilon}
\big)\mathcal{F}^{(0)}_{5;\mathcal{O}_d}(p_1,p_2\,|\,p_3,p_4,p_5)\,.
\end{align}

\section{Renormalization of dim-10 physical operators}
\label{app:nevtoeva}

In this appendix we give the renormalization of dim-10 length-4 and length-5 physical operators,
which contains nonzero mixing  to the evanescent ones.
As a review,
the renormalization of the dimension-four operator is well known, see \emph{e.g.}~\cite{Gehrmann:2011aa}, and one-loop renormalization for dimension 6 and 8 operators were considered in \cite{Gracey:2002he,Morozov:1984goy, Neill:2009tn, Harlander:2013oja, Dawson:2014ora}. The two-loop renormalization for dimension-6 operators (also with quark operators) were given in \cite{Jin:2018fak, Jin:2019opr}. The two-loop renormalization
for length-3 operators with dimension 8-16 are given in our recent work \cite{Jin:2020pwh}.

In Appendix~\ref{app:dim10oper4d-len4} and \ref{app:dim10oper4d-len5},
we give the expressions of  dim-10 basis physical operators of length four and length five.
In Appendix~\ref{app:Znon-eva}, we give the renormalization matrix of physical operators
as well as the anomalous dimensions.
In Appendix~\ref{app:finiteZ}, we consider the finite renormalization scheme which includes finite terms into the renormalization matrix
to absorb the finite mixing from evanescent operators to the physical ones.
In Appendix~\ref{app:Nequalto2}, we discuss the change in basis operators and
renormalization matrix when $N_c$ is reduced to 2.

For the convenience of notation, we use integer numbers to represent Lorentz indices
and abbreviate  $D_i D_j ....$ as $D_{ij...}$.
The same integers mean the indices are contracted, such as $F_{12}F_{12} = F_{\mu\nu}F^{\mu\nu}$.

\subsection{Dim-10 length-4  physical operators}

\label{app:dim10oper4d-len4}

For mass dimension 10,
there are 20 single-trace and 15 double-trace
physical basis operators with length four.
We can reorganize them so that
the form factor of each new basis operator is non-vanishing
only under one type of helicities, either $(-)^4$
or $(-)^3(+)$ or $(-)^2(+)^2$.
We call these three helicity types $\alpha,\beta,\gamma$ sectors respectively.

In Table \ref{tab:countphys1}, we give the counting of the dim-10 length-4 operators, where the
 evanescent ones have been summarized in Section \ref{sec:dim10full}.
The concept of $C$-even (odd) has been introduced
in Section \ref{sec:step2}, and it stands for the operator has
a sign change $+1\ (-1)$ under the reflection of traces.

\begin{table}[!t]
 \centering
 \caption{\label{tab:countphys1}Counting of dim-10 length-4 operators}
 \vspace{0.4cm}
\begin{tabular}{|c|c|c|c|c|}
\hline
 & $\alpha$ & $\beta$ & $\gamma$ & evanescent\\
\hline
single, C-even & 5 & 4 & 6 & 3\\
\hline
single, C-odd & 1 & 2 & 2 & 1\\
\hline
double, C-even & 5 & 4 & 6 & 3\\
\hline
\end{tabular}

\end{table}

We list single-trace and double-trace operators separately.
The operators are arranged according to descending orders of total derivatives, \emph{i.e.}
we first list the second order   derivatives (denoted by $D^{\mu\nu}T_{\mu\nu}$),
then the first order ones (denoted by $D^\mu V_{\mu}$), then the zero order ones (denoted by $S$).
We  omit the $(-\imI g)^2$ in front of each length-4 operator for simplicity.

\subsubsection*{Single-trace, $C$-even}

\begin{table}[!t]
 \centering
 \caption{Counting of dim-10 length-4 single-trace C-even operators}
 \vspace{0.4cm}
\begin{tabular}{|c|c|c|c|c|}
\hline
 & $\alpha$ & $\beta$ & $\gamma$ & evanescent\\
\hline
$D^{\mu\nu}T_{\mu\nu}$ & 2 & 2 & 3 & 2\\
\hline
$D^\mu V_\mu$ & 1 & 1 & 0 & 0\\
\hline
$S$ & 2 & 1 & 3 & 1\\
\hline
total & 5 & 4 & 6 & 3 \\
\hline
\end{tabular}

\end{table}
The $\alpha$-sector:
\begin{small}
\begin{align}
&\mathcal{O}_{10;\alpha;\mathrm{s}+;1}=D^2\Big(
\frac{1}{8} \text{tr}(F_{12} F_{12} F_{34} F_{34})-\frac{1}{8}
   \text{tr}(F_{12} F_{23} F_{34} F_{14})+\frac{1}{16}
   \text{tr}(F_{12} F_{34} F_{12} F_{34})
   +\frac{3}{8} \text{tr}(F_{12} F_{34} F_{23} F_{14})
\Big)\,,
\nonumber\\
&\mathcal{O}_{10;\alpha;\mathrm{s}+;2}=D^2\Big(
-\frac{1}{2}
   \text{tr}(F_{12} F_{12} F_{34} F_{34})-\text{tr}(F_{12} F_{23} F_{34} F_{14})-\frac{1}{4}
   \text{tr}(F_{12} F_{34} F_{12} F_{34})
\Big)\,,
\nonumber\\
&\mathcal{O}_{10;\alpha;\mathrm{s}+;3}=
D^\mu \mathrm{tr}(1234)\circ \Big(
\mathcal{Q}^{\alpha1}_{\mu}+\frac{1}{4}D^\nu\mathcal{P}^{e2}_{\mu\nu}
\Big)\,,
\nonumber\\
&\mathcal{O}_{10;\alpha;\mathrm{s}+;4}=
\frac{3}{4} \text{tr}(F_{12} F_{12} D _5 F_{34} D _5 F_{34})
+\frac{1}{4} \text{tr}(F_{12} D _5 F_{12} D _5 F_{34} F_{34})
+2 \text{tr}(F_{12} F_{23} D _5 F_{14} D _5 F_{34})
\nonumber\\
&+\text{tr}(F_{12} F_{23} D _5 F_{34} D _5 F_{14})
+\frac{3}{4} \text{tr}(F_{12} F_{34} D _5 F_{12} D _5 F_{34})
+2 \text{tr}(F_{12} D _1 F_{34} D _5 F_{23} F_{45})
-\text{tr}(F_{13} F_{23} D _2 F_{45} D _1 F_{45})\,,
\nonumber\\
&\mathcal{O}_{10;\alpha;\mathrm{s}+;5}=
-\frac{1}{2} \text{tr}(F_{12} F_{12} D _5 F_{34} D _5 F_{34})
-\frac{1}{2} \text{tr}(F_{12} D _5 F_{12} D _5 F_{34} F_{34})
-3 \text{tr}(F_{12} F_{23} D _5 F_{14} D _5 F_{34})
\nonumber\\
&+\text{tr}(F_{12} D _5 F_{23} D _5 F_{14} F_{34})
-\frac{1}{2} \text{tr}(F_{12} F_{34} D _5 F_{12} D _5 F_{34})\,,
\end{align}
\end{small}
where $\mathcal{P}_{\mu\nu}^{e2}$ is given in (\ref{eq:P01P02}) and
\begin{small}
\begin{align}
\label{eq:Q11}
&\mathcal{Q}^{\alpha1}_\mu=
-\frac{1}{2}  \lfloor F_{12},F_{12},F_{34},D_4 F_{3\mu}\rfloor
- \lfloor F_{13},F_{24},F_{3\mu},D_4 F_{12}\rfloor
+ \lfloor F_{13},F_{2\mu},F_{34},D_4 F_{12}\rfloor
- \lfloor F_{13},F_{3\mu},F_{24},D_4 F_{12}\rfloor
\nonumber\\
&+ \lfloor F_{24},F_{13},F_{3\mu},D_4 F_{12}\rfloor
- \lfloor F_{24},F_{3\mu},F_{13},D_4 F_{12}\rfloor
-\frac{1}{2}  \lfloor F_{34},F_{12},F_{12},D_4 F_{3\mu}\rfloor
+ \lfloor F_{34},F_{12},F_{13},D_4 F_{2\mu}\rfloor
\nonumber\\
&+ \lfloor F_{3\mu},F_{13},F_{24},D_4 F_{12}\rfloor  \,.
\end{align}
\end{small}
The $\beta$-sector:
\begin{small}
\begin{align}
&\mathcal{O}_{10;\beta;\mathrm{s}+;1}=
D^{\mu\nu}\mathrm{tr}(1234)\circ \mathcal{P}^{\beta1}_{\mu\nu}\,,
\nonumber\\
&\mathcal{O}_{10;\beta;\mathrm{s}+;2}=
D^{\mu\nu}\mathrm{tr}(1234)\circ \Big(
-\frac{2}{3}\mathcal{P}^{\beta1}_{\mu\nu}
+\mathcal{P}^{\beta2}_{\mu\nu}
-\mathcal{P}^{e2}_{\mu\nu}
\Big)\,,
\nonumber\\
&\mathcal{O}_{10;\beta;\mathrm{s}+;3}=
D^\mu \mathrm{tr}(1234)\circ\Big(
\mathcal{Q}^{\beta1}_{\mu}
+D^\nu\big(
-\frac{1}{3} \mathcal{P}^{\beta1}_{\mu\nu}
+\frac{1}{2} \mathcal{P}^{\beta4}_{\mu\nu}
\big)
\Big)\,,
\nonumber\\
&\mathcal{O}_{10;\beta;\mathrm{s}+;4}=
-\frac{1}{3} \text{tr}(F_{12} D _5 F_{34} F_{12} D _5 F_{34})
+\frac{2}{3} \text{tr}(F_{13} D _1 F_{45} F_{23} D _2 F_{45})
-\frac{2}{3} \text{tr}(F_{13} D _{12} F_{45} F_{23} F_{45})
\nonumber\\
&+\frac{2}{3} \text{tr}(F_{13} D _2 F_{45} F_{23} D _1 F_{45})\,,
\end{align}
\end{small}
where
\begin{small}
\begin{align}
&\mathcal{P}^{\beta1}_{\mu\nu}=
\frac{1}{2}  \lfloor F_{12},F_{3\nu},F_{12},F_{3\mu}\rfloor
+ \lfloor F_{13},F_{2\mu},F_{3\nu},F_{12}\rfloor
- \lfloor F_{3\nu},F_{12},F_{13},F_{2\mu}\rfloor
+\frac{1}{2}  \lfloor F_{3\nu},F_{12},F_{3\mu},F_{12}\rfloor
\nonumber\\
&-\frac{1}{4}g_{\mu\nu}  \lfloor F_{34},F_{12},F_{34},F_{12}\rfloor
\,,
\nonumber\\
&\mathcal{P}^{\beta2}_{\mu\nu}=
 \lfloor F_{2\mu},F_{12},F_{13},F_{3\nu}\rfloor
+ \lfloor F_{2\mu},F_{3\nu},F_{13},F_{12}\rfloor
+ \lfloor F_{3\nu},F_{13},F_{12},F_{2\mu}\rfloor
+ \lfloor F_{3\nu},F_{2\mu},F_{12},F_{13}\rfloor
\nonumber\\
&+g_{\mu\nu} \lfloor F_{14},F_{12},F_{23},F_{34}\rfloor
\,,
\nonumber\\
&\mathcal{P}^{\beta4}_{\mu\nu}=
-\frac{1}{2}
    \lfloor F_{12},F_{12},F_{3\nu},F_{3\mu}\rfloor
   + \lfloor F_{13},F_{3\nu},F_{12},F_{2\mu}\rfloor
   - \lfloor F_{3\nu},F_{13},F_{2\mu},F_{12}\rfloor
   +\frac{1}{2}  \lfloor F_{3\nu},F_{3\mu},F_{12},F_{12}\rfloor
   \,,
\nonumber\\
&\mathcal{Q}^{\beta1}_\mu=
 \lfloor D_4 F_{12},F_{34},F_{13},F_{2\mu}\rfloor
+ \lfloor D_4 F_{12},F_{3\mu},F_{13},F_{24}\rfloor
-  \lfloor D_4 F_{2\mu},F_{13},F_{34},F_{12}\rfloor
\,.
\end{align}
\end{small}
The $\gamma$-sector:
\begin{small}
\begin{align}
&\mathcal{O}_{10;\gamma;\mathrm{s}+;1}=D^2\Big(
\frac{1}{8} \text{tr}(F_{12} F_{34} F_{12} F_{34})
+\frac{1}{2} \text{tr}(F_{12} F_{34} F_{23} F_{14})
\Big)\,,
\nonumber\\
&\mathcal{O}_{10;\gamma;\mathrm{s}+;2}=D^2\Big(
-\frac{1}{4} \text{tr}(F_{12} F_{12} F_{34} F_{34})
-\frac{1}{2} \text{tr}(F_{12} F_{23} F_{34} F_{14})
+\frac{1}{8} \text{tr}(F_{12} F_{34} F_{12} F_{34})
\Big)\,,
\nonumber\\
&\mathcal{O}_{10;\gamma;\mathrm{s}+;3}=
D^{\mu\nu}\mathrm{tr}(1234)\circ  \Big(
\mathcal{P}^{\gamma1}_{\mu\nu}+\mathcal{P}^{e1}_{\mu\nu}
-\frac{1}{2}\mathcal{P}^{e2}_{\mu\nu}
\Big)\,,
\nonumber\\
&\mathcal{O}_{10;\gamma;\mathrm{s}+;4}=
\frac{1}{4} \text{tr}(F_{12} F_{12} D _5 F_{34} D _5 F_{34})
+\frac{1}{4} \text{tr}(F_{12} D _5 F_{12} D _5 F_{34} F_{34})
+\text{tr}(F_{12} F_{23} D _5 F_{34} D _5 F_{14})
\nonumber\\
&-\frac{1}{4} \text{tr}(F_{12} F_{34} D _5 F_{12} D _5 F_{34})\,,
\nonumber\\
&\mathcal{O}_{10;\gamma;\mathrm{s}+;5}=
\text{tr}(F_{12} F_{23} D _5 F_{14} D _5 F_{34})
+\text{tr}(F_{12} D _5 F_{23} D _5 F_{14} F_{34})
+\frac{1}{2} \text{tr}(F_{12} F_{34} D _5 F_{12} D _5 F_{34})\,,
\nonumber\\
&\mathcal{O}_{10;\gamma;\mathrm{s}+;6}=
-\frac{7}{4}  \text{tr}(F_{12} F_{12} D _5 F_{34} D _5 F_{34})
+\frac{5}{4} \text{tr}(F_{12} D _5 F_{12} D _5 F_{34} F_{34})
-4 \text{tr}(F_{12} F_{23} D _5 F_{14} D _5 F_{34})
\nonumber\\
& -\text{tr}(F_{12} F_{23} D _5 F_{34} D _5 F_{14})
+2 \text{tr}(F_{12} D _5 F_{23} D _5 F_{14} F_{34})
-\frac{3}{4} \text{tr}(F_{12} F_{34} D _5 F_{12} D _5 F_{34})
\nonumber\\
&-2 \text{tr}(F_{12} D _1 F_{34} D _5 F_{23} F_{45})
+\text{tr}(F_{13} F_{23} D _1 F_{45} D _2 F_{45})\,.
\end{align}
\end{small}
where $\mathcal{P}_{\mu\nu}^{e1}$ is given in (\ref{eq:P01P02}) and
\begin{small}
\begin{align}
&\mathcal{P}^{\gamma1}_{\mu\nu}=
-\frac{1}{4}  \lfloor F_{12},F_{12},F_{3\mu},F_{3\nu}\rfloor
-\frac{1}{4}  \lfloor F_{12},F_{12},F_{3\nu},F_{3\mu}\rfloor
+ \lfloor F_{12},F_{2\mu},F_{13},F_{3\nu}\rfloor
-\frac{1}{4}  \lfloor F_{12},F_{3\mu},F_{12},F_{3\nu}\rfloor
\nonumber\\
&+\frac{1}{4}  \lfloor F_{12},F_{3\mu},F_{3\nu},F_{12}\rfloor
+\frac{1}{4}  \lfloor F_{12},F_{3\nu},F_{12},F_{3\mu}\rfloor
-\frac{1}{4}  \lfloor F_{12},F_{3\nu},F_{3\mu},F_{12}\rfloor
- \lfloor F_{2\mu},F_{3\nu},F_{12},F_{13}\rfloor
-\frac{1}{4}  \lfloor F_{3\mu},F_{12},F_{12},F_{3\nu}\rfloor
\nonumber\\
&+\frac{1}{4}  \lfloor F_{3\mu},F_{12},F_{3\nu},F_{12}\rfloor
+\frac{1}{4}  \lfloor F_{3\mu},F_{3\nu},F_{12},F_{12}\rfloor
+\frac{1}{4}   \lfloor F_{3\nu},F_{12},F_{12},F_{3\mu}\rfloor
 -\frac{1}{4} \lfloor F_{3\nu},F_{12},F_{3\mu},F_{12}\rfloor
+ \lfloor F_{3\nu},F_{13},F_{2\mu},F_{12}\rfloor
\nonumber\\
&+ \lfloor F_{3\nu},F_{2\mu},F_{13},F_{12}\rfloor
-\frac{3}{4}  \lfloor F_{3\nu},F_{3\mu},F_{12},F_{12}\rfloor
+g_{\mu\nu}\Big(
\frac{1}{8}  \lfloor F_{34},F_{12},F_{12},F_{34}\rfloor
-\frac{1}{8}  \lfloor F_{34},F_{12},F_{34},F_{12}\rfloor
\nonumber\\
&+\frac{1}{8}  \lfloor F_{34},F_{34},F_{12},F_{12}\rfloor
\Big)\,.
\end{align}
\end{small}

\subsubsection*{Single-trace, C-odd}

\begin{table}[!h]
 \centering
 \caption{Counting of dim-10 length-4 single-trace $C$-odd operators}
 \vspace{0.4cm}
\begin{tabular}{|c|c|c|c|c|}
\hline
 & $\alpha$ & $\beta$ & $\gamma$ & evanescent\\
\hline
$D^{\mu\nu}T_{\mu\nu}$ & 0 & 1 & 0 & 0\\
\hline
$D^{\mu}V_{\mu}$ & 1 & 1 & 2 & 1\\
\hline
$S$ & 0 & 0 & 0 & 0\\
\hline
total & 1 & 2 & 2 & 1 \\
\hline
\end{tabular}

\end{table}
The $\alpha$-sector:
\begin{align}
\mathcal{O}_{10;\alpha;\mathrm{s}-;1}=
D^\mu\mathrm{tr}(1234)\circ\Big(
2 \mathcal{Q}^{\alpha1}_{\mu}
+2 \mathcal{Q}^{\alpha2}_{\mu}
+2\mathcal{Q}^{e1}_\mu
-D^\nu\big(
\frac{1}{2}\mathcal{P}^{\alpha1}_{\mu\nu}
+\frac{3}{2}\mathcal{P}^{e1}_{\mu\nu}
\big)
\Big)\,,
\end{align}
where $\mathcal{Q}_{\mu}^{e1}$ is given in (\ref{eq:Q01}) and
\begin{small}
\begin{align}
&\mathcal{P}^{\alpha1}_{\mu\nu}=
-\frac{1}{2}  \lfloor F_{12},F_{12},F_{3\mu},F_{3\nu}\rfloor
+ \lfloor F_{12},F_{13},F_{2\mu},F_{3\nu}\rfloor
+\frac{1}{2}  \lfloor F_{12},F_{3\mu},F_{3\nu},F_{12}\rfloor
-\frac{1}{2}  \lfloor F_{12},F_{3\nu},F_{12},F_{3\mu}\rfloor
\nonumber\\
&-\frac{1}{2}  \lfloor F_{12},F_{3\nu},F_{3\mu},F_{12}\rfloor
+ \lfloor F_{13},F_{3\nu},F_{12},F_{2\mu}\rfloor
+ \lfloor F_{2\mu},F_{3\nu},F_{12},F_{13}\rfloor
-\frac{1}{2}  \lfloor F_{3\mu},F_{3\nu},F_{12},F_{12}\rfloor
\nonumber\\
&-\frac{1}{2}  \lfloor F_{3\nu},F_{12},F_{3\mu},F_{12}\rfloor
+ \lfloor F_{3\nu},F_{13},F_{2\mu},F_{12}\rfloor
-\frac{1}{4} g_{\mu\nu} \lfloor F_{34},F_{12},F_{12},F_{34}\rfloor \,,
\nonumber\\
&\mathcal{Q}^{\alpha2}_\mu=
 \lfloor D_4 F_{12},F_{13},F_{24},F_{3\mu}\rfloor
+ \lfloor D_4 F_{12},F_{13},F_{3\mu},F_{24}\rfloor
- \lfloor D_4 F_{12},F_{24},F_{13},F_{3\mu}\rfloor
+ \lfloor D_4 F_{12},F_{2\mu},F_{13},F_{34}\rfloor
\nonumber\\
&- \lfloor D_4 F_{12},F_{3\mu},F_{13},F_{24}\rfloor
- \lfloor D_4 F_{12},F_{3\mu},F_{24},F_{13}\rfloor
- \lfloor D_4 F_{2\mu},F_{34},F_{12},F_{13}\rfloor
+\frac{1}{2}  \lfloor D_4 F_{3\mu},F_{34},F_{12},F_{12}\rfloor \,.
\end{align}
\end{small}
The $\beta$-sector:
\begin{align}
&\mathcal{O}_{10;\beta;\mathrm{s}-;1}=
D^{\mu\nu}\mathrm{tr}(1234)\circ  \mathcal{P}^{\beta4}_{\mu\nu}\,,
\nonumber\\
&\mathcal{O}_{10;\beta;\mathrm{s}-;2}=
D^\mu\mathrm{tr}(1234)\circ\Big(
-2\mathcal{Q}^{\beta1}_{\mu}
-2\mathcal{Q}^{\beta2}_{\mu}
+D^\nu\big(
- \mathcal{P}^{\beta4}_{\mu\nu}
+\frac{1}{2} \mathcal{P}^{\beta2}_{\mu\nu}
+ \mathcal{P}^{e1}_{\mu\nu}
-\frac{1}{2} \mathcal{P}^{e2}_{\mu\nu}
\big)
\Big)\,,
\end{align}
where
\begin{small}
\begin{align}
&\mathcal{Q}^{\beta2}_\mu=
 \lfloor D_4 F_{12},F_{13},F_{2\mu},F_{34}\rfloor
- \lfloor D_4 F_{12},F_{3\mu},F_{24},F_{13}\rfloor
- \lfloor D_4 F_{2\mu},F_{34},F_{12},F_{13}\rfloor
+\frac{1}{2}   \lfloor D_4 F_{3\mu},F_{12},F_{34},F_{12}\rfloor \,.
\end{align}
\end{small}
The $\gamma$-sector:
\begin{align}
&\mathcal{O}_{10;\gamma;\mathrm{s}-;1}=
D^\mu\mathrm{tr}(1234)\circ\Big(
2\mathcal{Q}_{\mu}^{\gamma1}+
\frac{1}{2}D^\nu\big(
\mathcal{P}_{\mu\nu}^{\gamma1}-\mathcal{P}_{\mu\nu}^{\gamma2}
-\mathcal{P}_{\mu\nu}^{e1}
\big)\Big)\,,
\nonumber\\
&\mathcal{O}_{10;\gamma;\mathrm{s}-;2}=
D^\mu\mathrm{tr}(1234)\circ
\Big(
2\mathcal{Q}_\mu^{\gamma2}+2\mathcal{Q}_\mu^{e1}
+\frac{1}{2}D^\nu
\big(
-\mathcal{P}_{\mu\nu}^{\gamma1}+\mathcal{P}_{\mu\nu}^{\gamma2}
-3\mathcal{P}_{\mu\nu}^{e1}
\big)
  \Big)\,,
\end{align}
where
\begin{small}
\begin{align}
&\mathcal{P}^{\gamma 2}_{\mu\nu}=
-\frac{1}{4}  \lfloor F_{12},F_{12},F_{3\mu},F_{3\nu}\rfloor
+\frac{1}{4}  \lfloor F_{12},F_{12},F_{3\nu},F_{3\mu}\rfloor
+ \lfloor F_{12},F_{13},F_{2\mu},F_{3\nu}\rfloor
-\frac{1}{4}  \lfloor F_{12},F_{3\mu},F_{12},F_{3\nu}\rfloor
\nonumber\\
&-\frac{1}{4}  \lfloor F_{12},F_{3\mu},F_{3\nu},F_{12}\rfloor
-\frac{1}{4}  \lfloor F_{12},F_{3\nu},F_{12},F_{3\mu}\rfloor
+\frac{1}{4}  \lfloor F_{12},F_{3\nu},F_{3\mu},F_{12}\rfloor
+ \lfloor F_{2\mu},F_{3\nu},F_{12},F_{13}\rfloor
+\frac{1}{4}  \lfloor F_{3\mu},F_{12},F_{12},F_{3\nu}\rfloor
\nonumber\\
&-\frac{1}{4}  \lfloor F_{3\mu},F_{12},F_{3\nu},F_{12}\rfloor
-\frac{1}{4}  \lfloor F_{3\mu},F_{3\nu},F_{12},F_{12}\rfloor
-\frac{1}{4}  \lfloor F_{3\nu},F_{12},F_{12},F_{3\mu}\rfloor
-\frac{1}{4}  \lfloor F_{3\nu},F_{12},F_{3\mu},F_{12}\rfloor
+\frac{1}{4}  \lfloor F_{3\nu},F_{3\mu},F_{12},F_{12}\rfloor
\nonumber\\
&+g_{\mu\nu}\Big(
\frac{1}{8}  \lfloor F_{34},F_{12},F_{12},F_{34}\rfloor
+\frac{1}{8}  \lfloor F_{34},F_{12},F_{34},F_{12}\rfloor
-\frac{1}{8}  \lfloor F_{34},F_{34},F_{12},F_{12}\rfloor
\Big)\,,
\nonumber\\
&\mathcal{Q}^{\gamma 1}_{\mu}=
- \lfloor D_4 F_{12},F_{13},F_{3\mu},F_{24}\rfloor
- \lfloor D_4 F_{12},F_{24},F_{13},F_{3\mu}\rfloor
+ \lfloor D_4 F_{12},F_{24},F_{3\mu},F_{13}\rfloor
+ \lfloor D_4 F_{12},F_{34},F_{2\mu},F_{13}\rfloor
\nonumber\\
& + \lfloor D_4 F_{12},F_{3\mu},F_{13},F_{24}\rfloor
- \lfloor D_4 F_{12},F_{3\mu},F_{24},F_{13}\rfloor
- \lfloor D_4 F_{2\mu},F_{13},F_{34},F_{12}\rfloor
+\frac{1}{2}  \lfloor D_4 F_{3\mu},F_{12},F_{34},F_{12}\rfloor \,,
\nonumber\\
&\mathcal{Q}^{\gamma 2}_{\mu}=
- \lfloor D_4 F_{12},F_{24},F_{3\mu},F_{13}\rfloor
+ \lfloor D_4 F_{12},F_{2\mu},F_{13},F_{34}\rfloor
- \lfloor D_4 F_{2\mu},F_{34},F_{12},F_{13}\rfloor
+\frac{1}{2}  \lfloor D_4  F_{3\mu},F_{34},F_{12},F_{12}\rfloor \,,
\end{align}
\end{small}

\subsubsection*{Double-trace, C-even}

\begin{table}[!h]
 \centering
 \caption{Counting of dim-10 length-4 double-trace operators, which are all
C-even}
\vspace{0.4cm}
\begin{tabular}{|c|c|c|c|c|}
\hline
 & $\alpha$ & $\beta$ & $\gamma$ & evanescent\\
\hline
$D^{\mu\nu}T_{\mu\nu}$ & 2 & 2 & 3 & 2\\
\hline
$D^\mu V_\mu$ & 1 & 1 & 0 & 0\\
\hline
$S$ & 2 & 1 & 3 & 1\\
\hline
total & 5 & 4 & 6 & 3 \\
\hline
\end{tabular}

\end{table}
The $\alpha$-sector:
\begin{small}
\begin{align}
&\mathcal{O}_{10;\alpha;\mathrm{d}+;1}=D^2\Big(
\frac{1}{16} \text{tr}(F_{12} F_{34}){}^2-\frac{1}{16} \text{tr}(F_{12} F_{34})
   \text{tr}(F_{23} F_{14})+\frac{3}{16} \text{tr}(F_{12} F_{23})
   \text{tr}(F_{34} F_{14})+\frac{1}{32} \text{tr}(F_{12} F_{12}) \text{tr}(F_{34} F_{34})
\Big)\,,
\nonumber\\
&\mathcal{O}_{10;\alpha;\mathrm{d}+;2}=D^2\Big(
-\frac{1}{4} \text{tr}(F_{12} F_{34}){}^2-\frac{1}{2} \text{tr}(F_{12} F_{34})
   \text{tr}(F_{23} F_{14})-\frac{1}{8} \text{tr}(F_{12} F_{12}) \text{tr}(F_{34} F_{34})
\Big)\,,
\nonumber\\
&\mathcal{O}_{10;\alpha;\mathrm{d}+;3}=
D^\mu\mathrm{tr}(12)\mathrm{tr}(34)\circ
\Big(
\frac{1}{2}\mathcal{Q}^{\alpha 2}_\mu
+\frac{1}{8}D^\nu\big(
\mathcal{P}^{\alpha1}_{\mu\nu}+\mathcal{P}^{e1}_{\mu\nu}
-\mathcal{P}^{e2}_{\mu\nu}
\big)\Big)\,,
\nonumber\\
&\mathcal{O}_{10;\alpha;\mathrm{d}+;4}=
\frac{3}{8} \text{tr}(F_{12} D _5 F_{34})^2
+\frac{1}{8} \text{tr}(F_{12} D _5 F_{34}) \text{tr}(D_5 F_{12} F_{34})
+\frac{1}{2} \text{tr}(F_{12} D _5 F_{34}) \text{tr}(F_{23} D _5 F_{14})
\nonumber\\
&+\text{tr}(F_{12} D _5 F_{14}) \text{tr}(F_{23} D _5 F_{34})
-\frac{1}{2} \text{tr}(F_{13} D _2 F_{45}) \text{tr}(F_{23} D _1 F_{45})
+\frac{3}{8} \text{tr}(F_{12} D _5 F_{12}) \text{tr}(F_{34} D _5 F_{34})
\nonumber\\
&+\text{tr}(F_{12} D _5 F_{23}) \text{tr}(D_1 F_{34} F_{45})\,,
\nonumber\\
&\mathcal{O}_{10;\alpha;\mathrm{d}+;5}=
-\frac{1}{4} \text{tr}(F_{12} D _5 F_{34})^2
-\frac{1}{4} \text{tr}(F_{12} D _5 F_{34}) \text{tr}(D_5 F_{12} F_{34})
-\frac{3}{2} \text{tr}(F_{12} D _5 F_{14}) \text{tr}(F_{23} D _5 F_{34})
\nonumber\\
&+\frac{1}{2} \text{tr}(F_{12} D _5 F_{23}) \text{tr}(F_{34} D _5 F_{14})
-\frac{1}{4} \text{tr}(F_{12} D _5 F_{12}) \text{tr}(F_{34} D _5 F_{34})\,.
\end{align}
\end{small}
The $\beta$-sector:
\begin{small}
\begin{align}
&\mathcal{O}_{10;\beta;\mathrm{d}+;1}=
D^{\mu\nu} \frac{1}{2}\mathrm{tr}(12)\mathrm{tr}(34)\circ
\mathcal{P}^{\beta5}_{\mu\nu}\,,
\nonumber\\
&\mathcal{O}_{10;\beta;\mathrm{d}+;2}=
 D^{\mu\nu}\mathrm{tr}(12)\mathrm{tr}(34)\circ
\Big(
-\mathcal{P}^{\beta3}_{\mu\nu}-\frac{1}{3}\mathcal{P}^{\beta5}_{\mu\nu}
\Big)\,,
\nonumber\\
&\mathcal{O}_{10;\beta;\mathrm{d}+;3}=
D^\mu\mathrm{tr}(12)\mathrm{tr}(34)\circ
\Big(
-\frac{1}{2}\mathcal{Q}^{\beta3}_{\mu}
+\frac{1}{4}D^\nu\big(
- \mathcal{P}^{\beta3}_{\mu\nu}
-\frac{1}{6}\mathcal{P}^{\beta5}_{\mu\nu}
+\mathcal{P}^{e2}_{\mu\nu}
\big)\Big)\,,
\nonumber\\
&\mathcal{O}_{10;\beta;\mathrm{d}+;4}=
-\frac{1}{6} \text{tr}(F_{12} F_{12}) \text{tr}(D_5 F_{34} D _5 F_{34})
+\frac{2}{3} \text{tr}(F_{13} F_{23}) \text{tr}(D_1 F_{45} D _2 F_{45})
-\frac{1}{3} \text{tr}(F_{13} F_{23}) \text{tr}(D_{12} F_{45} F_{45})\,.
\end{align}
\end{small}
where
\begin{small}
\begin{align}
&\mathcal{P}^{\beta3}_{\mu\nu}=
 \lfloor F_{12},F_{3\nu},F_{13},F_{2\mu}\rfloor
-\frac{1}{2}  \lfloor F_{12},F_{3\nu},F_{3\mu},F_{12}\rfloor
- \lfloor F_{2\mu},F_{13},F_{3\nu},F_{12}\rfloor
-\frac{1}{2}  \lfloor F_{3\nu},F_{12},F_{12},F_{3\mu}\rfloor
\nonumber\\
&+\frac{1}{4} g_{\mu\nu} \lfloor F_{34},F_{12},F_{12},F_{34}\rfloor
\,,
\nonumber\\
&\mathcal{P}^{\beta5}_{\mu\nu}=\mathcal{P}^{21}_{\mu\nu}
\big|_{\lfloor\mathcal{W}_1,\mathcal{W}_2,\mathcal{W}_3,\mathcal{W}_4 \rfloor
\rightarrow \lfloor\mathcal{W}_1,\mathcal{W}_3,\mathcal{W}_2,\mathcal{W}_4 \rfloor }\,,
\nonumber\\
&\mathcal{Q}^{\beta3}_{\mu}=
 \lfloor D_4 F_{12},F_{2\mu},F_{34},F_{13}\rfloor
- \lfloor D_4 F_{12},F_{3\mu},F_{13},F_{24}\rfloor
+ \lfloor D_4 F_{12},F_{3\mu},F_{24},F_{13}\rfloor
-\frac{1}{2}  \lfloor D_4 F_{2\mu},F_{12},F_{13},F_{34}\rfloor
\nonumber\\
&-\frac{1}{2}  \lfloor D_4 F_{2\mu},F_{12},F_{34},F_{13}\rfloor
-\frac{1}{2}  \lfloor D_4 F_{2\mu},F_{13},F_{12},F_{34}\rfloor
+\frac{1}{2}  \lfloor D_4 F_{2\mu},F_{13},F_{34},F_{12}\rfloor
+\frac{1}{2}  \lfloor D_4 F_{2\mu},F_{34},F_{12},F_{13}\rfloor
\nonumber\\
&-\frac{1}{2}  \lfloor D_4 F_{2\mu},F_{34},F_{13},F_{12}\rfloor
+\frac{1}{4}  \lfloor D_4 F_{3\mu},F_{12},F_{12},F_{34}\rfloor
-\frac{1}{4}  \lfloor D_4 F_{3\mu},F_{12},F_{34},F_{12}\rfloor
+\frac{1}{4} \lfloor D_4 F_{3\mu},F_{34},F_{12},F_{12}\rfloor \,.
\end{align}
\end{small}
The $\gamma$-sector:
\begin{small}
\begin{align}
&\mathcal{O}_{10;\gamma;\mathrm{d}+;1}=D^2\Big(
\frac{1}{4} \text{tr}(F_{12} F_{23}) \text{tr}(F_{34} F_{14})
+\frac{1}{16} \text{tr}(F_{12} F_{12}) \text{tr}(F_{34} F_{34})
\Big)\,,
\nonumber\\
&\mathcal{O}_{10;\gamma;\mathrm{d}+;2}=D^2\Big(
-\frac{1}{8} \text{tr}(F_{12} F_{34})^2
-\frac{1}{4} \text{tr}(F_{12} F_{34}) \text{tr}(F_{23} F_{14})
+\frac{1}{16} \text{tr}(F_{12} F_{12}) \text{tr}(F_{34} F_{34})
\Big)\,,
\nonumber\\
&\mathcal{O}_{10;\gamma;\mathrm{d}+;3}=
D^{\mu\nu}\mathrm{tr}(12)\mathrm{tr}(34)\circ
\Big(
\frac{1}{2}\mathcal{P}^{\gamma2}_{\mu\nu}-\frac{1}{4}\mathcal{P}^{e1}_{\mu\nu}
+\frac{1}{2}\mathcal{P}^{e2}_{\mu\nu}
\Big)\,,
\nonumber\\
&\mathcal{O}_{10;\gamma;\mathrm{d}+;4}=
\frac{1}{8} \text{tr}(F_{12} D _5 F_{34})^2
+\frac{1}{8} \text{tr}(F_{12} D _5 F_{34}) \text{tr}(D_5 F_{12} F_{34})
+\frac{1}{2} \text{tr}(F_{12} D _5 F_{34}) \text{tr}(F_{23} D _5 F_{14})
\nonumber\\
&-\frac{1}{8} \text{tr}(F_{12} D _5 F_{12}) \text{tr}(F_{34} D _5 F_{34})\,,
\nonumber\\
&\mathcal{O}_{10;\gamma;\mathrm{d}+;5}=
\frac{1}{2} \text{tr}(F_{12} D _5 F_{14}) \text{tr}(F_{23} D _5 F_{34})
+\frac{1}{2} \text{tr}(F_{12} D _5 F_{23}) \text{tr}(F_{34} D _5 F_{14})
+\frac{1}{4} \text{tr}(F_{12} D _5 F_{12}) \text{tr}(F_{34} D _5 F_{34})\,,
\nonumber\\
&\mathcal{O}_{10;\gamma;\mathrm{d}+;6}=
\frac{1}{8} (-7) \text{tr}(F_{12} D _5 F_{34})^2
+\frac{5}{8} \text{tr}(F_{12} D _5 F_{34}) \text{tr}(D_5 F_{12} F_{34})
-\frac{1}{2} \text{tr}(F_{12} D _5 F_{34}) \text{tr}(F_{23} D _5 F_{14})
\nonumber\\
&-2 \text{tr}(F_{12} D _5 F_{14}) \text{tr}(F_{23} D _5 F_{34})
+\frac{1}{2} \text{tr}(F_{13} D _1 F_{45}) \text{tr}(F_{23} D _2 F_{45})
+\text{tr}(F_{12} D _5 F_{23}) \text{tr}(F_{34} D _5 F_{14})
\nonumber\\
&-\frac{3}{8} \text{tr}(F_{12} D _5 F_{12}) \text{tr}(F_{34} D _5 F_{34})
-\text{tr}(F_{12} D _5 F_{23}) \text{tr}(D_1 F_{34} F_{45})\,.
\end{align}
\end{small}

\subsection{Dim-10 length-5  physical operators}

\label{app:dim10oper4d-len5}

For mass dimension 10, there are  4 single-trace and  3 double-trace
physical  operators with  length five. All of them are $C$-even operators.
Similar to the length-4 operators, they can be chosen to belong to
certain helicity sector.
For length-5 operators, we still use subscript $\alpha$, $\beta$, $\gamma$ to
label the helicity sector, but in length-5 case they correspond to
$(-)^5$, $(-)^4(+)$, $(-)^3(+)^2$, not as in length-4 case where they refer to
$(-)^4$, $(-)^3(+)$, $(-)^2(+)^2$.
We  omit the $(-\imI g)^3$ in front of each length-5 operator for simplicity.

\begin{align}
&\Xi_{10;\alpha;\mathrm{s}+;1}=
5 \text{tr}(F_{12} F_{12} F_{34} F_{35} F_{45})
-5 \text{tr}(F_{12} F_{13} F_{24} F_{35} F_{45})
+\text{tr}(F_{12} F_{13} F_{34} F_{45} F_{25})\,,
\nonumber\\
&\Xi_{10;\alpha;\mathrm{s}+;2}=
\frac{5}{2} \text{tr}(F_{12} F_{12} F_{34} F_{35} F_{45})
-3 \text{tr}(F_{12} F_{13} F_{24} F_{35} F_{45})
+\text{tr}(F_{12} F_{13} F_{24} F_{45} F_{35})\,,
\nonumber\\
&\Xi_{10;\gamma;\mathrm{s}+;1}=
\frac{3}{2} \text{tr}(F_{12} F_{12} F_{34} F_{35} F_{45})
-2 \text{tr}(F_{12} F_{13} F_{24} F_{35} F_{45})
-\text{tr}(F_{12} F_{13} F_{24} F_{45} F_{35})
\nonumber\\
&+\text{tr}(F_{12} F_{13} F_{34} F_{45} F_{25})\,,
\nonumber\\
&\Xi_{10;\gamma;\mathrm{s}+;2}=
\text{tr}(F_{12} F_{12} F_{34} F_{35} F_{45})
-2 \text{tr}(F_{12} F_{13} F_{24} F_{35} F_{45})
-2 \text{tr}(F_{12} F_{13} F_{24} F_{45} F_{35})\,.
\end{align}

\begin{align}
&\Xi_{10;\alpha;\mathrm{d}+;1}=
2 \text{tr}(F_{12} F_{34}) \text{tr}(F_{12} F_{35} F_{45})
-4 \text{tr}(F_{12} F_{13}) \text{tr}(F_{24} F_{35} F_{45})
+\text{tr}(F_{12} F_{12}) \text{tr}(F_{34} F_{35} F_{45})\,,
\nonumber\\
&\Xi_{10;\alpha;\mathrm{d}+;2}=
-4 \text{tr}(F_{12} F_{13}) \text{tr}(F_{24} F_{35} F_{45})
+\text{tr}(F_{12} F_{12}) \text{tr}(F_{34} F_{35} F_{45})\,,
\nonumber\\
&\Xi_{10;\alpha;\mathrm{d}+;3}=
-\frac{1}{2} \text{tr}(F_{12} F_{34}) \text{tr}(F_{12} F_{35} F_{45})
+\text{tr}(F_{12} F_{13}) \text{tr}(F_{24} F_{35} F_{45})
-\frac{1}{12} \text{tr}(F_{12} F_{12}) \text{tr}(F_{34} F_{35} F_{45})\,.
\nonumber\\
\end{align}

\subsection{One-loop renormalization of complete dim-10 operators}

\label{app:Znon-eva}

In this appendix, we provide the renormalization matrices and anomalous dimensions for physical operators of mass dimension 10.

The complete basis operators of mass dimension 10 can be separated into $C$-even and
$C$-odd sectors. Since $C$-even and $C$-odd sectors do not mix to each other,
the general form of one-loop renormalization matrix (\ref{eq:UVZ1}) can be written
for these two sectors separately:
\begin{align}
\label{eq:Z10+-}
&Z^{(1)}_{10;+}=
\left(\begin{array}{c|c}
Z^{(1) }_{10;\mathrm{p}+\rightarrow \mathrm{p}+} $\,$ &
$\,$ Z^{(1) }_{10;\mathrm{p}+\rightarrow \mathrm{e}+} \\[0.1em]
\hline\rule{0pt}{1\normalbaselineskip}
0 & $\,$ Z^{(1) }_{10;\mathrm{e}+\rightarrow \mathrm{e}+}
\end{array}
\right)\,,\quad
Z^{(1)}_{10;-}=
\left(\begin{array}{c|c}
Z^{(1) }_{10;\mathrm{p}-\rightarrow \mathrm{p}-} $\,$ &
$\,$ Z^{(1) }_{10;\mathrm{p}-\rightarrow \mathrm{e}-} \\[0.1em]
\hline\rule{0pt}{1\normalbaselineskip}
0 & $\,$ Z^{(1) }_{10;\mathrm{e}-\rightarrow \mathrm{e}-}
\end{array}
\right)\,.
\end{align}
The values of
$Z^{(1) }_{10;\mathrm{e}+\rightarrow \mathrm{e}+}$
and
$Z^{(1) }_{10;\mathrm{e}-\rightarrow \mathrm{e}-}$, characterizing the
renormalization of evanescent operators,
have been given in (\ref{eq:dim10evaZeven}) and (\ref{eq:dim10evaZodd}).
In this subsection we show the values of other sub-matrices
$Z^{(1) }_{10;\mathrm{p}+\rightarrow \mathrm{p}+}$,
$ Z^{(1) }_{10;\mathrm{p}+\rightarrow \mathrm{e}+}$,
$Z^{(1) }_{10;\mathrm{p}-\rightarrow \mathrm{p}-}$,
$ Z^{(1) }_{10;\mathrm{p}-\rightarrow \mathrm{e}-}$,
which are obtained from the renormalization of physical operators.

Similar to the notation of Section \ref{sec:calc-full},
the operators are understood as carrying a coupling factor,
for example,
$\mathcal{O}^e_{10;\mathrm{s}+;1}$ refers to $(-ig)^2\mathcal{O}^e_{10;\mathrm{s}+;1}$,
and $\Xi^e_{5;\mathrm{s}+;1}$ refers to $(-ig)^3\Xi^e_{5;\mathrm{s}+;1}$.

First, let us see the renormalization of $C$-odd sector, which is relatively simple.
We arrange  $C$-odd physical operators in following order:
\begin{align}
\label{eq:nevorder2}
\{\mathcal{O}_{10;\alpha;\mathrm{s}-;1},
\mathcal{O}_{10;\beta;\mathrm{s}-;1},
\mathcal{O}_{10;\beta;\mathrm{s}-;2},
\mathcal{O}_{10;\gamma;\mathrm{s}-;1},
\mathcal{O}_{10;\gamma;\mathrm{s}-;2}
\}\,.
\end{align}
As mentioned in Section \ref{sec:Zmatrix},
there is only one $C$-odd evanescent operator $\mathcal{O}^e_{10;\mathrm{s}-;1}$.
The renormalization matrix of the $C$-odd sector can be given as
\begin{align}
\label{eq:Z10-}
\left(\begin{array}{c|c}
Z^{(1)}_{10;\mathrm{p}-\rightarrow \mathrm{p}-} $\,$ &
$\,$ Z^{(1)}_{10;\mathrm{p}-\rightarrow \mathrm{e}-} \\[0.1em]
\hline\rule{0pt}{1\normalbaselineskip}
0 & $\,$ Z^{(1)}_{10;\mathrm{e}-\rightarrow \mathrm{e}-}
\end{array}
\right)
=
\frac{N_c}{\epsilon}
\left(
\begin{array}{c;{2pt/2pt}cc;{2pt/2pt}cc|c}
 \frac{16}{3} $\,$ & $\,$ 0 & 0 & $\,\,$ 0 & 0 & $\,\,$ \frac{1}{12} \\[0.3em]
 \hdashline[2pt/2pt] \rule{0pt}{0.9\normalbaselineskip}
 0 $\,$  & $\,$ \frac{17}{4} & 0 $\,$ & $\,\,$ 0 & 0 & $\,\,$ 0 \\[0.2em]
 0 $\,$ &$\,$  -1 $\,$ & \frac{25}{4} $\,$ & $\,\,$ 0 & 0 & $\,$ -\frac{1}{16} \\[0.3em]
 \hdashline[2pt/2pt] \rule{0pt}{1\normalbaselineskip}
 0 $\,$ & $\,$ 0 & 0 & $\,\,\,$ \frac{37}{10} & -\frac{1}{5} $\,$ & $\,\,$ \frac{19}{120} \\[0.3em]
 0 $\,$ & $\,$ 0 & 0 & $\,$ -\frac{3}{10} $\,$ & $\,$ \frac{82}{15}  & $\,\,$ \frac{13}{40} \\[0.3em]
 \hline \rule{0pt}{0.9\normalbaselineskip}
 0 $\,$ & $\,$ 0 & 0 & $\,\,$ 0 & 0 & $\,\,$ 5 \\[0.2em]
\end{array}
\right)\,.
\end{align}

Second, we consider the renormalization of $C$-even sector.
We divide the $C$-even physical operators into four groups:
length-4 $\alpha$-sector,
length-4 $\beta$-sector,
length-4 $\gamma$-sector, and length-5.
They are arranged in following order:
\begin{align}
\label{eq:nevorder1}
&4\alpha+:\quad \{\mathcal{O}_{10;\alpha;\mathrm{s}+;i},
\mathcal{O}_{10;\alpha;\mathrm{d}+;i}\} \,,
\nonumber\\
&4\beta+:\quad \{\mathcal{O}_{10;\alpha;\mathrm{s}+;i},
\mathcal{O}_{10;\alpha;\mathrm{d}+;i}\} \,,
\nonumber\\
&4\gamma+:\quad \{\mathcal{O}_{10;\alpha;\mathrm{s}+;i},
\mathcal{O}_{10;\alpha;\mathrm{d}+;i}\} \,,
\nonumber\\
&5p+:\quad \{ \Xi_{10;\alpha;\mathrm{s}+;i},
 \Xi_{10;\gamma;\mathrm{s}+;i},
 \Xi_{10;\alpha;\mathrm{d}+;i},
 \Xi_{10;\gamma;\mathrm{d}+;i},
\}\,.
\end{align}
The symbol $5p$ refers to length-5 physical operators, distinguished from
the length-5 evanescent operators listed in (\ref{eq:len5str}) and (\ref{eq:len5dtr}).
The arrangement of $C$-even evanescent operators is the same as given in Section \ref{sec:Zmatrix}:
\begin{align}
 \{\mathcal{O}^e_{10;\mathrm{s}+;i} ,\mathcal{O}^e_{10;\mathrm{d}+;i},
  \, \Xi^e_{10;\mathrm{s}+;i},  \, \Xi^e_{10;\mathrm{d}+;1}\} .
\end{align}

The full C-even renormalization matrix $Z^{(1)}_{10;+}$ in \eqref{eq:Z10+-} can be divided into sub-blocks:
\begin{align}
\label{eq:Z1sec+}
&Z^{(1)}_{10;\mathrm{p}+\rightarrow \mathrm{p}+}
=\left(
\begin{array}{c|c|c|c}
Z^{(1)}_{4\alpha +\rightarrow 4\alpha+} &
0 & 0 &
Z^{(1)}_{4\alpha+\rightarrow 5n+}
 \\[0.3em]
\hline \rule{0pt}{1\normalbaselineskip}
0 &
Z^{(1)}_{4\beta+\rightarrow 4\beta+} &
0 &
Z^{(1)}_{4\beta+\rightarrow 5n+}
 \\[0.3em]
\hline \rule{0pt}{1\normalbaselineskip}
0 & 0 &
Z^{(1)}_{4\gamma+\rightarrow 4\gamma+} &
Z^{(1)}_{4\gamma+\rightarrow 5n+}
 \\[0.3em]
\hline \rule{0pt}{1\normalbaselineskip}
0 & 0 & 0 & Z^{(1)}_{5n+\rightarrow 5n+}
 \\[0.3em]
\end{array}
\right),
\
Z^{(1)}_{10;\mathrm{p}+\rightarrow \mathrm{e}+}
=\left(
\begin{array}{ c}
Z^{(1)}_{4\alpha+\rightarrow e+}
 \\[0.3em]
Z^{(1)}_{4\beta+\rightarrow e+}
 \\[0.3em]
Z^{(1)}_{4\gamma+\rightarrow e+}
 \\[0.3em]
 Z^{(1)}_{5n+\rightarrow e+}
 \\[0.3em]
\end{array}
\right).
\end{align}
The  blocks of $Z^{(1)}_{10;\mathrm{p}+\rightarrow \mathrm{p}+}$ are
\begin{align}
\label{eq:Zaa}
Z^{(1)}_{4\alpha+\rightarrow 4\alpha+}=
\frac{N_c}{\epsilon}
\left(
\begin{array}{ccccc;{2pt/2pt}ccccc}
 0   & $\,\,$ \frac{5}{12} & 0 & $\,\,$ 0 & $\,\,$ 0 &
 0 & \frac{5}{N_c} & 0 & $\,\,$ 0 & 0 \\[0.2em]
 \frac{16}{3} & $\,\,$ \frac{17}{3} & 0 & $\,\,$ 0 & $\,\,$ 0 &
 $\,\,$ -\frac{64}{N_c} $\,\,$ & 0 & 0 & $\,\,$ 0 & 0 \\[0.2em]
 \frac{16}{3} & -\frac{5}{12} $\,\,$ & \frac{16}{3} & $\,\,$ 0 & $\,\,$ 0 &
  0 & -\frac{5}{N_c} $\,\,$ & 0 & $\,\,$ 0 & 0 \\[0.3em]
 0 & $\,\,$ \frac{1}{6} & \frac{2}{3} & $\,\,$ 8 & -\frac{2}{3} $\,$
 & $\,$ \frac{8}{N_c} & 0 & \frac{6}{N_c} & $\,\,$ 0 & -\frac{6}{N_c} \\[0.2em]
 \frac{8}{3} & -\frac{5}{12} $\,\,$ & \frac{2}{3} & -\frac{10}{3} & $\,\,$ \frac{14}{3}
 & 0 & 0 & 0 & $\,\,$ \frac{30}{N_c} & 0 \\[0.3em]
 \hdashline[2pt/2pt] \rule{0pt}{1\normalbaselineskip}
 -\frac{5}{N_c} $\,\,$ & -\frac{5}{4 N_c} & 0 & $\,\,$ 0 & $\,\,$ 0 &
  \frac{25}{3} & 0 & 0 & $\,\,$ 0 & 0 \\[0.3em]
$\,\,$ \frac{16}{N_c} & $\,\,$ \frac{4}{N_c} & 0 & $\,\,$ 0 & $\,\,$ 0 &
 0 & -\frac{11}{3} $\,\,$ & 0 & $\,\,$ 0 & 0 \\[0.3em]
$\,\,$ \frac{8}{N_c} & $\,\,$ \frac{5}{4 N_c} & \frac{3}{N_c} & $\,\,$ 0 & $\,\,$ 0 &
 0 & 0 & \frac{25}{3} & $\,\,$ 0 & 0 \\[0.3em]
 -\frac{2}{3 N_c} $\,$ & -\frac{7}{12 N_c} $\,$ & \frac{5}{6 N_c} & -\frac{25}{6 N_c} & -\frac{5}{6 N_c} $\,$
 & 0 & 0 & 0 & -\frac{11}{3} & 0 \\[0.3em]
$\,\,$ \frac{22}{3 N_c} & -\frac{5}{6 N_c} $\,$ & \frac{41}{6 N_c} & -\frac{115}{6 N_c} & -\frac{23}{6 N_c} $\,$
 & 0 & 0 & 0 & $\,\,$ 0 & \frac{25}{3}
   \\[0.2em]
\end{array}
\right)\,,
\end{align}
\begin{align}
\label{eq:Za5}
Z^{(1)}_{4\alpha+\rightarrow 5p+}=
\frac{N_c}{\epsilon}
\left(
\begin{array}{cccc;{2pt/2pt}ccc}
$\,\,$  0 & 0 & $\,\,$  0 & $\,\,\,$  0 $\,\,$ &
 $\,\,$  0 & 0 & $\,\,\,$  0 \\
$\,\,$  0 & 0 & $\,\,$  0 & $\,\,\,$  0 $\,\,$ &
 $\,\,$  0 & 0 & $\,\,\,$  0 \\
$\,\,$  0 & 0 & $\,\,$  0 & $\,\,\,$  0 $\,\,$ &
 $\,\,$  0 & 0 & $\,\,\,$  0 \\[0.2em]
 -\frac{17}{3} & 8 & $\,\,$ 0 & $\,\,\,$  0 $\,\,$
 & $\,\,$  -\frac{8}{N_c} & 0 & $\,\,\,$  0 \\[0.3em]
 -\frac{26}{3} & 10 & $\,\,$ 0 & $\,\,\,$  0 $\,\,$
 & $\,\,$  -\frac{15}{N_c} & 0 & $\,\,\,$  0 \\[0.4em]
\hdashline[2pt/2pt] \rule{0pt}{0.9\normalbaselineskip}
$\,\,$  0 & 0 & $\,\,$ 0 & $\,\,\,$  0 $\,\,$  &
 $\,\,$  0 & 0 & $\,\,\,$  0 \\
$\,\,$  0 & 0 & $\,\,$ 0 & $\,\,\,$  0 $\,\,$  &
 $\,\,$  0 & 0 & $\,\,\,$  0 \\
$\,\,$  0 & 0 & $\,\,$ 0 & $\,\,\,$  0 $\,\,$  &
 $\,\,$  0 & 0 & $\,\,\,$  0 \\[0.2em]
 -\frac{65}{6 N_c} & \frac{20}{N_c} & $\,\,$ 0 & $\,\,\,$  0 $\,\,$  &
 $\,\,$  0 & 0 & $\,\,\,$  0 \\[0.3em]
 -\frac{83}{6 N_c} & \frac{20}{N_c} & $\,\,$ 0 & $\,\,\,$  0 $\,\,$  &
 $\,\,$  0 & 0 & $\,\,\,$  0 \\[0.2em]
\end{array}
\right)
\end{align}
\begin{align}
\label{eq:Zbb}
Z^{(1)}_{4\beta+\rightarrow 4\beta+}=
\frac{N_c}{\epsilon}
\left(
\begin{array}{cccc;{2pt/2pt}cccc}
$\,\,$  5 & -\frac{3}{4} $\,\,$ & 0 $\,$ & $\,$ 0 $\,\,$
 &  $\,\,\,$ 0 & \frac{6}{N_c} & $\,$ 0 & $\,\,$ 0 \\[0.2em]
 -\frac{5}{3} $\,\,$ & $\,$ \frac{9}{4}  & 0 $\,$ & $\,$ 0 $\,\,$
 & $\,\,$ -\frac{40}{3 N_c} $\,$ & 0 & $\,$ 0 & $\,\,$ 0 \\[0.2em]
 -\frac{5}{24} $\,\,$ & -\frac{3}{8} $\,\,$ & \frac{21}{4} $\,$ & $\,$ 0 $\,\,$
 & $\,\,$ -\frac{5}{3 N_c} $\,$ & 0 & $\,$ 0 & $\,\,$ 0 \\[0.2em]
 $\,$ \frac{1}{6} & $\,$ \frac{1}{2} & -1 $\,\,$ & $\,$ 6 $\,\,$
 & $\,\,\,$ \frac{4}{3 N_c} & 0 & $\,$ 0 & $\,\,$ 0 \\[0.4em]
\hdashline[2pt/2pt] \rule{0pt}{0.9\normalbaselineskip}
 \frac{5}{3 N_c} & -\frac{5}{4 N_c} $\,\,$ & 0 $\,$ & $\,$ 0 $\,\,$
 & $\,\,$ 0 & 0 & $\,$ 0 & $\,\,$ 0 \\[0.2em]
 \frac{35}{9 N_c} & -\frac{35}{12 N_c} $\,\,$ & 0 $\,$ & $\,$ 0 $\,\,$
 &  $\,\,$ 0 & 6 & $\,$ 0 & $\,\,$ 0 \\[0.2em]
 \frac{35}{72 N_c} & -\frac{11}{24 N_c} $\,\,$ & $\,$ \frac{3}{4 N_c} $\,$ & $\,$ 0 $\,\,$
 &  $\,\,$ 0 & 0 & $\,$ 6 & $\,\,$ 0 \\[0.2em]
 -\frac{7}{18 N_c} $\,\,$ & $\,$ \frac{1}{6 N_c} & $\,$ \frac{1}{N_c} $\,$ & $\,$ 0  $\,\,$
 &  $\,\,$ 0 & 0 & $\,$ 0 & $\,\,$ 6 \\[0.2em]
\end{array}
\right)\,,
\end{align}
\begin{align}
\label{eq:Zb5}
Z^{(1)}_{4\beta+\rightarrow 5p+}=
\frac{N_c}{\epsilon}
\left(
\begin{array}{cccc;{2pt/2pt}ccc}
$\,\,$ 0 & 0 &  $\,$ 0 & $\,\,$ 0 $\,\,$ &
 $\,\,\,$ 0 & 0 &  $\,\,\,$ 0 \\
$\,\,$ 0 & 0 & $\,$ 0 & $\,\,$ 0 $\,\,$ &
 $\,\,\,$ 0 & 0 &  $\,\,\,$ 0 \\
$\,\,$ 0 & 0 & $\,$ 0 & $\,\,$ 0 $\,\,$ &
 $\,\,\,$ 0 & 0 &  $\,\,\,$ 0 \\[0.2em]
 -\frac{46}{9} & \frac{22}{3} & $\,$ \frac{2}{3} & $\,\,$ \frac{4}{9} $\,\,$ &
 $\,$ -\frac{10}{3 N_c} & \frac{49}{9 N_c} & -\frac{20}{N_c} \\[0.4em]
\hdashline[2pt/2pt] \rule{0pt}{0.9\normalbaselineskip}
$\,\,$ 0 & 0 & $\,$ 0 & $\,\,$ 0 $\,\,$ &
 $\,\,\,$ 0 & 0 &  $\,\,\,$ 0 \\
$\,\,$ 0 & 0 & $\,$ 0 & $\,\,$ 0 $\,\,$ &
 $\,\,\,$ 0 & 0 &  $\,\,\,$ 0 \\
$\,\,$ 0 & 0 & $\,$ 0 & $\,\,$ 0 $\,\,$ &
 $\,\,\,$ 0 & 0 &  $\,\,\,$ 0 \\[0.2em]
 -\frac{6}{N_c} & \frac{12}{N_c} & $\,$ 0 & $\,\,$ 0 $\,\,$ &
 $\,\,\,$ \frac{11}{6} & \frac{2}{3} &  $\,\,\,$ \frac{22}{3} \\[0.2em]
\end{array}
\right)\,,
\end{align}
\begin{align}
\label{eq:Zcc}
Z^{(1)}_{4\gamma+\rightarrow 4\gamma+}=
\frac{N_c}{\epsilon}
\left(
\begin{array}{cccccc;{2pt/2pt}cccccc}
$\,\,$ \frac{1}{2} & $\,\,$ \frac{1}{2} & 0 & 0 & 0 & 0 &
 $\,\,\,$ 0 & $\,\,$ \frac{22}{3 N_c} &  0 & $\,\,$ 0 & $\,\,$ 0 & 0 \\[0.2em]
$\,\,$ 1 & $\,\,$ \frac{14}{3} & 0 & 0 & 0 & 0 &
$\,\,$ -\frac{44}{3 N_c} & $\,\,$ 0 & 0 & $\,\,$ 0 & $\,\,$ 0 & 0 \\[0.2em]
$\,\,$ 4 & -\frac{1}{2} & $\,\,$ \frac{9}{2} & 0 & 0 & 0 &
$\,\,\,$ 0 & -\frac{22}{3 N_c} & 0 & $\,\,$ 0 & $\,\,$ 0 & 0 \\[0.3em]
 -\frac{1}{15} & $\,\,$ \frac{2}{5} & $\,\,$ \frac{1}{60} & \frac{112}{15} & -\frac{1}{6} & $\,\,$ \frac{1}{15} &
$\,\,\,$ \frac{44}{9 N_c} & $\,\,$ 0 & \frac{11}{18 N_c} & $\,\,$ 0 & -\frac{55}{9 N_c} & -\frac{20}{9 N_c} \\[0.3em]
 -\frac{26}{15} & $\,\,$ \frac{3}{20} & -\frac{1}{15} & -\frac{1}{5} & $\,\,$ \frac{31}{6} & -\frac{4}{15} $\,\,$ &
$\,\,\,$  0 & $\,\,$ \frac{11}{3 N_c} & 0 & $\,\,$ \frac{22}{3 N_c} & $\,\,$ 0 & $\,\,$ \frac{1}{N_c} \\[0.3em]
$\,\,$ \frac{64}{15} & $\,\,$ \frac{2}{5} & $\,\,$ \frac{14}{15} & $\,\,$ \frac{14}{5} & -\frac{28}{3} & $\,\,$ \frac{67}{30} &
$\,\,\,$ \frac{16}{N_c} & $\,\,$ \frac{40}{3N_c} & \frac{7}{2 N_c} & $\,\,$ \frac{280}{3 N_c} & -\frac{35}{N_c} & $\,\,$ 0 \\[0.4em]
\hdashline[2pt/2pt]  \rule{0pt}{0.9\normalbaselineskip}
 -\frac{25}{6 N_c} $\,$ & -\frac{25}{6 N_c} & 0 & 0 & 0 & 0 &
 $\,\,\,$ \frac{11}{3} & $\,\,$ 0 & 0 & $\,\,$ 0 & $\,\,$ 0 & $\,\,$ 0 \\[0.2em]
 -\frac{1}{N_c} & -\frac{1}{N_c} & 0 & 0 & 0 & 0 &
 $\,\,\,$ 0 & -\frac{11}{3} & 0 & $\,\,$ 0 & $\,\,$ 0 & $\,\,$ 0 \\[0.2em]
$\,\,$ \frac{10}{3 N_c} & $\,\,$ \frac{25}{6 N_c} & -\frac{5}{6 N_c} & 0 & 0 & 0 &
 $\,\,\,$ 0 & $\,\,$ 0 & \frac{11}{3} & $\,\,$ 0 & $\,\,$ 0 & $\,\,$ 0 \\[0.3em]
$\,\,$ \frac{7}{5 N_c} & $\,\,$ \frac{1}{10 N_c} & $\,\,$ \frac{11}{40 N_c} &
-\frac{3}{10 N_c} & -\frac{11}{4 N_c} & \frac{7}{20 N_c}  &
 $\,\,\,$ 0 & $\,\,$ 0 & 0 & -\frac{11}{3} & $\,\,$ 0 & $\,\,$ 0 \\[0.3em]
 -\frac{37}{45 N_c} $\,$ & -\frac{149}{60 N_c} & $\,\,$ \frac{29}{360 N_c} &
  -\frac{209}{30 N_c} & -\frac{59}{36 N_c} & -\frac{137}{180  N_c} $\,\,$ &
$\,\,\,$  0 & $\,\,$ 0 & 0 & $\,\,$ 0 & $\,\,$ \frac{11}{3} & $\,\,$ 0 \\[0.3em]
$\,\,$ \frac{188}{45 N_c} & -\frac{137}{30 N_c} & $\,\,$ \frac{329}{360 N_c}  &
-\frac{959}{30 N_c} & -\frac{329}{36 N_c} &   -\frac{587}{180 N_c} $\,\,$ &
   $\,\,\,$ 0 & $\,\,$ 0 & 0 & $\,\,$ 0 & $\,\,$ 0 & $\,\,$ \frac{29}{3} \\[0.2em]
\end{array}
\right)\,,
\end{align}
\begin{align}
\label{eq:Zc5}
Z^{(1)}_{4\gamma+\rightarrow 5p+}=
\frac{N_c}{\epsilon}
\left(
\begin{array}{cccc;{2pt/2pt}ccc}
 0 $\,\,\,$ & 0 & $\,\,$ 0 & $\,\,$ 0 & $\,\,\,$ 0 & $\,\,$ 0 & $\,\,$ 0 \\
 0 $\,\,\,$ & 0 & $\,\,$ 0 & $\,\,$ 0 & $\,\,\,$ 0 & $\,\,$ 0 & $\,\,$ 0 \\
 0 $\,\,\,$ & 0 & $\,\,$ 0 & $\,\,$ 0 & $\,\,\,$ 0 & $\,\,$ 0 & $\,\,$ 0 \\[0.2em]
 0 $\,\,\,$ & 0 & -\frac{1}{5} & -\frac{1}{15} &
 $\,\,\,$ 0 & -\frac{31}{6 N_c} & $\,\,$ 0 \\[0.3em]
 0 $\,\,\,$ & 0 & -\frac{6}{5} & -\frac{7}{30} &
 $\,\,\,$ 0 & -\frac{3}{2 N_c} & -\frac{8}{N_c} \\[0.3em]
 0 $\,\,\,$ & 0 & -\frac{197}{10} & -\frac{169}{60} &
 $\,\,\,$ 0 & -\frac{5}{6 N_c} & -\frac{110}{N_c} \\[0.4em]
\hdashline[2pt/2pt]  \rule{0pt}{0.9\normalbaselineskip}
 0 $\,\,\,$ & 0 & $\,\,$ 0 & $\,\,$ 0 & $\,\,\,$ 0 & $\,\,$ 0 & $\,\,$ 0 \\
 0 $\,\,\,$ & 0 & $\,\,$ 0 & $\,\,$ 0 & $\,\,\,$ 0 & $\,\,$ 0 & $\,\,$ 0 \\
 0 $\,\,\,$ & 0 & $\,\,$ 0 & $\,\,$ 0 & $\,\,\,$ 0 & $\,\,$ 0 & $\,\,$ 0 \\[0.2em]
 0 $\,\,\,$ & 0 & -\frac{111}{20 N_c} & -\frac{29}{40 N_c} &
 $\,\,\,$ 0 & $\,\,$ 0 & $\,\,$ 0 \\[0.3em]
 0 $\,\,\,$ & 0 & -\frac{31}{20 N_c} & $\,\,$ \frac{859}{360 N_c} &
 $\,\,\,$ 0 & $\,\,$ 0 & $\,\,$ 0 \\[0.3em]
 0 $\,\,\,$ & 0 & -\frac{701}{20 N_c} & -\frac{671}{360 N_c} $\,\,$ &
 $\,\,\,$ 0 & $\,\,$ 0 & $\,\,$ 0 \\[0.2em]
\end{array}
\right)\,,
\end{align}
\begin{align}
\label{eq:Z55}
Z^{(1)}_{5p+\rightarrow 5p+}
= \frac{N_c}{\epsilon}
\left(
\begin{array}{cc;{2pt/2pt}cc|c;{2pt/2pt}cc}
 -\frac{11}{3} & 10 & $\,\,$ 0 & 0 & $\,\,$ -\frac{35}{N_c} $\,\,$ & $\,\,$ 0 & 0 \\[0.3em]
 -9 & \frac{49}{3} & $\,\,$ 0 & 0 & $\,\,$ -\frac{25}{N_c} $\,\,$ & $\,\,$ 0 & 0 \\[0.4em]
\hdashline[2pt/2pt]  \rule{0pt}{0.9\normalbaselineskip}
 0 & 0 & $\,\,$ \frac{9}{2} & \frac{1}{4} & $\,\,$ 0 & $\,\,$ \frac{1}{2 N_c} & \frac{10}{N_c} \\[0.3em]
 0 & 0 & $\,\,$ 1 & \frac{37}{6} & $\,\,$ 0 & $\,\,$ -\frac{13}{3 N_c} & 0 \\[0.3em]
 \hline \rule{0pt}{1\normalbaselineskip}
 \frac{12}{N_c} & -\frac{24}{N_c} $\,$ & $\,\,$ 0 & 0 & $\,\,$ \frac{7}{3} & $\,\,$ 0 & 0 \\[0.4em]
\hdashline[2pt/2pt]  \rule{0pt}{1\normalbaselineskip}
 0 & 0 & $\,\,$ -\frac{20}{3 N_c} & -\frac{10}{3 N_c} $\,\,$ & $\,\,$ 0 & $\,\,$ \frac{20}{3} & 0 \\[0.3em]
 0 & 0 & $\,\,$ -\frac{1}{2 N_c} & -\frac{1}{4 N_c} $\,\,$ & $\,\,$ 0 & $\,\,$ 0 &  \frac{7}{3}   \\[0.3em]
\end{array}
\right)\,.
\end{align}
%
The  blocks of $Z^{(1)}_{10;\mathrm{p}+\rightarrow \mathrm{e}+}$ are
\begin{align}
\label{eq:Zae}
Z^{(1)}_{4\alpha+\rightarrow e+}=
\frac{N_c}{\epsilon}
\left(
\begin{array}{ccc;{2pt/2pt}ccc|cc;{2pt/2pt}c}
$\,\,$ 0 & $\,\,$ 0 & 0 & $\,\,$ 0 & $\,\,$ 0 & 0 &
 $\,\,\,\,$ 0 &  $\,\,\,$ 0 &  $\,\,\,$ 0 \\
$\,\,$ 0 & $\,\,$ 0 & 0 & $\,\,$ 0 & $\,\,$ 0 & 0 &
 $\,\,\,\,$  0 &  $\,\,\,$ 0 &  $\,\,\,$ 0 \\[0.2em]
 -\frac{1}{3} & $\,\,$ -\frac{1}{4} & 0 & $\,\,$ -\frac{5}{6 N_c} & $\,\,$ \frac{4}{3 N_c} & 0 &
  $\,\,\,\,$ 0 &  $\,\,\,$ 0 &  $\,\,\,$ 0 \\[0.3em]
 -\frac{41}{72} & -\frac{23}{36} & -\frac{7}{12} $\,\,$ & $\,$ -\frac{11}{12 N_c} & $\,\,$ \frac{5}{12 N_c} & -\frac{3}{2 N_c} &
 $\,\,\,\,$ \frac{65}{18} & $\,\,\,$ \frac{31}{18} &  $\,\,\,$ \frac{17}{3 N_c} \\[0.3em]
$\,\,$ \frac{19}{72} & $\,\,$ \frac{13}{36} & \frac{5}{12} & $\,$ -\frac{37}{12 N_c} & -\frac{65}{12 N_c} & -\frac{15}{2 N_c} &
  $\,$ -\frac{11}{18} &  -\frac{23}{9} &  $\,\,\,$ \frac{41}{N_c} \\[0.4em]
 \hdashline[2pt/2pt]\rule{0pt}{0.9\normalbaselineskip}
$\,\,$ 0 & $\,\,$ 0 & 0 & $\,\,$ 0 & $\,\,$ 0 & 0 &
  $\,\,\,\,$ 0 &  $\,\,\,$ 0 &  $\,\,\,$ 0 \\
$\,\,$ 0 & $\,\,$ 0 & 0 & $\,\,$ 0 & $\,\,$ 0 & 0 &
  $\,\,\,\,$ 0 &  $\,\,\,$ 0 &  $\,\,\,$ 0 \\[0.2em]
 -\frac{1}{4 N_c} & -\frac{11}{12 N_c} & 0 & -\frac{1}{6} & $\,\,$ \frac{1}{6} & 0 &
  $\,\,\,\,$ 0 &  $\,\,\,$ 0 &  $\,\,\,$ 0 \\[0.3em]
$\,\,$ \frac{5}{18 N_c} & $\,\,$ \frac{1}{18 N_c} & \frac{4}{3 N_c} & $\,\,$ \frac{2}{3} & $\,\,$ \frac{7}{3} & 3 &
  $\,$ -\frac{10}{9 N_c} & $\,\,\,$ \frac{31}{36 N_c}  $\,$ &  $\,\,$ -12 \\[0.3em]
$\,\,$ \frac{37}{36 N_c} & $\,\,$ \frac{20}{9 N_c} & \frac{10}{3 N_c} & $\,\,$ \frac{1}{3} & -\frac{1}{3} $\,\,$ & 0 &
  $\,$ -\frac{46}{9 N_c} &  -\frac{491}{36 N_c} $\,$ &  $\,\,\,$ 0 \\[0.2em]
\end{array}
\right)\,,
\end{align}
\begin{align}
\label{eq:Zbe}
Z^{(1)}_{4\beta+\rightarrow e+}=
\frac{N_c}{\epsilon}
\left(
\begin{array}{ccc;{2pt/2pt}ccc|cc;{2pt/2pt}c}
 -\frac{1}{18} & $\,\,\,$ \frac{13}{72} & 0 & -\frac{1}{9 N_c} & -\frac{19}{18 N_c} & $\,\,$ 0 &
 $\,\,\,$ 0 & $\,\,\,$ 0   & 0 \\[0.3em]
 -\frac{1}{54} & $\,\,\,$ \frac{49}{216} & 0 & $\,\,$ \frac{5}{27 N_c} & -\frac{38}{27 N_c} &  $\,\,$  0 &
 $\,\,\,$  0 & $\,\,\,$ 0   & 0 \\[0.3em]
 -\frac{143}{864} & -\frac{41}{432} & 0 & $\,$ -\frac{85}{216 N_c} &  $\,\,$  \frac{53}{108 N_c} &  $\,\,$  0 &
 $\,\,\,$  0 &  $\,\,\,$ 0   & 0 \\[0.3em]
$\,\,$ \frac{43}{216} & $\,\,\,$ \frac{7}{27} & \frac{1}{4} &  $\,\,$ \frac{17}{54 N_c} & -\frac{59}{54 N_c} & -\frac{7}{6 N_c}  $\,$  &
 $\,\,\,$ \frac{10}{27} &  -\frac{46}{27}   &  $\,\,$ \frac{8}{9 N_c} \\[0.4em]
\hdashline[2pt/2pt]  \rule{0pt}{1\normalbaselineskip}
$\,\,$ 0 & $\,\,\,$ \frac{7}{8 N_c} & 0 & $\,\,\,$ 0 &  $\,\,\,$  0 &  $\,\,$  0 &
 $\,\,\,$ 0 &  $\,\,\,$ 0   & 0 \\[0.3em]
$\,\,$ 0 & $\,\,\,$ \frac{1}{8 N_c} & 0 & -\frac{1}{9} & -\frac{19}{18} &  $\,\,$  0 &
 $\,\,\,$ 0 &  $\,\,\,$ 0   & 0 \\[0.3em]
 -\frac{1}{32 N_c} & -\frac{19}{48 N_c} & 0 & $\,\,\,$  0 &  $\,\,\,$  0 &  $\,\,$  0 &
  $\,\,\,$ 0 &  $\,\,\,$ 0   & 0 \\[0.3em]
 -\frac{1}{24 N_c} & $\,\,\,$ \frac{5}{6 N_c} & \frac{11}{12 N_c} $\,$ & $\,\,\,$  0 &  $\,\,\,$  0 &  $\,\,$  0 &
  $\,\,\,$ 0 &  - \frac{13}{3 N_c}   $\,$ & 0 \\[0.2em]
\end{array}
\right)\,,
\end{align}
\label{eq:Zce}
\begin{align}
Z^{(1)}_{4\gamma+\rightarrow e+}=
\frac{N_c}{\epsilon}
\left(
\begin{array}{ccc;{2pt/2pt}ccc|cc;{2pt/2pt}c}
 $\,\,$  0 &  $\,\,$  0 &  $\,\,$  0 & 0 & 0 & 0 &
  $\,\,\,$  0 & $\,\,\,$ 0 &  $\,\,\,\,$  0 \\
 $\,\,$  0 &  $\,\,$  0 &  $\,\,$  0 & 0 & 0 & 0 &
 $\,\,\,$   0 & $\,\,\,$ 0 &  $\,\,\,\,$  0 \\[0.2em]
 -\frac{19}{18} & -\frac{29}{72} &  $\,\,$  0 & -\frac{29}{6 N_c} & $\,\,\,$ \frac{8}{3 N_c} & 0 &
 $\,\,\,$   0 & $\,\,\,$ 0 &  $\,\,\,\,$  0 \\[0.3em]
 -\frac{1}{240} & -\frac{1}{80} & -\frac{1}{60} & $\,\,$ -\frac{65}{216 N_c} & -\frac{19}{216 N_c} & -\frac{1}{12 N_c} $\,$ &
   -\frac{11}{20}  &  $\,\,\,\,$ \frac{1}{10} & $\,\,$ -\frac{17}{6 N_c} \\[0.3em]
 -\frac{1}{240} &  $\,\,\,$  \frac{1}{20} &  $\,\,\,$  \frac{1}{40} & $\,\,$ \frac{7}{72 N_c} & -\frac{1}{72 N_c} & \frac{1}{12 N_c} &
 $\,\,\,$   \frac{281}{180} &   -\frac{1}{90}     &  $\,\,\,\,\,$  \frac{3}{2 N_c} \\[0.3em]
 -\frac{27}{80} & -\frac{11}{80} & -\frac{19}{40} & -\frac{107}{72 N_c} & -\frac{145}{72 N_c} & \frac{7}{12 N_c} &
  $\,\,\,$   \frac{277}{60} & -\frac{63}{20} & $\,$ -\frac{149}{6 N_c} \\[0.4em]
\hdashline[2pt/2pt]  \rule{0pt}{1\normalbaselineskip}
 $\,\,$  0 &  $\,\,$  0 &  $\,\,$  0 & 0 & 0 & 0 &
 $\,\,\,$   0 & $\,\,$ 0 &  $\,\,\,\,$  0 \\
 $\,\,$  0 &  $\,\,$  0 &  $\,\,$  0 & 0 & 0 & 0 &
 $\,\,\,$   0 & $\,\,$ 0 &  $\,\,\,\,$  0 \\[0.2em]
 $\,\,$  \frac{5}{36 N_c} & -\frac{103}{72 N_c} &  $\,\,$  0 & \frac{3}{2} & \frac{5}{6} & 0 &
 $\,\,\,$  0 &  $\,\,$ 0 &  $\,\,\,\,$  0 \\[0.3em]
 -\frac{1}{60 N_c} & -\frac{1}{20 N_c} &  $\,\,$  \frac{1}{10 N_c} & 0 & 0 & 0 &
  $\,\,\,$  \frac{21}{20 N_c} & -\frac{227}{120 N_c} $\,$ &  $\,\,\,\,$ 0 \\[0.3em]
 -\frac{83}{1080 N_c} & -\frac{83}{2160 N_c} & -\frac{13}{120 N_c}  $\,$  & 0 & 0 & 0 &
$\,$  -\frac{17}{60 N_c} $\,$ & $\,\,$ \frac{1139}{360 N_c} $\,$ &  $\,\,\,\,$  0 \\[0.3em]
 -\frac{229}{540 N_c} &  $\,\,$  \frac{941}{1080 N_c} &  $\,\,$  \frac{47}{120 N_c} & \frac{8}{9} & -\frac{7}{18}  $\,\,$  & \frac{1}{2} &
  $\,\,\,$  \frac{151}{20 N_c} & -\frac{4471}{360 N_c} $\,$ & $\,\,$ -2 \\[0.2em]
\end{array}
\right)\,,
\end{align}
\begin{align}
\label{eq:Z5e}
Z^{(1)}_{5p+\rightarrow e+}
=\frac{N_c}{\epsilon}
\left(
\begin{array}{cccccc|cc;{2pt/2pt}c}
$\,$ 0 & $\,$ 0 & $\,$ 0 & $\,$ 0 & $\,$ 0 & $\,$ 0 $\,\,$ &
$\,\,\,$ \frac{25}{18} & -\frac{55}{18} & $\,\,\,$ \frac{115}{3 N_c} \\[0.3em]
$\,$ 0 & $\,$ 0 & $\,$ 0 & 0 & $\,$ 0 & $\,$ 0 $\,\,$ &
$\,\,\,$ \frac{10}{9} & - \frac{19}{9} & $\,\,\,$ \frac{65}{3 N_c} \\[0.3em]
$\,$ 0 & $\,$ 0 & $\,$ 0 & 0 & $\,$ 0 & $\,$ 0 $\,\,$ &
$\,\,\,\,$ \frac{1}{6} & -\frac{5}{6} & $\,\,\,$ \frac{53}{6 N_c} \\[0.3em]
$\,$ 0 & $\,$ 0 & $\,$ 0 & 0 & $\,$ 0 & $\,$ 0 $\,\,$ &
  -\frac{2}{9} & - \frac{4}{9} & $\,\,\,$ \frac{7}{N_c} \\[0.4em]
 \hline \rule{0pt}{0.9\normalbaselineskip}
$\,$ 0 & $\,$ 0 & $\,$ 0 & $\,$ 0 & $\,$ 0 & $\,$ 0 $\,\,$ &
$\,\,$  0 & -\frac{14}{3 N_c} $\,$ &  $\,\,\,$  \frac{14}{3} \\[0.3em]
$\,$ 0 & $\,$ 0 & $\,$ 0 & $\,$ 0 & $\,$ 0 & $\,$ 0 $\,\,$ &
$\,\,$ \frac{20}{9 N_c} & - \frac{65}{9 N_c} $\,$ &  $\,\,\,$  0 \\[0.3em]
$\,$ 0 & $\,$ 0 & $\,$ 0 & $\,$ 0 & $\,$ 0 & $\,$ 0 $\,\,$ &
$\,\,$ \frac{1}{6N_c} & $\,\,$ \frac{1}{  N_c}   & -\frac{7}{6} \\[0.2em]
\end{array}
\right)\,.
\end{align}

The anomalous dimensions are given as the eigenvalues of one-loop dilation operators
defined in (\ref{eq:Dila1}).
The existence of non-zero $Z^{(1)}_{\mathrm{p}+\rightarrow \mathrm{e}+}$
and $Z^{(1)}_{\mathrm{p}-\rightarrow \mathrm{e}-}$
do not effect the eigenvalues because at the order of $\mathcal{O}(\epsilon^{-1})$
there is no mixing in the opposite direction.
Similarly, the mixing from length-4 operators to length-5 operators do not
effect the eigenvalues.
The eigenvalues determined by sub-matrices
$Z^{(1)}_{\mathrm{e}+\rightarrow \mathrm{e}+}$
and $Z^{(1)}_{\mathrm{e}-\rightarrow \mathrm{e}-}$ are
given in (\ref{eq:einodd})-\eqref{eq:einevenL5}.
Here we give the eigenvalues determined by  sub-matrices
$Z^{(1)}_{\mathrm{p}+\rightarrow \mathrm{p}+}$ and
$Z^{(1)}_{\mathrm{p}-\rightarrow \mathrm{p}-}$.
Furthermore, at one-loop order, different helicity sectors
do not mix to each other, so the eigenvalues can be obtained
from the sub-matrices of each helicty sector separately.


Denote the eigenvalues of $C$-odd length-4 $\alpha,\beta,\gamma$-sector by
$\hat{\gamma}_{-;\alpha}^{(1)}$, $\hat{\gamma}_{-;\beta}^{(1)}$, $\hat{\gamma}_{-;\gamma}^{(1)}$.
They are determined by $Z^{(1)}_{10;\mathrm{p}-\rightarrow \mathrm{p}-}$ given in (\ref{eq:Z10-}):
\begin{align}
\hat{\gamma}^{(1)}_{-;\alpha}=\{\frac{32 N_c}{3}\},\quad
\hat{\gamma}^{(1)}_{-;\beta}=\{\frac{17 N_c}{2},\frac{25 N_c}{2}\},\quad
\hat{\gamma}^{(1)}_{-;\gamma}=\{11 N_c,\frac{22 N_c}{3}\}.
\end{align}


Denote the eigenvalues of $C$-even length-4 $\alpha,\beta,\gamma$-sector by
$\hat{\gamma}_{+;\alpha}^{(1)}$, $\hat{\gamma}_{+;\beta}^{(1)}$, $\hat{\gamma}_{+;\gamma}^{(1)}$.
Respectively they are determined by sub-matrices $Z^{(1)}_{10;4\alpha+\rightarrow 4\alpha+}$,
$Z^{(1)}_{10;4\beta+\rightarrow 4\beta+}$,
$Z^{(1)}_{10;4\gamma+\rightarrow 4\gamma+}$ given in (\ref{eq:Zaa}), (\ref{eq:Zbb}), (\ref{eq:Zcc}).
Denote the eigenvalues of $C$-even length-5 sector by $\hat{\gamma}_{+;5}^{(1)}$,
They are determined by $Z^{(1)}_{10;5p+\rightarrow 5p+}$ given in (\ref{eq:Z55}).
For convenience we introduce $\einv_h=\hat{\gamma}_{+,h}^{(1)}/N_c$ where
$h=\alpha,\beta,\gamma,5$.

For $C$-even length-4 $\alpha$-sector,
$\einv_\alpha$ are roots of equation
\begin{small}
\begin{align}
0&=(\omega_{\alpha }-\frac{50}{3}) (\omega_{\alpha }-\frac{32}{3})
\times \Big(\omega _{\alpha }^4-\frac{104 \omega_{\alpha }^3}{3}
+\omega_{\alpha }^2 (\frac{764}{3}-\frac{360}{N_c^2})
+\omega_{\alpha }(\frac{48208}{27}-\frac{2400}{N_c^2})
-(\frac{1390400}{81}-\frac{73760}{N_c^2})
\Big)
\nonumber\\
&\times\Big(\omega_{\alpha}^4-\frac{62 \omega_{\alpha }^3}{3}
-\omega_{\alpha }^2 (\frac{76}{3}+\frac{640}{N_c^2})
+\omega_{\alpha } (\frac{39640}{27}+\frac{14080}{3 N_c^2})
+(\frac{88000}{81}-\frac{486400}{9 N_c^2})
\Big)\,.
\end{align}
\end{small}
The order four factorized polynomial in the second line
is inherited from the $\alpha$-sector with mass dimension 8.

Expand eigenvalues up to $\mathcal{O}(N_c^{-1})$:
\begin{align}
\hat{\gamma}^{(1)}_{+;\alpha}=N_c
\Big\{ & (\frac{17}{3}\pm\sqrt{41})  -\frac{20 (41\pm19 \sqrt{41})}{41N_c^2},
(\frac{38}{3}\pm2\sqrt{5}) +\frac{18 (360\pm173 \sqrt{5})}{19 N_c^2},
\frac{32  }{3},
\nonumber\\
&\frac{50  }{3},\frac{50  }{3}-\frac{690}{N_c^2},
\frac{50 }{3}+\frac{80}{N_c^2},-\frac{22  }{3}-\frac{40}{N_c^2},
-\frac{22  }{3}+\frac{150}{19 N_c^2}
\Big\}.
\end{align}
The sub-leading $N_c$ correction breaks the triple degeneracy
of $50N_c/3$ and double degeneracy of $-22N_c/3$.


For $C$-even length-4 $\beta$-sector, $\einv_\beta$ are roots of equation
\begin{align}
0=(\omega _{\beta }-12){}^3 (\omega _{\beta }-\frac{21}{2})
\Big(\omega_{\beta }^4
-\frac{53 \omega_{\beta }^3}{2}
+\omega_{\beta }^2 (214-\frac{160}{N_c^2})
-\omega_{\beta } (480-\frac{1520}{N_c^2})
-\frac{6400}{N_c^2}
\Big).
\end{align}
Expand eigenvalues up to $\mathcal{O}(N_c^{-1})$:
\begin{align}
\hat{\gamma}^{(1)}_{+;\beta}&=N_c
\Big\{
\frac{1}{4} (29\pm\sqrt{201}) -\frac{40 (201\pm19 \sqrt{201})}{201 N_c^2},
\frac{21  }{2},12  ,12  ,12  ,12  +\frac{280}{3 N_c^2},-\frac{40}{3 N_c^2}
\Big\}\,.
\end{align}
The sub-leading $N_c$ correction breaks the quartet degeneracy
of $12N_c $ to triple degeneracy.


For $C$-even length-4 $\gamma$-sector,
$\einv_\gamma$ are roots of equation
\begin{small}
\begin{align}
0&=
(\omega _{\gamma }-9) (\omega _{\gamma }-\frac{22}{3})
\times\Big(
\omega _{\gamma }^4-\frac{31 \omega _{\gamma }^3}{3}
-(\frac{418}{9}+\frac{1936}{9 N_c^2}) \omega_{\gamma }^2
+(\frac{15004}{27}-\frac{60016}{27 N_c^2}) \omega_{\gamma }
-(\frac{10648}{27}-\frac{42592}{27 N_c^2})
\Big)
\nonumber\\
&\times \Big(
\omega_{\gamma }^6-\frac{736 \omega_{\gamma }^5}{15}
+(\frac{3887}{5}-\frac{5170}{9 N_c^2}) \omega_{\gamma }^4
-(\frac{384932}{135}-\frac{42736}{3 N_c^2}) \omega _{\gamma }^3
-(\frac{13919744}{405}+\frac{7114772}{81 N_c^2}) \omega_{\gamma }^2
\nonumber\\
&+(\frac{119573168}{405}-\frac{13628032}{81 N_c^2}) \omega_{\gamma }
-(\frac{224969008}{405}-\frac{156611312}{81 N_c^2})
\Big)\,.
\end{align}
\end{small}
The order four factorized polynomial in the first line
is inherited from the $\gamma$-sector with mass dimension 8.

Expand eigenvalues up to $\mathcal{O}(N_c^{-1})$:
\begin{align}
\hat{\gamma}^{(1)}_{+;\gamma}&=N_c
\Big\{
9  ,
\frac{1}{6} (31\pm\sqrt{697})  +\frac{121 (21607\pm811 \sqrt{697})}{43911 N_c^2},
x_i   +\frac{y_i}{N_c^2},
\nonumber\\
&
\frac{22  }{3}-\frac{23260}{8559 N_c^2},
\frac{22  }{3},
\frac{22  }{3}-\frac{1100}{9 N_c^2},
\frac{58  }{3}+\frac{2176405}{32454 N_c^2},
-\frac{22  }{3}+\frac{22}{7 N_c^2},
-\frac{22  }{3}-\frac{897}{175 N_c^2}
\Big\}\,.
\end{align}
There are three pairs of $\{x_i,y_i\}$, where
$x_1,x_2,x_3$ and $y_1,y_2,y_3$ are respectively the roots of equations
\begin{align}
&
0=x^3-\frac{446 x^2}{15}+\frac{769 x}{3}-\frac{8014}{15},
\nonumber\\
&
0=y^3+\frac{35538354643 y^2}{600128550}+\frac{1310963227092295265 y}{2216037779059699}-\frac{40910844565828125}{633153651159914}\,.
\end{align}
The sub-leading $N_c$ correction breaks the triple degeneracy
of $22N_c/3$ and double degeneracy of $-22N_c/3$.


For $C$-even length-5 sector, $\einv_5$ are roots of equation
\begin{align}
0&=
\Big(
\omega_5^3-30 \omega_5^2
+\omega_5 (\frac{716}{3}-\frac{720}{N_c^2})
-(\frac{15176}{27}-\frac{12000}{N_c^2})
\Big)
\nonumber\\
&\times\Big(
\omega_5^4-\frac{118 \omega_5^3}{3}
+\omega_5^2 (\frac{5006}{9}-\frac{220}{9 N_c^2})
-\omega_5 (\frac{89300}{27}-\frac{680}{27 N_c^2})
+(\frac{61600}{9}+\frac{42400}{27 N_c^2})
\Big)
\end{align}
The order three factorized polynomial in the first line belongs to
the $\alpha$-sector of length-5, and
order four factorized polynomial in the second line belongs to
the $\gamma$-sector of length-5.
Expand eigenvalues up to $\mathcal{O}(N_c^{-1} )$:
\begin{align}
\hat{\gamma}^{(1)}_{+;5}&=N_c
\Big\{
(\frac{38}{3}\pm2 \sqrt{10})  +\frac{180\mp54 \sqrt{10}}{N_c^2},
\frac{14  }{3}-\frac{360}{N_c^2},
\nonumber\\
&\frac{1}{3} (32\pm\sqrt{34})  -\frac{5 (39304\pm6887 \sqrt{34})}{4437 N_c^2},
\frac{40  }{3}+\frac{760}{9 N_c^2},
\frac{14  }{3}+\frac{120}{29 N_c^2}
\Big\}\,.
\end{align}
The sub-leading $N_c$ correction breaks the double degeneracy of $14N_c/3$.

\subsection{Finite renormalization as a scheme change}
\label{app:finiteZ}

As mentioned in (\ref{eq:schemchange}),
One can choose another renormalization scheme by appending finite terms to $Z^{(1)}$
to absorb  the mixing from the evanescent operators
to the physical ones at the order of $\mathcal{O}(\epsilon^{0})$.
Discussing $C$-even and $C$-odd sectors separately,
the finite terms added to $Z^{(1)}_{10;+}$ and  $Z^{(1)}_{10;-}$ in (\ref{eq:Z10+-}) are
\begin{align}
\label{eq:Z10+-2}
&Z^{(1),\mathrm{fin}}_{10;+}=
\left(\begin{array}{c|c}
0 & $\,\,\,$ 0 $\,\,\,$ \\[0.1em]
\hline\rule{0pt}{1\normalbaselineskip}
 Z^{(1),\mathrm{fin}}_{10;\mathrm{e}+\rightarrow \mathrm{p}+} $\,$ & $\,\,\,$ 0 $\,\,\,$
\end{array}
\right)\,,\quad
Z^{(1),\mathrm{fin}}_{10;-}=
\left(\begin{array}{c|c}
0 & $\,\,\,$ 0 $\,\,\,$ \\[0.1em]
\hline\rule{0pt}{1\normalbaselineskip}
 Z^{(1),\mathrm{fin}}_{10;\mathrm{e}-\rightarrow \mathrm{p}-} $\,$ & $\,\,\,$ 0 $\,\,\,$
\end{array}
\right)\,.
\end{align}
In this subsection we give the values of $Z^{(1),\mathrm{fin}}_{10;\mathrm{e}+\rightarrow \mathrm{p}+}$
and $Z^{(1),\mathrm{fin}}_{10;\mathrm{e}-\rightarrow \mathrm{p}-}$.

\begin{align}
Z^{(1),\mathrm{fin}}_{10;\mathrm{e}-\rightarrow \mathrm{p}-}=
N_c
\left(
\begin{array}{ccccc}
 -\frac{8}{3} & 0 & -3 & \frac{4}{3} & -\frac{14}{3} \\
\end{array}
\right)\,,
\quad
Z^{(1),\mathrm{fin}}_{10;\mathrm{e}+\rightarrow \mathrm{p}+}=
\left(
\begin{array}{cccc}
 Z^{(1),\mathrm{fin}}_{4e+\rightarrow 4\alpha+} &
 Z^{(1),\mathrm{fin}}_{4e+\rightarrow 4\beta+}  &
 Z^{(1),\mathrm{fin}}_{4e+\rightarrow 4\gamma+} &
 Z^{(1),\mathrm{fin}}_{4e+\rightarrow 5p+} \\[0.2em]
 0 & 0 & 0 &
 Z^{(1),\mathrm{fin}}_{5e+\rightarrow 5p+} \\
\end{array}
\right)
\end{align}
where
\begin{align}
 Z^{(1),\mathrm{fin}}_{4e+\rightarrow 4\alpha+}=
 N_c
\left(
\begin{array}{ccccc;{2pt/2pt}ccccc}
$\,\,$ \frac{16}{3} & -\frac{2}{3} & $\,\,$ 0 & 0 & 0 &
 $\,\,$ -\frac{32}{3 N_c} & -\frac{8}{3 N_c} & 0 & 0 & 0 \\[0.3em]
$\,\,$ 0 & $\,\,\,$ 1 & $\,\,$ 0 & 0 & 0 &
$\,\,\,$ -\frac{32}{3 N_c} & $\,\,$ \frac{10}{3 N_c} & 0 & 0 & 0 \\[0.2em]
$\,\,$ 0 & -\frac{2}{3} & $\,\,$ 0 & -\frac{4}{3} $\,\,$ & \frac{4}{3} &
 $\,\,$ 0 & -\frac{2}{N_c} & -\frac{8}{N_c} &
 -\frac{8}{N_c} & \frac{8}{N_c} \\[0.4em]
 \hdashline[2pt/2pt] \rule{0pt}{0.9\normalbaselineskip}
$\,\,$ \frac{16}{3 N_c} & $\,\,$ \frac{4}{3 N_c} & $\,\,$ 0 & 0 & 0 &
 $\,\,$ 0 & -2 & 0 & 0 & 0 \\[0.3em]
 -\frac{16}{3 N_c} $\,\,$ & -\frac{4}{3 N_c} & $\,\,$ 0 & 0 & 0 &
 $\,\,$ 0 &  -2 & 0 & 0 & 0 \\[0.2em]
$\,\,$ 0 & $\,\,$ \frac{5}{3 N_c} & -\frac{4}{N_c} & \frac{40}{3 N_c} & \frac{8}{3 N_c} $\,$&
$\,\,$  0 & $\,\,\,$ 2 & 0 & 16 & 0 \\[0.2em]
\end{array}
\right)\,,
\end{align}
\begin{align}
 Z^{(1),\mathrm{fin}}_{4e+\rightarrow 4\beta+}=
 N_c
 \left(
\begin{array}{cccc;{2pt/2pt}cccc}
$\,\,$ \frac{10}{3} & -2 & $\,\,$ 0 &  0 &
$\,\,$ -\frac{160}{9 N_c} & $\,\,\,$ \frac{40}{3 N_c} & $\,\,$ 0 & 0 \\[0.3em]
 -\frac{1}{3} & $\,\,$ 0 & $\,\,$ 0 & 0 &
 $\,\,$ -\frac{10}{9 N_c} & -\frac{2}{3 N_c} & 0 & 0 \\[0.3em]
$\,\,$ \frac{2}{9} & -\frac{3}{2} & $\,\,$ \frac{10}{3} & -\frac{14}{3} $\,\,$ &
 -\frac{2}{N_c} & -\frac{26}{3 N_c} & \frac{112}{3 N_c} &  -\frac{28}{N_c} \\[0.4em]
 \hdashline[2pt/2pt] \rule{0pt}{1\normalbaselineskip}
$\,\,$ \frac{80}{9 N_c} & -\frac{20}{3 N_c} & $\,\,$ 0 & 0 &
 -\frac{50}{9} & \frac{14}{3} & 0 & 0 \\[0.3em]
 -\frac{52}{9 N_c} $\,\,$ & $\,\,\,$ \frac{13}{3 N_c} & $\,\,$ 0 & 0 &
  -\frac{50}{9} & \frac{14}{3} & 0 & 0 \\[0.3em]
$\,\,$ \frac{13}{3 N_c} & -\frac{17}{6 N_c} & -\frac{10}{3 N_c} & 0 & $\,\,\,$ \frac{50}{9}
 & 0 & -\frac{56}{3} $\,\,\,$ & \frac{28}{3} \\[0.2em]
\end{array}
\right)\,,
\end{align}
\begin{align}
 Z^{(1),\mathrm{fin}}_{4e+\rightarrow 4\gamma+}=
 N_c
 \left(
\begin{array}{cccccc;{2pt/2pt}cccccc}
$\,\,$ \frac{16}{3} & -\frac{14}{3} & $\,\,\,$ \frac{14}{3} & 0 & $\,\,$ 0 & $\,\,$  0 &
$\,\,$ -\frac{32}{3 N_c} & $\,\,$ \frac{52}{3 N_c} & -\frac{28}{N_c} & $\,\,$ 0 & $\,\,$ 0 &  0 \\[0.3em]
 -\frac{2}{3} & $\,\,\,$ \frac{7}{3} & -\frac{1}{3} & 0 & $\,\,$  0 & $\,\,$  0 &
 $\,\,$ -\frac{32}{3 N_c} & $\,\,$ \frac{16}{3 N_c} & $\,\,$ 0 & $\,\,$ 0 & $\,\,$ 0 & 0 \\[0.3em]
$\,\,$ \frac{26}{9} & $\,\,\,$ 0 & $\,\,\,$ \frac{1}{9} & \frac{28}{3} & -\frac{58}{9} & -\frac{14}{9} &
$\,\,\,\,$ \frac{248}{9 N_c} & $\,\,$ \frac{4}{N_c} & $\,\,$ \frac{40}{9 N_c} &
$\,\,$ \frac{160}{3 N_c} & -\frac{400}{9 N_c} & -\frac{80}{9 N_c} \\[0.4em]
 \hdashline[2pt/2pt] \rule{0pt}{0.9\normalbaselineskip}
$\,\,$ \frac{16}{3 N_c} & -\frac{26}{3 N_c} & $\,\,$ \frac{14}{N_c} & 0 & $\,\,$  0 & $\,\,$  0 &
 $\,\,\,$ 0 & $\,\,$ 4 & -\frac{28}{3} & $\,\,$ 0 & $\,\,$ 0 & 0 \\[0.3em]
 -\frac{14}{3 N_c} & $\,\,\,$ \frac{13}{3 N_c} & -\frac{9}{N_c} & 0 & $\,\,$  0 & $\,\,$  0 &
$\,\,\,$  0 & $\,\,$ 4 & -\frac{28}{3} & $\,\,$ 0 & $\,\,$ 0 & 0 \\[0.3em]
 -\frac{70}{9 N_c} & -\frac{2}{N_c} & $\,\,$ \frac{19}{9 N_c} &
 \frac{4}{3 N_c} & $\,$ \frac{194}{9 N_c} & -\frac{26}{9 N_c} $\,$ &
   -\frac{56}{3} & -4 & $\,\,$ 0 & -16 & $\,\,$ \frac{112}{3} & 0 \\[0.2em]
\end{array}
\right)\,,
 \end{align}
\begin{align}
Z^{(1),\mathrm{fin}}_{4e+\rightarrow 5p+}=
N_c
\left(
\begin{array}{cccc;{2pt/2pt}ccc}
$\,\,$ 0 & $\,\,$ 0 &   0 & 0   & $\,\,\,$ 0 & $\,\,$  0 & $\,\,$ 0 \\
$\,\,$ 0 & $\,\,$ 0 &   0 & 0 & $\,\,\,$ 0 & $\,\,$ 0 & $\,\,$ 0 \\[0.2em]
- \frac{182}{9} & $\,\,$ \frac{92}{3} & -\frac{50}{3} $\,$ & -4 $\,$ &
$\,$ - \frac{88}{3 N_c} & -\frac{2}{3 N_c} & -\frac{100}{N_c} \\[0.4em]
\hdashline[2pt/2pt] \rule{0pt}{0.9\normalbaselineskip}
$\,\,$ 0 & $\,\,$ 0 &   0 & 0 & $\,\,\,$ 0 & $\,\,$ 0 & $\,\,$ 0 \\
$\,\,$ 0 & $\,\,$ 0 &   0 & 0 & $\,\,\,$ 0 & $\,\,$ 0 & $\,\,$ 0 \\[0.2em]
$\,\,$ \frac{12}{N_c} & - \frac{56}{3 N_c} $\,$ &  \frac{110}{3 N_c} &  \frac{23}{9 N_c}  $\,$ &
 $\,\,\,$   \frac{35}{3} & $\,\,$ \frac{20}{3} & $\,\,$  \frac{140}{3} \\[0.2em]
\end{array}
\right)\,,
\end{align}
\begin{align}
Z^{(1),\mathrm{fin}}_{5e+\rightarrow 5p+}=
\left(
\begin{array}{cccc;{2pt/2pt}ccc}
$\,\,\,$  \frac{14}{3} & -\frac{17}{3} $\,\,$  & $\,$ \frac{10}{3} & -\frac{1}{2} $\,\,$ &
 $\,\,\,$ \frac{2}{3 N_c} & $\,\,\,$  \frac{59}{6 N_c} & $\,$ \frac{34}{N_c} \\[0.3em]
-  \frac{35}{6} & $\,\,$ \frac{25}{3} $\,\,$  & -\frac{5}{3} $\,$ & $\,\,\,$ \frac{5}{4} $\,\,$ &
$\,$ -\frac{25}{3 N_c} & -\frac{25}{6 N_c} & $\,$ \frac{25}{N_c} \\[0.4em]
\hdashline[2pt/2pt] \rule{0pt}{0.9\normalbaselineskip}
 -\frac{2}{3 N_c} $\,\,$  & \frac{4}{3 N_c} & \frac{4}{3 N_c} & $\,\,$  \frac{2}{3 N_c} $\,\,$ &
 $\,\,\,\,$ \frac{7}{4} & $\,\,\,$  4 &    7 \\[0.2em]
\end{array}
\right)\,.
\end{align}

\subsection{Discussion on $N_c=2$}

\label{app:Nequalto2}

So far our discussion are for general $N_c$, which is the rank of gauge group.
Here we discuss the special $N_c=2$ case, which was briefly mentioned in the end of Section \ref{sec:step2}.

In the $N_c=2$ case, single-trace color factors
and double-trace color factors have linear relation
\begin{align}
\label{eq:N2}
2\mathrm{tr}(jklm)=\mathrm{tr}(jk)\mathrm{tr}(lm)
-\mathrm{tr}(jl)\mathrm{tr}(km)+\mathrm{tr}(jm)\mathrm{tr}(kl)\,.
\end{align}
In this case all single-trace length-4 $C$-odd and length-5 basis operators become zero,
while the single-trace  length-4 $C$-even basis operators
are equal to combinations of double-trace ones
\begin{align}
\label{eq:N2b}
\mathcal{O}_{s+;i}\xlongrightarrow{N_c=2}
\sum_{k }  M^i_{\ k}\mathcal{O}_{d;k}\,,
\end{align}
where $M^i_{\ k}$ depends on the basis choice.
As a result, the independent set of the complete dim-10 evanescent operators
only contains three double-trace length-4 ones
which are given in (\ref{eq:dtr-dim10}).

The special property of operators basis in $N_c=2$  provides a consistency check of
the renormalization matrix obtained for general $N_c$.
One can first write out the renormalization of a single-trace operator $\mathcal{O}_{s+;i}$ and then
replace the  single-trace operators  $\mathcal{O}_{s+;j}$  appearing in the mixing pattern with the double-trace ones
according to  (\ref{eq:N2b}):
\begin{align}
\mathcal{O}_{s+;i;R}-\mathcal{O}_{s+;i;B}
&\xlongrightarrow{N_c=2}
\alpha_s\sum_{j}
(Z^{(1)}_{s\rightarrow s})^i_{\ j} \sum_{k}M^j_{\ k}\mathcal{O}_{d;k}
+\alpha_s\sum_{k}
(Z^{(1)}_{s\rightarrow d})^i_{\ k} \mathcal{O}_{d;k}
+\mathcal{O}(\alpha_s^2)
\nonumber
\end{align}
On the other hand, one can first rewrite $\mathcal{O}_{s+;i}$ in terms of
double-trace operators $\mathcal{O}_{d;k}$ according to  (\ref{eq:N2b})
 and then write out the renormalization of $\mathcal{O}_{d;k}$:
\begin{align}
\mathcal{O}_{s+;i;R}-\mathcal{O}_{s+;i;B}
&\xlongrightarrow{N_c=2}
\sum_k M^i_{\ k}\Big(
\alpha_s
\sum_j (Z^{(1)}_{d\rightarrow s})^k_{\ j} \sum_l M^j_{\ l}
\mathcal{O}_{d;l}
+\alpha_s
\sum_l(Z^{(1)}_{d\rightarrow d})^k_{\ l} \mathcal{O}_{d;l}
+\mathcal{O}(\alpha_s^2)
\Big)\,.
\nonumber
\end{align}
The two processes should yield the same result, which means
\begin{align}
\label{eq:N2check}
N_c=2:\quad
(Z^{(1)}_{s\rightarrow s})\cdot M+(Z^{(1)}_{s\rightarrow d})
=M\cdot (Z^{(1)}_{d\rightarrow s})\cdot M+M\cdot (Z^{(1)}_{d\rightarrow d})\,.
\end{align}
This provides a consistency condition.
Take the dim-10 length-4 evanescent sector as an example,
under the basis choice given by (\ref{eq:changebase}),
the matrix $M$ in (\ref{eq:N2b}) is given as
\begin{align}
\begin{pmatrix}
\mathcal{O}^e_{10;\mathrm{s}+;1} \\
\mathcal{O}^e_{10;\mathrm{s}+;2}\\
\mathcal{O}^e_{10;\mathrm{s}+;3}
\end{pmatrix}
\xlongrightarrow{N_c=2}
\begin{pmatrix}
1 & 0 & 0\\
0 & -2 & 0\\
0 & 0 & -2
\end{pmatrix}
\begin{pmatrix}
\mathcal{O}^e_{10;\mathrm{d}+;1} \\
\mathcal{O}^e_{10;\mathrm{d}+;2}\\
\mathcal{O}^e_{10;\mathrm{d}+;3}
\end{pmatrix}\,,
\end{align}
Here we  skip the $C$-odd operators because as mentioned before they vanishes when $N_c=2$.
One can see this $M$ matrix together with
$Z^{(1)}_{10;\mathrm{eva};+}$ given in (\ref{eq:dim10evaZeven})
satisfies equation (\ref{eq:N2check}).


\providecommand{\href}[2]{#2}\begingroup\raggedright\endgroup

\end{document}